\documentclass[acmsmall]{acmart}

\usepackage{amsmath,amsfonts}
\usepackage[utf8]{inputenc}
\usepackage{algorithmic}
\usepackage{todonotes}
\usepackage{graphicx}
\usepackage{textcomp}
\usepackage{svg}
\usepackage{xcolor}
\usepackage{longtable}
\usepackage{listings}
\usepackage{multirow}
\usepackage{lineno}
\def\BibTeX{{\rm B\kern-.05em{\sc i\kern-.025em b}\kern-.08em
    T\kern-.1667em\lower.7ex\hbox{E}\kern-.125emX}}
    
\usepackage{flushend}
\usepackage{tcolorbox}
\usepackage{color}
\usepackage{xspace}
\usepackage{tabularx}
\usepackage{booktabs}
\usepackage{longtable}
\usepackage{subcaption}
\usepackage{url}
\usepackage{tikz}
\usepackage[inline, shortlabels]{enumitem}
\usepackage[edges]{forest}
\usetikzlibrary{shadows,arrows.meta}
\tikzset{
  parent/.style={align=center,text width=4cm,fill=gray!50,rounded corners=2pt},
  child/.style={align=center,text width=2.5cm,fill=gray!20,rounded corners=6pt},
  grandchild/.style={fill=white,text width=2.3cm}
}

\usepackage{hyperref}

\usepackage{makecell}
\usepackage{pifont}%
    
\newboolean{showcomments}
\setboolean{showcomments}{true}
\ifthenelse{\boolean{showcomments}}
 { \newcommand{\mynote}[2]{
      \fbox{\bfseries\sffamily\scriptsize#1}
        {\small$\blacktriangleright$\textsf{\emph{#2}}$\blacktriangleleft$}}}
        { \newcommand{\mynote}[2]{}}

\definecolor{codegreen}{rgb}{0,0.6,0}
\definecolor{codegray}{rgb}{0.5,0.5,0.5}
\definecolor{codepurple}{rgb}{0.58,0,0.82}
\definecolor{backcolour}{rgb}{0.95,0.95,0.92}

\setcopyright{none}

\begin{document}
\author{Li Li}
\affiliation{%
  \institution{Beihang University}
  \country{China}}

\author{Xiang Gao}
\affiliation{%
  \institution{Beihang University}
  \country{China}}

\author{Hailong Sun}
\affiliation{%
  \institution{Beihang University}
  \country{China}}

\author{Chunming Hu}
\affiliation{%
  \institution{Beihang University}
  \country{China}}

\author{Xiaoyu Sun}
\affiliation{%
 \institution{The Australian National University}
 \country{Australia}
 }

\author{Haoyu Wang}
\affiliation{%
 \institution{Huazhong University of Science and Technology}
 \country{China}
 }

 \author{Haipeng Cai}
\affiliation{%
 \institution{Washington State University, Pullman}
 \country{USA}
 }

\author{Ting Su}
\affiliation{%
 \institution{East China Normal University}
 \country{China}
 }

\author{Xiapu Luo}
\affiliation{%
 \institution{The Hong Kong Polytechnic University}
 \country{China}
 }

 \author{Tegawend\'e F. Bissyand\'e}
\affiliation{
  \institution{University of Luxembourg}
  \country{Luxembourg}
}

 \author{Jacques Klein}
\affiliation{
  \institution{University of Luxembourg}
  \country{Luxembourg}
}

\author{John Grundy }
\affiliation{%
  \institution{Monash University}
  \country{Australia}
}

\author{Tao Xie}
\affiliation{%
  \institution{Peking University}
  \country{China}
}

\author{Haibo Chen}
\affiliation{%
  \institution{Shanghai Jiao Tong University}
  \country{China}
}

\author{Huaimin Wang}
\affiliation{%
  \institution{National University of Defense Technology}
  \country{China}
}

\title{Software Engineering for OpenHarmony: A Research Roadmap}

\begin{abstract}
Mobile software engineering has been a hot research topic for decades.
Our fellow researchers have proposed various approaches (with over 7,000 publications for Android alone) in this field that essentially contributed to the great success of the current mobile ecosystem.
Existing research efforts mainly focus on popular mobile platforms, namely Android and iOS.
OpenHarmony, a newly open-sourced mobile platform, has rarely been considered, although it is the one requiring the most attention as OpenHarmony is expected to occupy one-third of the market in China (if not in the world).
To fill the gap, we present to the mobile software engineering community a research roadmap for encouraging our fellow researchers to contribute promising approaches to OpenHarmony.
Specifically, we start by presenting a literature review of mobile software engineering, attempting to understand what problems have been targeted by the mobile community and how they have been resolved.
We then summarize the existing (limited) achievements of OpenHarmony and subsequently highlight the research gap between Android/iOS and OpenHarmony.
This research gap eventually helps in forming the roadmap for conducting software engineering research for OpenHarmony.
\end{abstract}

\maketitle

\noindent

\section{Introduction}

Mobile Software Engineering has been a hot topic for many years. It concerns all the aspects of software engineering in mobile, including the design, development, validation, execution, and evolution of mobile applications.
This has been considered extremely important as nowadays our lives have been empowered by the massive increase in the use of mobile apps. 
Indeed, the number of mobile devices will reach 7 billion in 2023.
The number of mobile apps that can be run on each mobile device (for both Android and iOS) has exceeded the 2 million mark.
Furthermore, these figures are constantly increasing, thanks to app stores and marketplaces that allow users to effortlessly download and install applications. 

Mobile platforms are rapidly evolving as well in order to continuously integrate diverse and powerful capabilities, including various sensors, cameras, wireless communication channels, as well as on-device memory and disk capacities.
As a result of ingeniously applying these technological developments, developers of mobile software are pushing the boundaries with innovative mobile services and exciting mobile applications. Consequently, due to the rapid development and evolution of mobile software, developers face new software engineering challenges. 

To address these challenges, researchers in the software engineering community have explored various research directions and developed lots of novel tools supported by formally grounded methods.
Indeed, researchers have proposed various static program analysis approaches (i.e., by just scanning the code without actually running mobile apps) for characterizing issues (including ones related to mobile security, compatibility, energy consumption, etc.) of mobile apps~\cite{li2017static}.
For example, Arzt et al.~\cite{arzt2014flowdroid} have designed and developed the famous FlowDroid approach that performs static taint analysis of Android apps for pinpointing privacy leaks.
Except static analysis approaches, researchers have also invented various dynamic testing approaches (i.e., by actually running mobile apps on devices) for detecting potential defects of mobile apps at runtime~\cite{kong2018automated}.
For example, Amalfitano et al.~\cite{amalfitano2012using} have proposed an GUI ripping approach for automated testing of Android apps. Su et al.~\cite{su2017guided} have proposed to achieve the same purpose through a model-based approach.
The aforementioned research approaches have contributed to the huge success of the current flourishing mobile ecosystem, including both Android and iOS.

Unfortunately, these approaches cannot directly benefit OpenHarmony\footnote{https://www.openharmony.cn}, which is a new open-sourced mobile platform launched by the OpenAtom Foundation after receiving a donation of the open-source code from Huawei.
These approaches, theoretically, should be generic and hence should also work for OpenHarmony.
However, there still require significant engineering efforts to achieve that due to the following reasons (more details will be given in the background section):
(1) the Openharmony platform empowers a new framework supported by layered architecture,
(2) Openharmony apps are written by a newly designed language called ArkTS. 

Unlike Android and iOS, which have been mature for many years and each has a healthy ecosystem to support their growth, the development of the Openharmony ecosystem is still at an earlier stage.
We, therefore, argue that OpenHarmony requires more help from the software engineering research community.\footnote{This could be regarded as new opportunities for the mobile software engineering community.}
We call on actions for conducting software engineering research for OpenHarmony.
As our initial attempt, we decide to present to the community an initial research roadmap for guiding our mobile software engineering community to achieve that.
We start by conducting a lightweight literature review of Mobile Software Engineering to understand the current achievements.
We then conduct a comparative study to locate the technical gaps between mature platforms and OpenHarmony.
Based on that, we summarize the technical deficiencies of OpenHarmony and propose a roadmap for our research community to complete.


\section{Background of OpenHarmony}

\subsection{Overview}

\begin{figure}[!t]
  \centering
  \includegraphics[width=\linewidth]{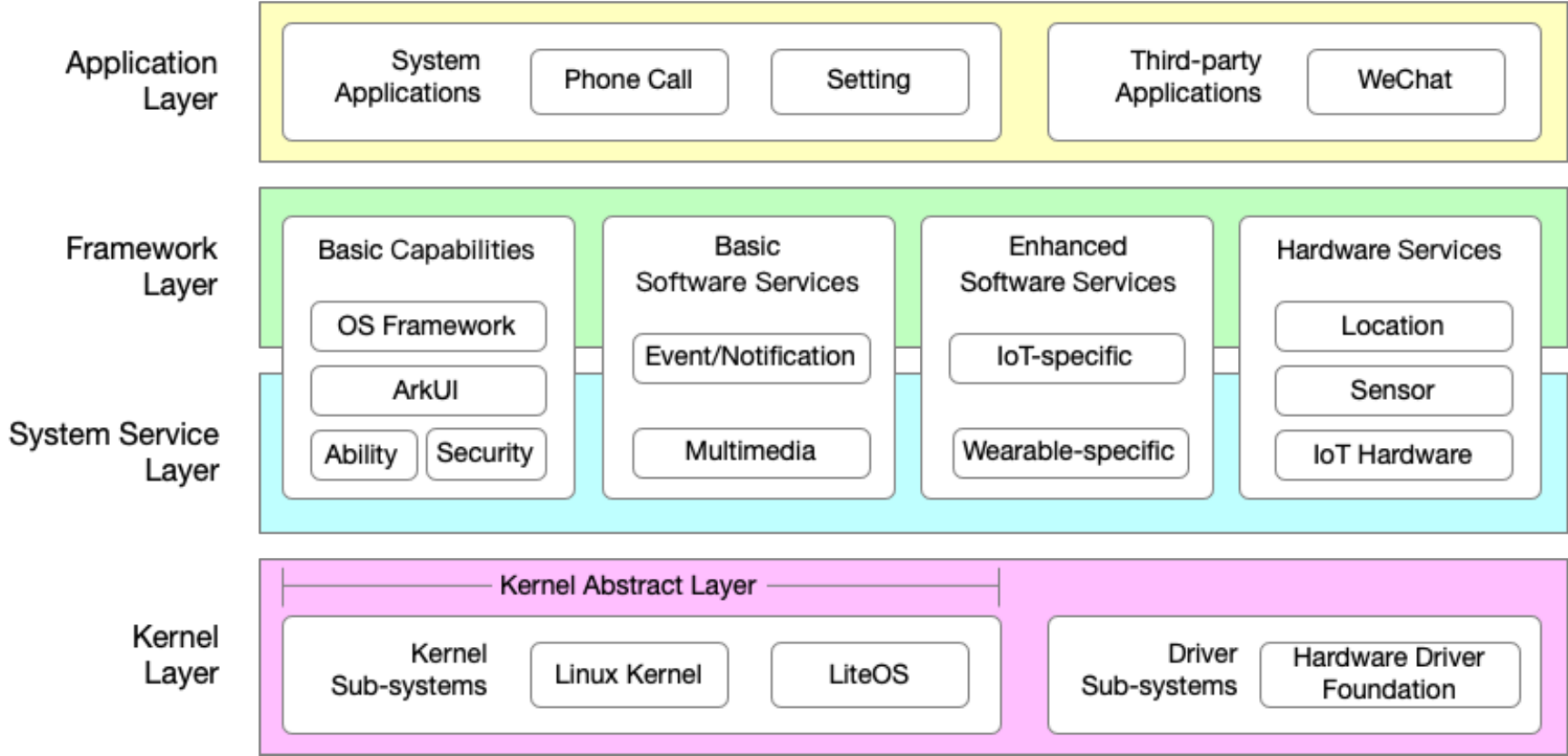}
  \caption{The Software Architecture of OpenHarmony.}
  \label{fig:openharmony_overview}
\end{figure}

OpenHarmony is designed with a layered architecture.
As illustrated in Fig.~\ref{fig:openharmony_overview}, it consists of four layers.
From bottom to top, the four layers are (1) the Kernel Layer, (2) the System Service Layer, (3) the Framework Layer, and (4) the Application Layer. We now briefly detail these four layers to help readers better understand this work.

\textbf{Kernel Layer.}
The kernel layer of OpenHarmony contains two main sub-systems, namely a kernel sub-system that powers an operating system kernel (such as the Linux Kernel) for scheduling the software execution of the whole system and a driver sub-system that is responsible for connecting the software stack with the various hardware.
Observant readers may have noticed that, unlike other systems, there is a special component called Kernel Abstract Layer (KAL) in the Kernel Layer of OpenHarmony.
This component is indeed a special OpenHarmony feature that is designed to support multi-kinds of mobile devices.
For different devices, OpenHarmony may select different OS kernels (e.g., Linux or LiteOS~\cite{cao2006liteos}) to power the system.
KAL is proposed to mitigate such a difference, aiming at offering the same capabilities for the upper software layers.

\textbf{System Service Layer.}
The system service layer is the core part of OpenHarmony that provides the actual implementation of all the system services required to run OpenHarmony apps.
Except for supporting basic capabilities such as the ones related to security control or providing intelligent functions, it also includes components related to common software services such as Events and Notifications, device-specific software services such as the ones dedicated to IoT devices or wearable devices, as well as hardware-related services such as sensors and location services.

\textbf{Framework Layer.}
The framework layer provides an interface for developers to implement OpenHarmony applications and such an interface is often provided within a Software Development Kit (SDK).
As shown in Fig.~\ref{fig:openharmony_overview}, generally speaking, this layer provides similar capabilities as the system service layer.
However, this layer is specifically required as it keeps app code from directly accessing system services, which might be abused by third-party developers if not controlled.
Indeed, through the framework layer, system services do not need to be exposed to third-party developers and how they should be called or scheduled can be pre-defined.
This layer is also very important as it defines the set of APIs needed to be seen by third-party apps.
This set of APIs needs to be appropriate as defining fewer APIs may cause the implementation of OpenHarmony apps difficult while defining more APIs would increase the complexity and subsequently the maintainability of the framework.

\textbf{Application Layer.}
The application layer is the place where OpenHarmony apps are located. There are two types of apps: system apps and third-party apps.
The former one should be provided by OpenHarmony itself, covering the basic functions that allow the OpenHarmony system to be practically usable.
The latter ideally should be supported by third-party developers that help the Openharmony system to good user experience, which is the key to the success of the OpenHarmony ecosystem.

\subsection{The App Development Framework}

\begin{figure}[!t]
  \centering
  \includegraphics[width=.8\linewidth]{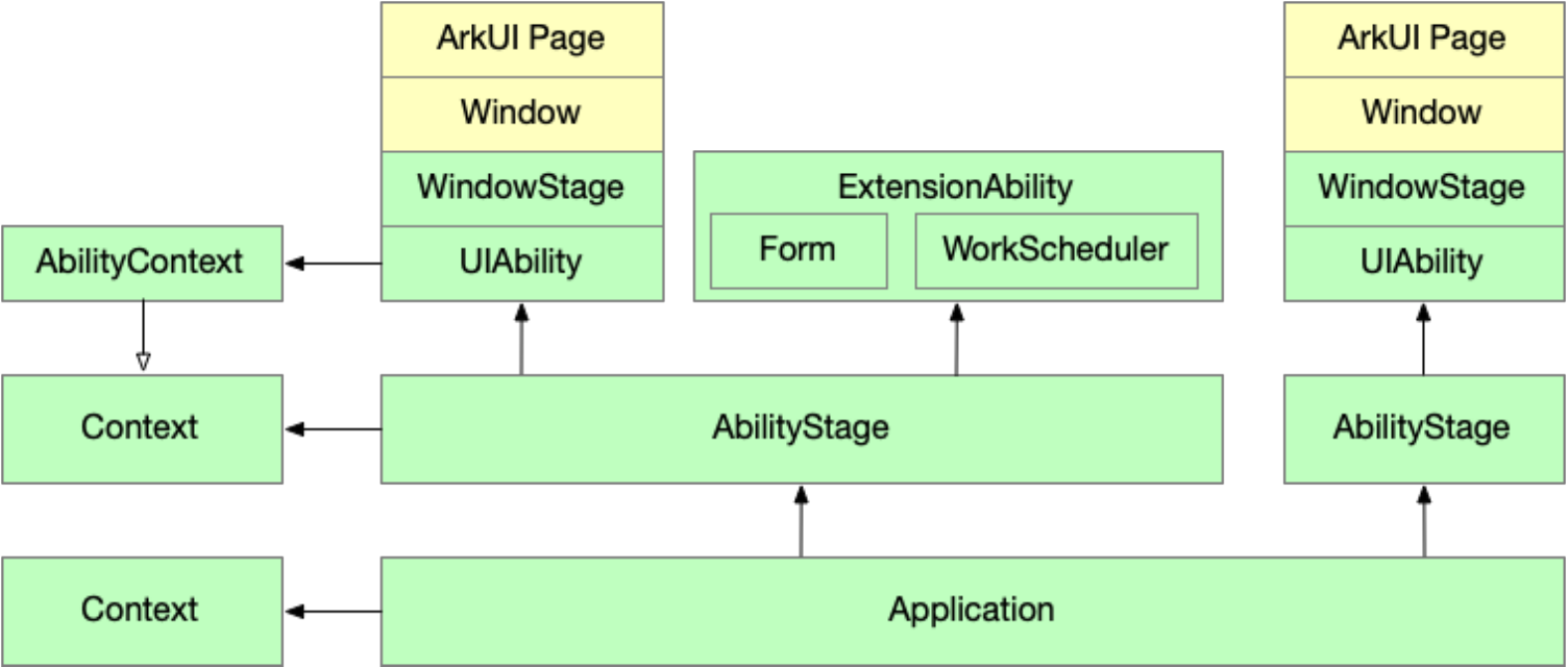}
  \caption{The Architecture of ArkTS-based App Development Framework (Stage Model) of OpenHarmony.}
  \label{fig:openharmony_arkts}
\end{figure}

We now briefly introduce OpenHarmony's app development framework.
There are actually two versions of app development frameworks supported by OpenHarmony to develop third-party apps: one based on Java program language and another based on ArkTS program language.\footnote{ArkTS (also known as eTS) is the preferred programming language introduced by Huawei to develop OpenHarmony applications. It is extended from the famous TypeScript language.}
Since the Java version will be gradually replaced by the ArkTS version, in this work, we will only focus on the ArkTS version.
Fig.~\ref{fig:openharmony_arkts} highlights the core components of ArkTS-based OpenHarmony app development framework.
OpenHarmony actually supports two ways of app logic (named Ability) developments, namely the FA (Feature Ability) model and the Stage model. The Stage model is newly introduced (since API version 9) to replace the FA model. Hence, in this work, we will only focus on the Stage model.

\textbf{Stage-based Ability Framework.}
As shown in Fig.~\ref{fig:openharmony_arkts}, in the Stage model, an OpenHarmony app is made up of \emph{AbilityStage} components. Each \emph{AbilityStage} should contain one or more \emph{Ability} components.
In OpenHarmony, there are two types of Ability components: UIAbility and ExtensionAbility.
\emph{UIAbility}, like \emph{Activity} in Android, is responsible for implementing the app's visual parts (i.e., GUI pages).
This is the reason why \emph{UIAbility} component will include a \emph{WindowStage} component that further contains a \emph{Window} module with an ArkUI page attached to it.

For other functions that are not directly relevant to the app's UI pages, OpenHarmony has introduced the so-called \emph{ExtensionAbility} component to support their implementation.
Normally, in Android, such functions should be implemented in one of the following three types of components: \emph{Service}, \emph{Broadcast Receiver}, and \emph{Content Provider}.
In OpenHarmony, the \emph{ExtensionAbility} mechanism provides a more fine-grained way to implement such functions.
For example, \emph{ServiceExtensionAbility}, a sub-class of \emph{ExtensionAbility}, is designed to support background tasks, providing equivalent functions as that of \emph{Service} in Android.
Another sub-class of \emph{ExtensionAbility}, namely \emph{DataShareExtensionAbility}, is designed to support data sharing, providing equivalent functions as that of \emph{Content Provider} in Android.

\begin{figure}[!h]
  \centering
  \includegraphics[width=.8\linewidth]{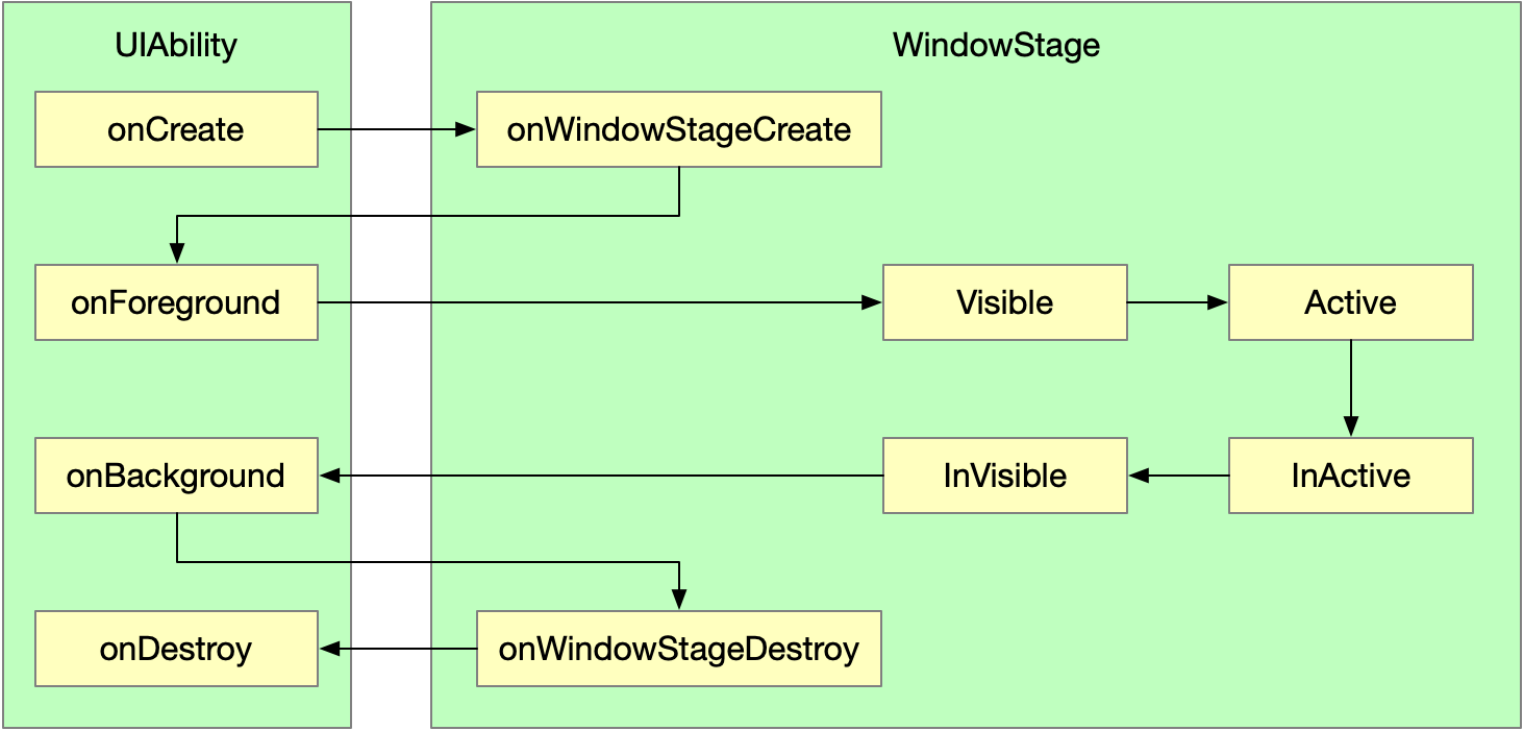}
  \caption{The Lifecycle of OpenHarmony's \emph{UIAbility} component.}
  \label{fig:openharmony_lifecycle}
\end{figure}

Like what has been designed in Android's components, there are lifecycle methods designed in OpenHarmony's ability components.
Fig.~\ref{fig:openharmony_lifecycle} illustrates the lifecycle of OpenHarmony's UIAbility component, which by itself contains four states, namely Create, Foreground, Background, and Destroy.
\emph{Create} state is at the stage when an UIAbility is started. At that time, the system will call the corresponding \emph{onCreate()} callback method, in which certain resources could be initiated.
After the \emph{onCreate()} method is called, the state moves to \emph{Foreground} and the \emph{onForeground()} callback method will be invoked.
At this stage, the UI page of the UIAbility becomes visible and will be displayed to users.
Once the UI page becomes invisible (e.g., other UI pages become visible), the state will be moved from \emph{Foreground} to \emph{Background}.
At this time, the \emph{onBackground} callback method will be called.
When the UIAbility is going to be terminated, the \emph{onDestroy()} callback method will be invoked and this is the place to store relevant data and free requested resources.
Observant readers may have noticed that the lifecycle of UIAbility is associated with a \emph{WindowStage} component, which per se has a sequence of lifecycle methods to be invoked as the UIAbility's state goes by.

\textbf{ArkUI Module.}
As shown in Fig.~\ref{fig:openharmony_overview} (with yellow background), the actual UI pages are implemented through the so-called ArkUI framework.
ArkUI is a core module of ArkTS that is newly introduced to support UI developments.
Fig.~\ref{fig:openharmony_arkui} illustrates the architecture of the ArkUI module.
This module supports two ways of UI implementation. The first way is to leverage Web-based technicals (e.g., HTML, CSS, Javascript) and the other way is through the so-called declarative programming (specifically designed to support the implementation of OpenHarmony apps).
This module also includes an \emph{UI Engine} module to provide common UI-related functions and other modules to allow visual display of UI pages.

\begin{figure}[!t]
  \centering
  \includegraphics[width=.8\linewidth]{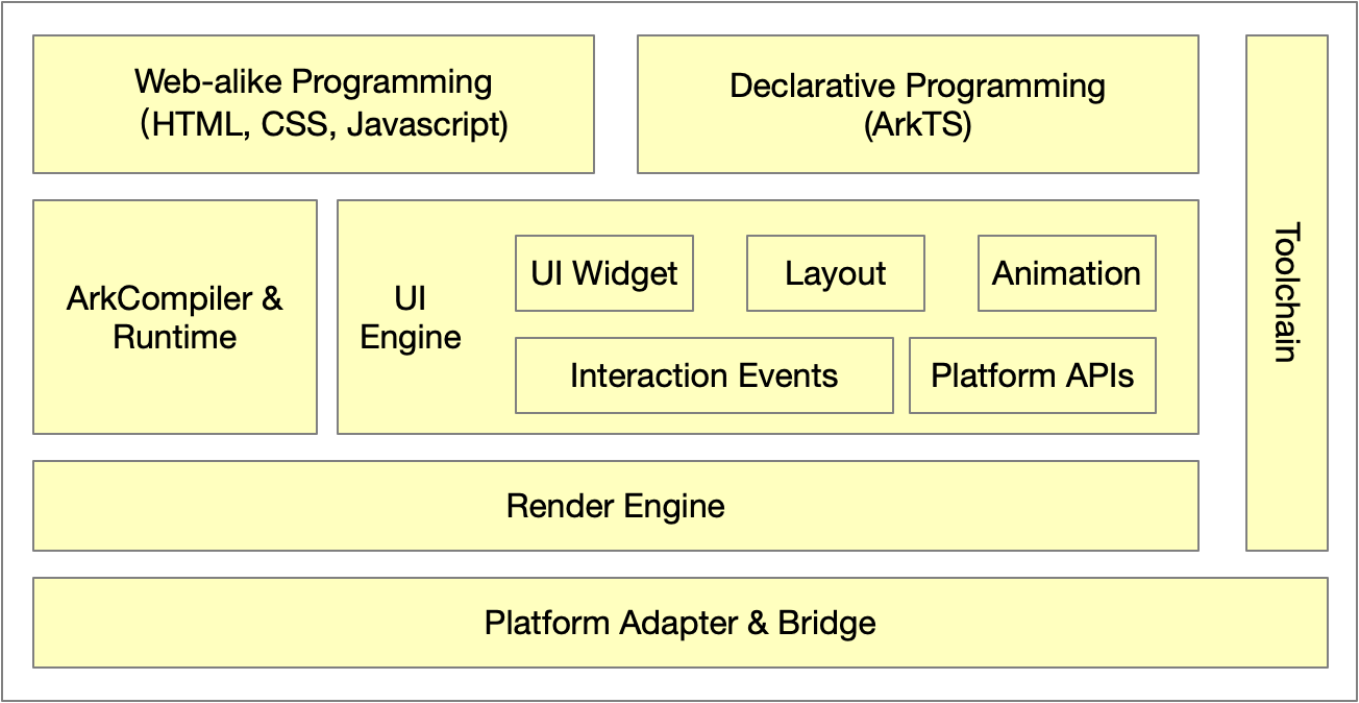}
  \caption{The Architecture of OpenHarmony's ArkUI Module.}
  \label{fig:openharmony_arkui}
\end{figure}

\section{The State of OpenHarmony Ecosystem}

As revealed in the previous literature review, despite Mobile Software Engineering has been a longstanding and hot topic, the efforts spent by our fellow researchers for exploring OpenHarmony have been limited.
Indeed, there is almost no contribution made to OpenHarmony in the current MSE community.
Therefore, as our initial attempt towards bringing OpenHarmony research to the Mobile Software Engineering (MSE) community, we summarize the current achievements of OpenHarmony to help readers better understand the state of the OpenHarmony ecosystem.
Specifically, in this section, we first briefly introduce the OpenAtom OpenHarmony initiative and subsequently highlight the existing toolchains and datasets available in the community.
After that, we go one step further to also summarize other existing resources that may not be directly related to OpenHarmony but could still be beneficial to grow the ecosystem of OpenHarmony.

\subsection{The OpenAtom OpenHarmony Initiative}

Recall that OpenHarmony is now a fully open-sourced project, which is currently incubated by the OpenAtom Foundation.
At the moment, the OpenHarmony project is run by an operating committee, and its technical part is mainly run by a Technical Steering Committee (TSC). 
To support the technical success of OpenHarmony, the TSC has further set up various domain-specific Technical Supporting Groups (TSG).
At the moment, there are six TSGs and the number is growing.
The current supporting groups cover the following six domains: (1) Program Language, (2) Cross-platform App Development Framework, (3) Security and Privacy Computing, (4) Web3 Standard, (5) Robot, and (6) Integrated Development Environment (IDE).
These domain-specific TSGs are responsible for understanding the domain-related requirements, summarizing the technical map of the domain, and preparing the fundamental technicals for helping domain-specific technical-related decision-making and eventually supporting the overall success of OpenHarmony.

\subsection{Existing Toolchains}
\label{subsec:toolchains}

We then look at the existing toolchains offered by the official OpenHarmony framework to support app developments and these toolchains are considered important and essential.
Indeed, these tools could provide fundamental capabilities to support the implementation of more advanced OpenHarmony-specific toolchains.
Ideally, these toolchains should cover the full lifecycle of app development, including development, build, testing, debugging, code review, and publishing.
Table~\ref{tab:toolchains} summarizes some of the tools provided by OpenHarmony.
The second column demonstrates the software engineering phase that the tool intends to support.
At the moment, these toolchains have covered almost all the aforementioned lifecycle phases, e.g., including app development-related ones (e.g., IDE, Emulator, Device Manager), app building tools (e.g., hvigor), app testing tools (e.g., jsunit, uitest), debugging tools (e.g., HiLog, profiler), code reviewing tools (e.g., Code Linter), command line tools (e.g., hdc), and package management tool (e.g., ohpm).
The only exception is the phase of publishing. 
At the moment, there is no such tool offered for OpenHarmony.
It is nonetheless understandable as there is no app market available for hosting OpenHarmony apps yet.
We believe such a tool will be provided once a dedicated app market is offered.

It is worth noting that, at the moment, we only conducted a high-level overview and did not check in detail to what extent are the required functions in each phase covered by these tools. For example, there is a tool called \emph{Monkey} in Android that supports random exploration of Android apps, it is not clear to us if the existing toolchains of OpenHarmony provide equivalent functions.
As for our future work, we plan to have a more detailed look at these tools and provide to the community a clearer overview of these toolchains.

\begin{table}[!h]
\caption{A selected list of OpenHarmony Toolchains.}
\label{tab:toolchains}
\resizebox{\linewidth}{!}{
\begin{tabular}{r c p{0.8\linewidth}}
\hline
Tool                        &  SE Phase     & Function \\
\hline
DevEco Studio               &  Development   & The recommended integrated development environment for implementing OpenHarmony apps.       \\
Device Manager              &  Development   & This tool provides an interface for developers to manage OpenHarmony devices, including both emulator-based and real-world devices. \\
Emulator                    &  Development   & This tool can set up OpenHarmony emulators (either remotely or locally) that allow developers to install, run, and test their apps on an emulator instead of real-world OpenHarmony devices.      \\

hvigor                      &  Build   & The recommended tool for building OpenHarmony source code project to runnable apps. \\

arkXtest/jsunit            &  Test     & This tool allows developers to run unit tests when implementing OpenHarmony apps.       \\
arkXtest/uitest            &  Test     & This tool allows developers to search and update certain widgets in a given GUI page, which is essential for supporting automated OpenHarmony app testing.  \\

HiLog                       &  Debug   & The default tool that is designed to log information such as user operations or system running statuses for the system framework, services, and OpenHarmony apps.  \\

profiler       &  Debug   & This tool provides a visual interface for developers to quickly check the profiling information such as the currently used system and memory resources, including the heap and stack memories of each task. \\

Code Linter                 &  Code Review   & This tool is responsible for  grammatically checking the correctness of ArkTS code, which is the default programming language for implementing OpenHarmony apps.  \\

hdc                         &  Other   & The OpenHarmony Device Connector tool allows developers to connect their PC-side development machine to a given OpenHarmony device.  \\

ohpm        &   Other  & OpenHarmony Package Manager. \\
\hline
\end{tabular}}
\end{table}

\subsection{Existing Datasets}
\label{subsec:datasets}

As shown in Section~\ref{sec:mse}, the datasets targeted by our MSE community can be mainly divided into four types: (1) Mobile apps (including both open-sourced and closed-sourced apps), (2) Mobile App Development Framework, (3) Third-party Libraries, (4) App Store Info (including app reviews).
We now respectively summarize the current situation of these types of datasets in OpenHarmony, respectively.
We further go one step deeper to harvest the relevant datasets, if possible, and make them publicly available for supporting our fellow researchers to conduct OpenHarmony-related software engineering research.

\begin{table}[!h]
\caption{A selected list of OpenHarmony Toolchains.}
\label{tab:framework}
\resizebox{0.85\linewidth}{!}{
\begin{tabular}{rll}
\hline
Type                                 & OpenHarmony       & 
Android                    \\ \hline
Name                                 & OpenHarmony/interface\_sdk-js & aosp-mirror/platform\_frameworks\_base  \\
Platform                             & Gitee             & Github                     \\
\#. Branches                         & 105               & 471                        \\
\#. Tags                             & 30                & 1,850                      \\
\#. Forks                            & 1,400              & 6,300                      \\
\#. Stars                            & 57                & 10,500                     \\
\#. Commits                          & 7,882             & 822,906                    \\
\multicolumn{1}{l}{\#. Contributors} & 627               & 
1,399                     \\
\hline
\end{tabular}}
\end{table}

\textbf{OpenHarmony Framework.}
Recall that OpenHarmony is a fully open-sourced system, its app development framework is open-sourced.
The framework is the first gate that OpenHarmony apps need to interact with before running into the system.
The interaction is mainly through APIs provided by the app development SDK, as part of the OpenHarmony framework.
Some of the meta-data of the OpenHarmony framework are shown in Table~\ref{tab:framework}.
The current framework is open-sourced at the \emph{interface\_sdk-js} repository\footnote{We remind the readers that the framework and the SDK are not exactly the same as the framework may contain more capabilities that are reserved for system apps while SDK is only supposed to be used by third-party apps.
For simplicity, in this work, we will not differentiate this as there is no direct repository provided for hosting the framework code of OpenHarmony.} on Gitee and it currently has 105 branches, 30 tags, 1,400 forks, 57 stars, 7,833 commits, and 627 contributors.
As a comparison, the last column of Table~\ref{tab:framework} shows the meta-data of the Android framework repository, respectively.
It is obvious that OpenHarmony has a big step to go in order to catch up with Android, which poses lots of opportunities for our MSE community to mitigate the gap between the OpenHarmony framework and the Android framework.

\begin{figure}[!h]
  \centering
  \includegraphics[width=\linewidth]{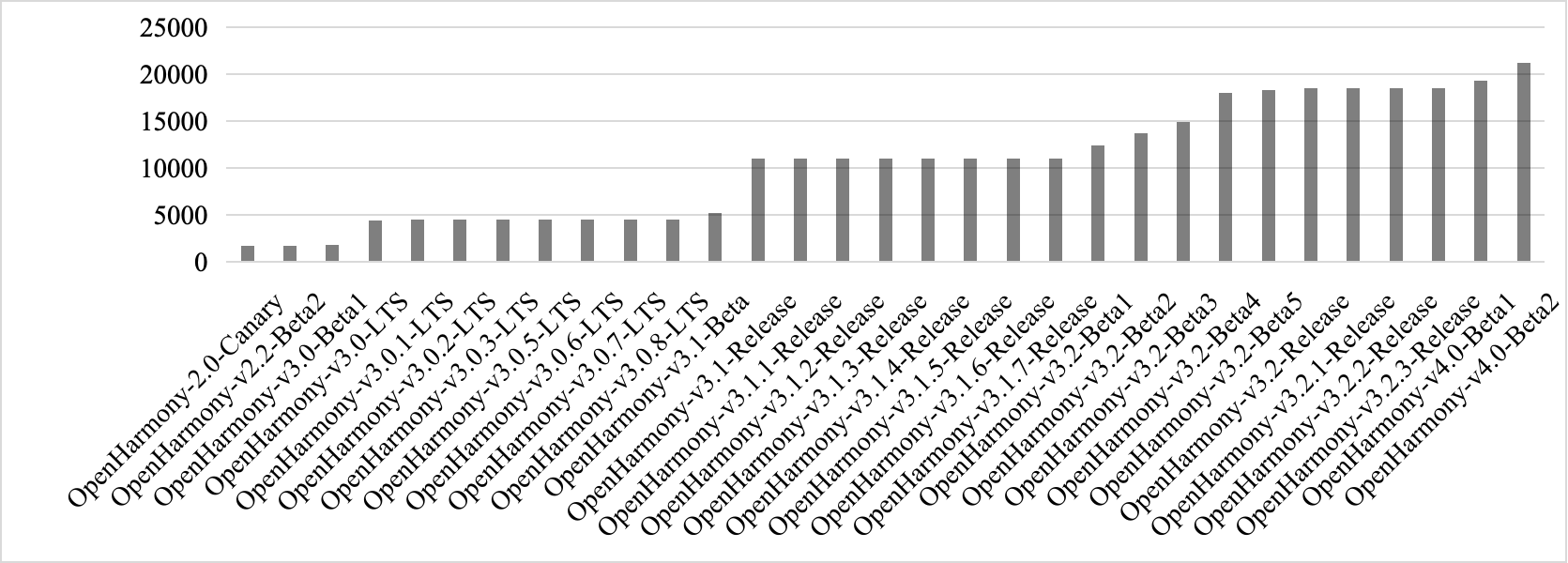}
  \caption{The evolution of the number of APIs offered by the OpenHarmony framework. The X-axis includes all the tags (ranked based on their released time, the earlier, the former) available in the OpenHarmony repository.}
  \label{fig:openharmony_api}
\end{figure}

We further look into the number of APIs offered by the OpenHarmony framework. Since there is no such information directly provided on the web, we decide to write a parser to directly harvest that from the open-source repository.
We select the latest version (i.e., OpenHarmony 4.0) and only count the number of functions (including static and non-static functions).
In the latest version, there are 10,435 APIs. This number is also significantly smaller than that of the Android framework, which already has over 30,000 APIs in 2018 (i.e., API version 28~\cite{li2020cda}).
Nonetheless, as illustrated in Fig.~\ref{fig:openharmony_api}, the number of APIs (again, any functions are considered) is continuously increasing, showing that the capabilities of OpenHarmony are keeping maturing.
We believe as time goes by, such a difference between the APIs of Android and OpenHarmony will be much smaller.

\textbf{OpenHarmony Apps.}
One of the most important reasons that make mobile software engineering (especially for the Android community) a longstanding hot topic is due to the existence of a large number of mobile apps~\cite{joorabchi2013real,francese2017mobile}.
Indeed, there are over 2 million Android apps (there is a similar number for the iOS community) available on the official Google Play store.
In the famous AndroZoo dataset~\cite{li2017androzoo++}, there are over 23 million Android apps collected from various sources (e.g., the official Google Play store and over 10 third-party markets such as PlayDrone, AppChina, etc.) spanning various years.
Liu et al. have subsequently harvested the open-sourced Android apps and formed them as a dataset called AndroZooOpen~\cite{liu2020androzooopen}.
This dataset is also made publicly available to the software engineering community and has been demonstrated to be useful in supporting Android research tasks.
Inspired by this, we hypothesize that OpenHarmony apps will be one of the most important resources for supporting OpenHarmony research.
We, therefore, take our initial attempt to harvest existing OpenHarmony apps.
Since there is no app market available for OpenHarmony yet, we solely focus on open-sourced OpenHarmony apps.
Specifically, we take OpenHarmony as well as HarmonyOS as the search keyword and apply it to two famous cloud-based software version control websites, namely GitHub and Gitee, which are the most famous sites of such in the world and in China, respectively.

Our initial search results in 3,804 repositories\footnote{The full list is made available on GitHub (https://github.com/SMAT-Lab/SE4OpenHarmony).}, for which 910 of them are from GitHub while the remaining 2,894 from Gitee.
We remind the readers that these identified repositories may not always be OpenHarmony apps.
Therefore, we resort to a Shell script (with manually identified features of OpenHarmony apps considered) to select such repositories that indeed contain OpenHarmony apps.
Our experiment has eventually discovered 174 such repositories, with 147 and 27 from Gitee and Github, respectively.
To facilitate further research, we have also made this list publicly available on the same site.\footnote{Our further investigation finds that most of these apps are not comprehensive ones (i.e., might be toy apps or demonstrating the usability of certain libraries). We hence commit to keep updating this list toward forming a more useful dataset for supporting OpenHarmony-based software engineering research.}

\textbf{Third-party Libraries}
We remind the readers that OpenHarmony takes a newly introduced language called ArkTS to support app implementations.
In this work, we also look at the existing third-party libraries that are available for supporting the development of OpenHarmony apps.
Specifically, we would like to understand to what extent are ArkTS-based libraries available in our community and what are they designed for.
In OpenHarmony, the official team has introduced a tool called \emph{ohpm} (as also shown in Table~\ref{tab:toolchains}) for managing all the third-party libraries designed for developing OpenHarmony apps.
In the current central registry\footnote{https://ohpm.openharmony.cn}, there are already 96 libraries and the number is growing.
The functions of these libraries can be divided into 10 categories.
Table~\ref{tab:libraries} enumerates some of the representative ones for each category.
Generally, we show one or two libraries for each category.
The two are randomly chosen if there are more libraries available for the given category.

\begin{table}[!h]
\caption{A sample list of OpenHarmony's third-party libraries (available in OpenHarmony's central registry).}
\label{tab:libraries}
\resizebox{\linewidth}{!}{
\begin{tabular}{rc l p{0.65\linewidth}}
\hline
Category        & Count     &   Repo &  \\ 
\hline
\multirow{2}{*}{UI}              &     \multirow{2}{*}{2}      & @ohos/pulltorefresh & Pull-to-refresh and pull-up loading component  \\
    &   & @ohos/mpchart & Support the implementation of various types of charts such as Pie chart, Candle chart, etc.\\
\hline

\multirow{2}{*}{Animation}       &     \multirow{2}{*}{2}     & @ohos/lottie & The animation library for OpenHarmony. Similar to Java's lottie, AndroidViewAnimations, and Leonids libraries. \\
& & @ohos/svg & SVG-formatted image parser and render library.\\ 
\hline

Network         &      1     & @ohos/axios & The promise-based HTTP Client implementation library for OpenHarmony. \\
\hline

\multirow{2}{*}{Image}           &      \multirow{2}{*}{2}     & @ohos/imageknife & An efficient, lightweight, and simple image loading cache library \\
&& @ohos/xmlgraphicsbatik & For working with images in SVG format \\
\hline

Multimedia      &     1      & @ohos/ijkplayer & FFmpeg-based video player\\

\hline

\multirow{2}{*}{Data Storage}    &      \multirow{2}{*}{2}     & @ohos/disklrucache & Support cache functions for accessing disks \\
&& @ohos/mmkv & A lightweight key-value storage framework \\
\hline

\multirow{2}{*}{Event}   &   \multirow{2}{*}{2}        & @ohos/mqtt & Support MQTT-based actions such as message subscription\\
&& @ohos/liveeventbus & Support inter-process and inter-app message broadcast\\

\hline

Security        &     1      & @ohos/crypto-js & Support the implementation of cryptographic functions such as MD5, SHA256, etc. \\

\hline

\multirow{2}{*}{Utility}            &     \multirow{2}{*}{2}      & @ohos/zxing & Support read or generate QR Code for OpenHarmony \\
&& @ohos/pinyin4js & Translating Chinese characters to pinyin \\
\hline

\multirow{2}{*}{Other}           &     \multirow{2}{*}{2}      & @ohos/arouteronactivityresult & Support message transmission when performing inter-page or inter-app communications. \\
&& @ohos/coap & Support Constrained Application Protocol (CoAP) capabilities. \\

\hline
\end{tabular}}
\end{table}

Furthermore, as mentioned previously, ArkTS is not entirely new. It actually extends Typescript, which has been a popular programming language for more than 10 years.
Typescript is Javascript with syntax for types, i.e., adding static typing with optional type annotations to Javascript.
Theoretically, existing Typescript code (as well as Javascript code) can be directly reused for developing OpenHarmony apps. 
Those Typescript/Javascript implementations could be regarded as third-party libraries as well.
By taking Typescript and Javascript as the search keyword, Github returns 513,000 and 1.7 million repositories for Typescript and Javascript, respectively.
Such a large number of repositories (despite not all of them being code-related repositories) indicates that there are already a lot of potential third-party libraries available for OpenHarmony.\footnote{We hypothesize that this is one of the major reasons why ArkTS is proposed as the default programming language for developing OpenHarmony apps.}
Those libraries could be leveraged (either directly or with additional efforts contributed by our fellow researchers) to facilitate the development of OpenHarmony apps and its broad ecosystem.

\textbf{App Store Info.}
Our software engineering researchers have leveraged app store info (such as the app's author information, description, user rating, user reviews, etc.) to support various studies.
For example, Gorla et al.~\cite{gorla2014checking} have leveraged the app's description to check against the app's behavior.
Obie et al.~\cite{obie2022violation} have leveraged the app's review data to investigate the violation of honesty in mobile apps.
To the best of our knowledge, there is barely any app store hosting OpenHarmony apps at the moment. Therefore, there is no such dataset that can be collected so far.
Nonetheless, the OpenHarmony version of a given app will also share much of such information as that available in Android or iOS.
This information could also be helpful when mining OpenHarmony-specific app store information.

\subsection{Existing OpenHarmony Research}
\label{subsec:oh_research}

As our initial attempt towards building the research roadmap for guiding our software engineering researchers to contribute to OpenHarmony, we start by conducting a lightweight literature review about OpenHarmony.
Our method is straightforward. 
We use \emph{OpenHarmony} and \emph{HarmonyOS} as the search keywords and we apply them separately to search for relevant publications on both \emph{Google Scholar} and \emph{DBLP}, respectively.
At this step, when applied to Google Scholar, we will only consider the top 100 results.
Table~\ref{tab:papers} enumerates the list of OpenHarmony-related publications. 
In total, we only found 8 papers and among which only one (i.e., the one published at the APWeb conference) can be found on DBLP, while all of them can be found on Google Scholar.
At this step, we only consider a paper relevant if and only if it directly contributes to the OpenHarmony project or if it takes OpenHarmony as its dataset to evaluate their approaches.
There are several other papers that are not included in this review although they do involve OpenHarmony/HarmonyOS systems.
They are excluded because they do not contribute anything to OpenHarmony as they only involve running their approaches on OpenHarmony/HarmonyOS systems.
For example, the work proposed by Qiu et al.~\cite{qiu2022re} is not included in this paper because it only leverages HarmonyOS to support their model implementation about supporting distributed user interfaces to be dynamically configured on multiple IoT devices based on user preferences.

\begin{table}[!h]
\caption{The list of OpenHarmony-related primary publications.}
\label{tab:papers}
\resizebox{\linewidth}{!}{
\begin{tabular}{c p{0.5\linewidth} p{0.3\linewidth} l  l}
\hline
Year & Title  & Relevance & Venue & CORE-Rank  \\
\hline
2023 & CiD4OhOs: A Solution to HarmonyOS Compatibility Issues & API-induced compatability issues & Industry Challenge Track of ASE & A\\
2023 & HiLog: A High Performance Log System of OpenHarmony  & Targeted OpenHarmony's log system & Journal of Software & - \\
2023 & Design and Implementation of HiLog, the high performance log system of OpenHarmony  & Targeted OpenHarmony's log system & Journal of Software & - \\
2023 & Breaking the Trust Circle in HarmonyOS by Chaining Multiple Vulnerabilities  & Investigated the security of HarmonyOS's trust circle service  & ACCTCS & -\\
2023 & Unveiling the Landscape of Operating System Vulnerabilities  & Studied HarmonyOS's vulnerabilities & Future Internet & - \\
2022 & A Deep Looking at the Code Changes in OpenHarmony  & Studied OpenHarmony's code changes & APWeb & B \\
2022 & Cross Platform API Mappings based on API Documentation Graphs & Studied HarmonyOS's API documentation & QRS & B \\
2021 & SparrowHawk: Memory Safety Flaw Detection
via Data-driven Source Code Annotation & Applied to detect vulnerabilities in OpenHarmony & Inscrypt & National\\
\hline
\end{tabular}}
\end{table}

As shown in Table~\ref{tab:papers}, there are only eight OpenHarmony-related papers published in the community.
The efforts could be neglected if compared to those for Android, where there are over 7,000 papers published as recorded in DBLP (searching by taking Android as the keyword).
This evidence confirms our previous argument that there is still a huge gap between OpenHarmony and Android.
This, however, also demonstrates that there are huge opportunities open for our community.
Ideally, the research methods applied to Android or iOS could also be applied to OpenHarmony.
Despite there being only eight papers published, it is motivating to find that the number of relevant papers keeps growing.
The venues where the current papers are published are generally not in reputed journals or conferences.
Indeed, among the eight papers, only four of them are published at venues recorded by CORE and only three of them are ranked.
We would hope that our community could spend more effort in developing software engineering approaches for OpenHarmony and publish more papers at mainstream venues.

\section{Overview of Mobile Software Engineering}
\label{sec:mse}

In this work, we are interested in building a research roadmap for conducting software engineering research for OpenHarmony.
Unfortunately, since OpenHarmony is still in its early stages, there is not much work proposed for that.
Nevertheless, we believe all the research efforts contributed to improving the Android and iOS ecosystem could be also conducted for OpenHarmony.
Therefore, in this section, we first resort to a systematic literature review to understand the status quo of mobile software engineering research.
We will then leverage the empirical observations to form our research roadmap dedicated to OpenHarmony. 

\begin{figure}[!t]
  \centering
  \includegraphics[width=.8\linewidth]{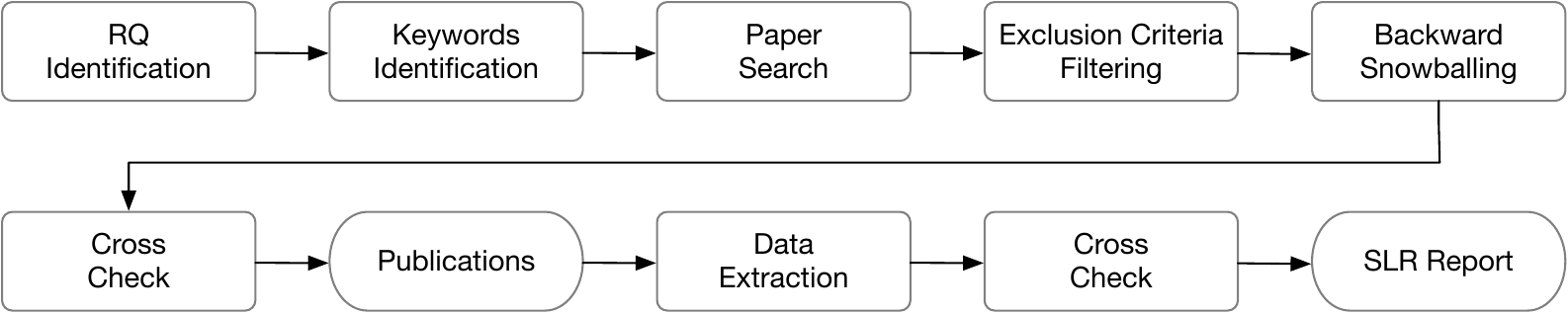}
  \caption{The working process of our systematic literature review.}
  \label{fig:slr_process}
\end{figure}

In this work, we conduct the systematic literature review following the methodologies outlined by Brereton et al.~\cite{brereton2007lessons} and Li et al.~\cite{li2019rebooting}.
Fig.~\ref{fig:slr_process} highlights the working process.
In \textbf{Step 1}, we plan to investigates the latest research advancements in the area of mobile software engineering by answering the following research question.

\begin{center}
    \textbf{RQ: What problems are targeted by our fellow researchers in the MSE community and how they are resolved? }
\end{center}

Then, in \textbf{Step 2}, we identify the search keywords that could be used to find all the relevant publications, in order to answer the pre-defined research questions. 
When we started to do that, we immediately realized that the number of primary publications was too huge, it is literally impossible to manually read all of them. 
Indeed, taking Android-related research alone, there are already over 7,500 papers recorded on DBLP. 
If one is able to read a paper in 2 hours, it would still require years to complete.
To mitigate this problem, we resort to only considering the existing survey and literature review papers, for which our fellow researchers have already systematically reviewed the different aspects of mobile software engineering.
We believe these survey papers are representative of the status quo of mobile software engineering research.
To this end, we identify the search keywords based on these concerns and list of included keywords are summarized in Table~\ref{tab:keywords}.
In total, we have identified two groups of keywords, represented as G1 and G2, respectively.
We then form the query based on this rule\footnote{($g1_1$ OR $...$ OR $g1_x$) AND ($g2_1$ OR $...$ OR $g2_y$), where $g1_i \in G1$, $g2_j \in G2$, and $1 \le i \le x$, $1 \le j \le y$, for which $x$ and $y$ are the number of keywords in G1 and G2, respectively} for which we require it to contain at least one keyword from each group.



\begin{table}[!h]
\centering
\caption{Repository Search Keywords.}
\label{tab:keywords}
\begin{tabular}{| c | l |}
\hline
Group (and) & Keywords (or) \\
\hline
G1 & Mobile, Android, iOS, *phone*\\
\hline
G2 & survey, literature review, mapping study, handbook, overview\\
\hline
\end{tabular}
\end{table}

After the query is formed, in \textbf{Step 3}, we directly applied to search relevant studies in the Computer Science Bibliography (DBLP) Database.
This step gives us 65 papers that are potentially relevant to out study.
After that, in \textbf{Step 4}, we refine the gathered list of relevant papers by manually to ensure their relevance to mobile software engineering (i.e., could indeed be helpful for answering the aforementioned research question).
Specifically, we filter out the less relevant papers based on the following set of exclusion criteria.

\begin{enumerate}
    \item Since we only consider survey or literature review papers, all the non-survey papers are simply excluded from our study.
    \item Although there are some papers that meet our selection criteria (i.e., whose title contains the group keywords in Table~\ref{tab:keywords}), their topics may not strictly fall into the software engineering category. We initially eliminate these papers by manually reviewing the abstracts and identifying those that have the potential to provide guidance for OpenHarmony.
\end{enumerate}

Once the irrelevant papers are filtered out, we then conduct a backward snowballing (i.e., \textbf{Step 5}) by scanning all the referenced papers and checking if they should also be considered for our study.
We remind the readers that we have cross-checked the results (i.e., \textbf{Step 6}) in all the previous two steps (i.e., exclusion criteria filtering and backward snowballing) to ensure the reliability of the results.

We are able to eventually collect 39 papers for answering our research question defined at the beginning of this study. 
Table~\ref{tab:targeted_paper} enumerates the list of selected papers, including their publication year and venue.
Once the relevant papers are collected, we carefully read all of them and attempt to extract the relevant data (i.e., \textbf{Step 7}) from each paper to answer the research question.
Specifically, we aim to extract the following two types of information:
(1) Targeted Problems, which involve understanding the issues within the Android/iOS ecosystem that have been identified by our MSE researchers as problems needing resolution to create a more user-friendly mobile ecosystem, and (2) Fundamental Techniques, 
aimed at discovering the techniques required to address the various challenges in the mobile community.
Considering that OpenHarmony may encounter similar issues to those faced by Android and iOS, we argue that insights gained from exploring these two aspects could prove valuable in shaping the roadmap for conducting software engineering research for OpenHarmony. Furthermore, similar to our approach in identifying relevant papers, we have conducted cross-checks of our observations, involving at least two authors, to ensure the reliability of these observations, thereby enhancing the trustworthiness of the research roadmap.

\begin{table}[!htbp]
\centering
\setlength{\belowcaptionskip}{0pt}
\setlength{\abovecaptionskip}{0pt}
\caption{The List of Selected Publications.}
\label{tab:targeted_paper}
\resizebox{\linewidth}{!}{
\begin{tabular}{l | p{0.9\linewidth} | c | c }\hline
Authors  & Title & Year & Venue  \\ \hline
\citet{10.1145/3556974} &   Android Source Code Vulnerability Detection: A Systematic Literature Review
& 2023  & CSUR  \\ \hline
\citet{wu2023systematic} & A systematic literature review on Android-specific smells & 2023 & JSS  \\ \hline
\citet{10.1145/3544968} &   Deep Learning for Android Malware Defenses: A Systematic Literature Review
& 2022  & CSUR  \\ \hline
\citet{10.1145/3507903} &  Dynamic Testing Techniques of Non-Functional Requirements in Mobile Apps: A Systematic Mapping Study
& 2022 & CSUR  \\ \hline
\citet{10.1145/3510579} &   A Survey of Privacy Vulnerabilities of Mobile Device Sensors
& 2022 & CSUR  \\ \hline
\citet{10.1145/3530814} &   A Systematic Survey on Android API Usage for Data-Driven Analytics with Smartphones & 2022 & CSUR  \\ \hline
\citet{nakamura2022factors} & What factors affect the UX in mobile apps? A systematic mapping study on the analysis of app store reviews
& 2022 & JSS  \\ \hline
\citet{wimalasooriya2022systematic} & A systematic mapping study addressing the reliability of mobile applications: The need to move beyond testing reliability
& 2022 & JSS  \\ \hline
\citet{zhan2021research} & Research on Third-Party Libraries in Android Apps: A Taxonomy and Systematic Literature Review
& 2021 & TSE  \\ \hline
\citet{shamsujjoha2021developing} & Developing Mobile Applications Via Model Driven Development: A Systematic Literature Review
& 2021 & IST  \\ \hline
\citet{EBRAHIMI2021106466} & Mobile app privacy in software engineering research: A systematic mapping study
& 2021 & IST  \\ \hline
\citet{de2021measurement} & Measurement-based Experiments on the Mobile Web: A Systematic Mapping Study
& 2021 & EASE  \\ \hline
\citet{10.1145/3368961} &  Autonomous Visual Navigation for Mobile Robots: A Systematic Literature Review & 2020 & CSUR  \\ \hline
\citet{10.1145/3372788} &   A Survey of Context Simulation for Testing Mobile Context-Aware Applications
& 2020 & CSUR  \\ \hline
\citet{10.1145/3417986} &   Energy Diagnosis of Android Applications: A Thematic Taxonomy and Survey
& 2020 & CSUR  \\ \hline
\citet{10.1145/3417978} &  A Survey of Android Malware Detection with Deep Neural Models
& 2020 & CSUR  \\ \hline
\citet{li2019rebooting} & Rebooting Research on Detecting Repackaged Android Apps: Literature Review and Benchmark
& 2019 & TSE  \\ \hline
\citet{8606261} & App store effects on software engineering practices & 2019 & TSE \\
\hline
\citet{kaur2019investigation} & Investigation on test effort estimation of mobile applications: Systematic literature review and survey
& 2019 & IST  \\ \hline
\citet{barmpatsalou2018current} &  Current and Future Trends in Mobile Device Forensics: A Survey & 2018 & CSUR  \\ \hline
\citet{biorn2018survey} &  A Survey and Taxonomy of Core Concepts and Research Challenges in Cross-Platform Mobile Development
& 2018 & CSUR  \\ \hline
\citet{jabangwe2018software} &  Software engineering process models for mobile app development: A systematic literature review & 2018 & JSS  \\
\hline
\citet{ahmad2018perspectives} & Perspectives on usability guidelines for smartphone applications: An empirical investigation and systematic literature review
& 2018 & IST  \\
\hline
\citet{8330039} & A Survey on Recent OS-Level Energy Management Techniques for Mobile Processing Units & 2018 & TPDS  \\
\hline
\citet{8453877} &  Automated Testing of Android Apps: A Systematic Literature Review
& 2018 & TRel  \\
\hline

\citet{genc2017systematic} & A systematic literature review: Opinion mining studies from mobile app store user reviews
& 2017 & JSS  \\
\hline
\citet{li2017static} & Static analysis of android apps: A systematic literature review
& 2017 & IST  \\
\hline
\citet{xu2016toward} &  Toward Engineering a Secure Android Ecosystem: A Survey of Existing Techniques
& 2016 & CSUR  \\
\hline
\citet{martin2016survey} & A survey of app store analysis for software engineering & 2016 & TSE \\ \hline
\citet{DBLP:journals/jss/ZeinSG16} &  A systematic mapping study of mobile application testing techniques
& 2016 & JSS \\
\hline
\citet{sufatrio2015securing} &  Securing Android: A Survey, Taxonomy, and Challenges
& 2015 & CSUR  \\
\hline
\citet{hoseini2013survey} &   A survey on smartphone-based systems for opportunistic user context recognition & 2013 & CSUR  \\
\hline
\citet{pereira2013survey} &   Survey and analysis of current mobile learning applications and technologies & 2013 & CSUR  \\
\hline

\citet{shahzad2017socio} & Socio-technical challenges and mitigation guidelines in developing mobile healthcare applications & 2017 & JMIHI \\
\hline 

\citet{wang2022runtime} & Runtime permission issues in android apps: Taxonomy, practices, and ways forward & 2022 & TSE \\
\hline 

\citet{ali2021self} & Self-adaptation in smartphone applications: Current state-of-the-art techniques, challenges, and future directions & 2021 & DKE \\
\hline 

\citet{autili2021software} & Software engineering techniques for statically analyzing mobile apps: research trends, characteristics, and potential for industrial adoption & 2021 & JISA \\
\hline 

\citet{silva2022mapping} & A mapping study on mutation testing for mobile applications & 2022 & STVR \\
\hline

\citet{hort2021survey} & A survey of performance optimization for mobile applications & 2021 & TSE \\
\hline

\end{tabular}}
\end{table}

\subsection{Problem}

\begin{figure}[!t]
  \centering
  \includegraphics[width=0.7\linewidth]{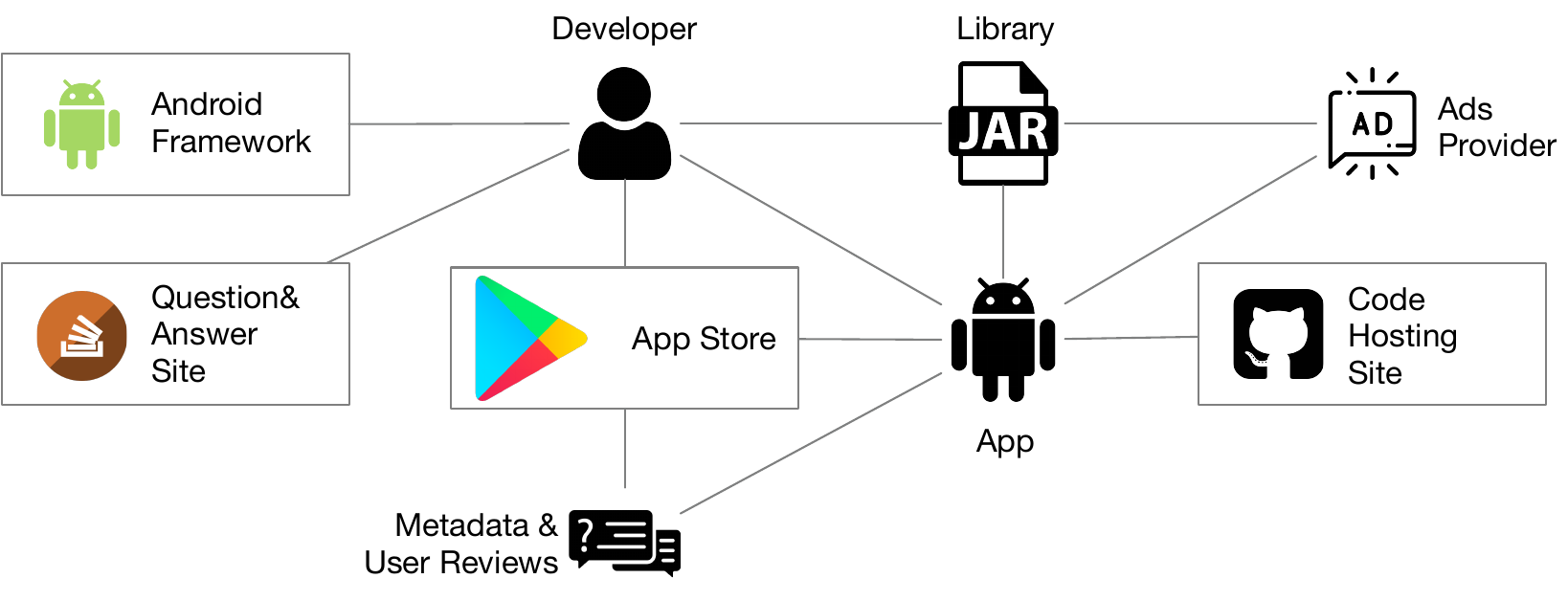}
  \caption{Overview of the Major Participants (or Artifacts) Involved in MSE Research.}
  \label{fig:fig_mse_participants}
\end{figure}

Before going into the details in summarizing the top problems targeted by our fellow researchers in MSE, we first present the major participants (or artifacts) involved in MSE research.
These participants have been closely associated with the top problems identified and handled in MSE.
As illustrated in Fig.~\ref{fig:fig_mse_participants}, \textbf{developers} play a core role in MSE, who contribute to the ecosystem by implementing \textbf{mobile apps} based on the \textbf{Android framework}  (also known as the SDK) provided by Google, along with various \textbf{third-party libraries} that are pre-developed for facilitating app developments.
The libraries also include the ones used to provide \textbf{advertisements}, which also play a crucial role in Android as they are the major source for app developers to make profits.\footnote{Indeed, app developers often cannot make profits directly from the apps per se as they are often made available to users as free apps.}
When there are problems encountered while developing an app, developers frequently resort to \textbf{question and answer website} (such as Stack Overflow) to search for solutions.
The app's source code is often managed on code hosting websites such as Github, which is also one of the most important resources leveraged by mining software repository researchers to learn for improving Android apps.
Once the apps are developed, they will be uploaded to \textbf{app stores} such as the official Google Play store, on which various \textbf{metadata} associated with the app (such as app's description, name, authors, etc.) will also be provided.
The app stores are the main portal for users to find and install apps.
Except for searching and installing apps, app stores also provide a platform for users to leave feedback (i.e., \textbf{user comments}, which could be complaints about defects or suggestions regarding new app features) for their apps on dedicated pages.

We now highlight the top problems targeted by our fellow researchers (cf. Table~\ref{tab:top_problems}). These top problems could be applied to any of the aforementioned participants highlighted in Fig.~\ref{fig:fig_mse_participants}.
The problems are mainly grouped into nine categories, including app development, app deployment, user experience, security and privacy, quality, reliability, performance, energy, and socio-technical issues.
To help readers better understand each of the categories (i.e., the actual problems handled by our fellow researchers), we also provide various problem examples in the second column of the table.

\begin{table}[!htbp]
\centering
\setlength{\belowcaptionskip}{0pt}
\setlength{\abovecaptionskip}{0pt}
\caption{The top problems targeted by the examined papers.}
\label{tab:top_problems}
\resizebox{\linewidth}{!}{
\begin{tabular}{c | p{0.9\linewidth}}\hline
Category & Problem Examples  \\ 
\hline

App Development  &  Representative problems include (1) Learning new requirements by analyzing user comments, (2) Facilitating app developments by recommending third-party libraries, APIs, code snippets, (3) Generating code for GUI components, (4) Facilitating app testing by automatically generating test cases, etc.  \\ \hline

App Deployment & Problems related to app deployment include (1) Supporting code obfuscation, (2) Supporting app hardening, (3) Supporting obfuscation for AI models inside apps. \\ \hline

User Experience & Example problems include (1) Optimizing user experience by analyzing end-user perception, (2) Understanding user satisification by analyzing user reviews (feedback on app stores), (3) Characterizing human-centric issues related to the success of apps. \\ \hline

Security and Privacy & 
Representative problems include (1) Detecting privacy leaks, (2) Discovering sensitive hidden behaviors, (3) Exploiting component hijack attacks, (4) Exploring privilege escalation attacks, (5) Uncovering cryptographic API misuses, (6) Predicting malware and its families, etc. \\ \hline

Quality &  
Representative problems include 
(1) Detecting and fixing concurrency errors in mobile apps, 
(2) Characterizing the app's maintainability by understanding the evolution of deprecated APIs, the usage of incompatible APIs,
(3) Improving effectiveness and efficiency of app testing approaches by automatically generating better test cases, estimating test efforts and prioritizing test cases.\\ \hline




Reliability & Targeted problems include understanding, locating, and automatically repairing app crashes (caused by API misuses, compatibility issues), failures, exceptions, and runtime errors. \\ \hline

Performance & 
Performance related problems include
(1) Assuring the app's efficiency by detecting and refactoring code smells and  
(2) Summarizing performance anti-patterns and their potential improving counterparts.\\ \hline

Energy & Energy Management problems include (1) Adjusting power states of processing units and (2) Exploiting computing resources, and (3) Characterizing and detecting energy issues (e.g., bugs, leaks, hogs, hotspots, wakelock, sensors, network, and display).\\ \hline

Socio-technical issues & Targeted problems include 
(1) Understanding why mobile app users do not adopt security precautions in the smartphone context and studying how to use media campaigns to raise user awareness of security issues and 
(2) Identifying the common risks that hinder mobile application development in the healthcare domain and the mitigating strategies against those risks. \\ \hline

\hline
\end{tabular}}
\end{table}

\subsection{Technique}
\begin{figure}[!t]
  \centering
  \includegraphics[width=0.8\linewidth]{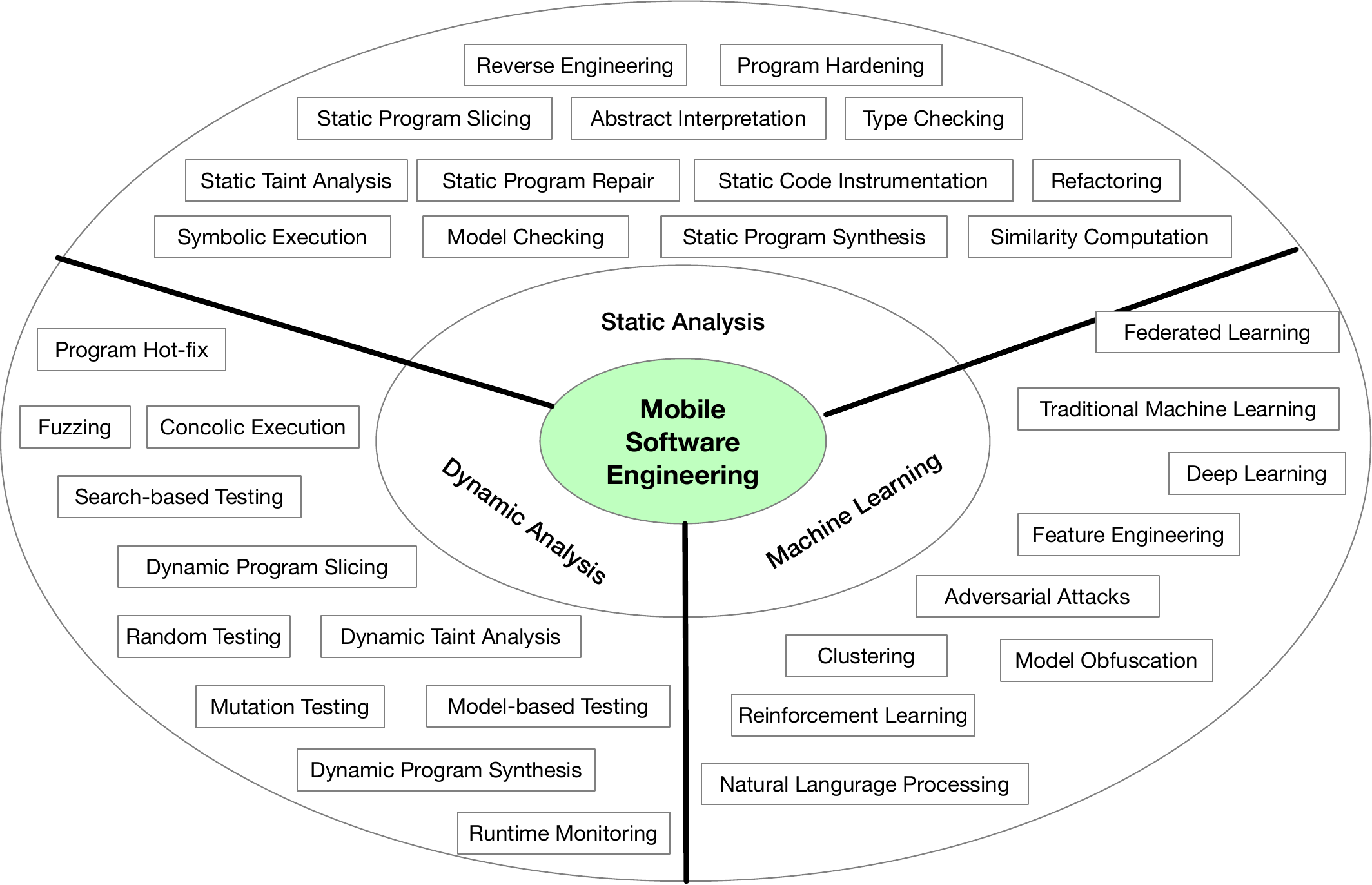}
  \caption{Overview of the Representative Techniques Adopted in MSE.}
  \label{fig:fig_mse_technique}
\end{figure}

To solve the above software engineering problems, researchers have proposed various kinds of techniques.
Note that, while there are more techniques designed to solve the above problems, e.g., trust environment execution (TEE) for increasing mobile application security, we will not include them but only consider the software engineering techniques in this work.
Also, resolving software engineering tasks often involves manual efforts, such as confirming the warnings yielded by static analyzers or labeling datasets for training machine learning models, etc. In this work, we will not take into account those manual approaches.
For the remaining techniques, after discussing them among co-authors, we preliminarily categorize them as static-based, dynamic-based, and learning-based approaches.
Fig.~\ref{fig:fig_mse_technique} highlights the represented ones.

\textbf{Static Approaches.}
Static approaches are the analysis of programs performed without executing them.
The widely used static approaches are listed in Fig.~\ref{fig:fig_mse_technique}.
These static approaches have been applied to the SE problems of mobile applications, Android frameworks and mobile operating systems.
Specifically, static approaches (e.g., taint analysis, symbolic execution, code instrumentation, model checking) are widely used to detect application bugs, including functional errors, code smells, security weaknesses/vulnerabilities, energy and performance bugs, permission escalations, and etc.
Beyond bug detection, static approaches (e.g., application hardening, code sign) are also used to increase the security and reliability of mobile applications.
Moreover, with the rapid development of machine/deep learning, we have observed an trend to use static approaches to extract program features, which are then provided to learning approaches.

\textbf{Dynamic approaches.}
In contrast with static approaches, dynamic approaches are performed on programs during their execution.
Similar to static approaches, the dynamic approaches are also applied for program testing.
Widely used dynamic testing techniques include search-based testing, black-box/random testing, grey-box fuzzing, concolic execution, event-driven test generation, mutation testing, and etc.
Dynamic program analysis are also applied for security analysis (e.g., dynamic taint analysis and runtime monitoring) and automated program repair.


\textbf{Learning-based approaches}
Beyond the traditional static and dynamic approaches, we have seen a increasing trend that applies machine/deep learning techniques to solve mobile software engineering problem.
Learning techniques train models by extracting features from large program artifacts and have achieved significant success in the field of code analysis.
Learning-based techniques have been applied to solve many mobile software engineering tasks, including vulnerability detection, privacy issues detection, program testing, code smell checking and etc.
Moreover, it has recently garnered considerable research attention to employ deep learning techniques to thwart Android malware attacks.



\section{The Research Roadmap}
\label{sec:roadmap}

As our initial attempt to prompt software engineering research for OpenHarmony, we now present the preliminary research roadmap by summarizing the research gaps between Android/iOS and OpenHarmony.
When detailing the gaps, we also present example works that we believe should be also proposed for OpenHarmony.
We hope these works could be contributed by our fellow researchers so as to fill the aforementioned gaps, making OpenHarmony a popular mobile platform and a popular research topic in the mobile software engineering field.

\begin{figure}[!t]
  \centering
  \includegraphics[width=0.8\linewidth]{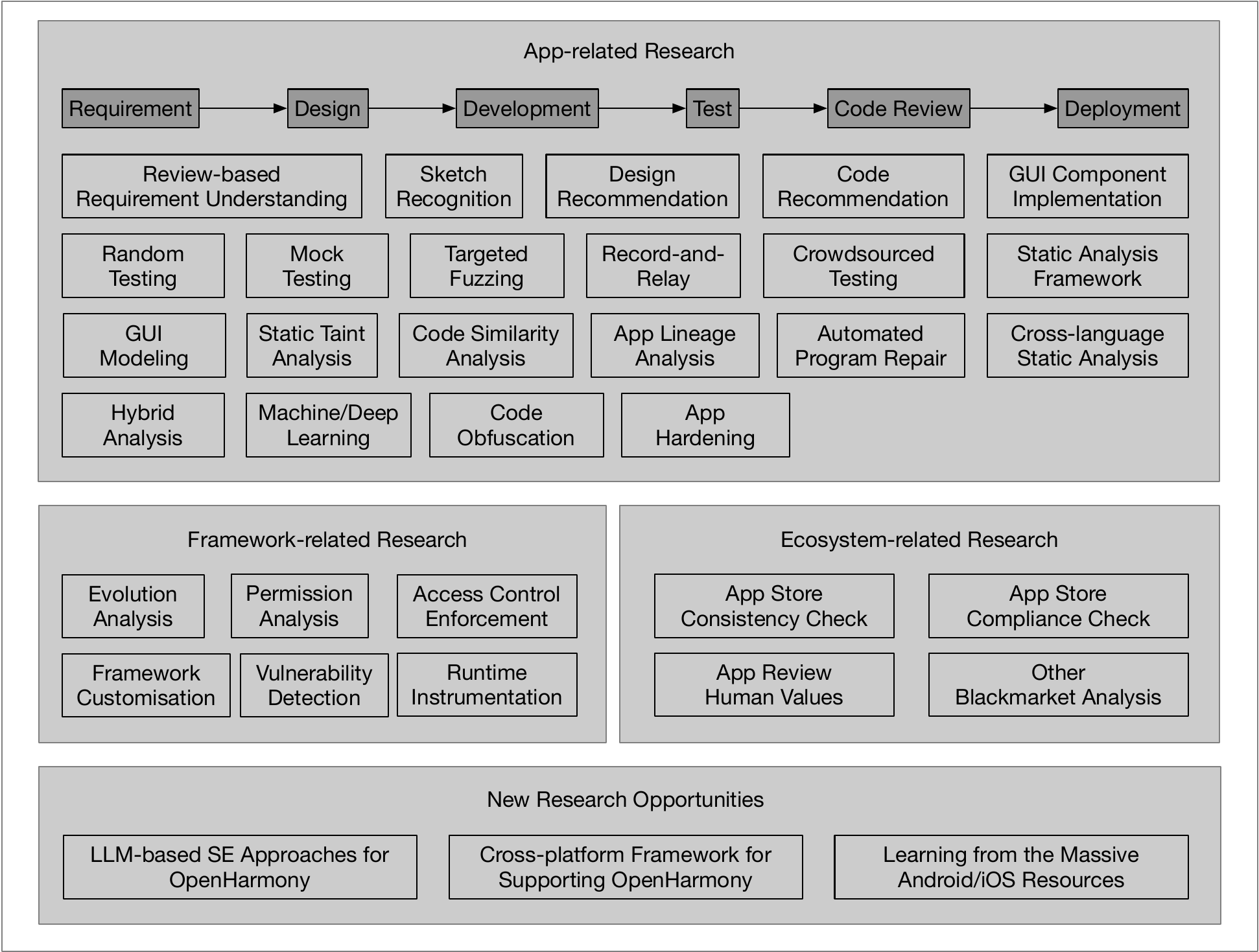}
  \caption{Overview of Research Gaps between OpenHarmony and MSE.}
  \label{fig:openharmony_arkui}
\end{figure}

\subsection{Gap in App-related Research}

As highlighted in Section~\ref{sec:mse}, the majority of MSE studies focus on mobile apps.
At the beginning of this section, we first summarize some of the representative works that propose software engineering techniques to support their studies.
Specifically, we summarize them based on the general software development processes, including Requirement, Design, Development, Test, Code Review, and Deployment.
As expected, there are fewer works that target the phases before app development.
Indeed, most of the works are proposed to examine mobile apps once they are developed.

\begin{enumerate}

\item \textbf{[Requirement] Mining User Reviews for Requirement Analysis.}
Since it is generally not possible to obtain the original requirements of mobile apps (e.g., what functions to offer and how should they be interacted with users), which are often considered confidential, the research community mainly focuses on mining user reviews for requirement understanding.
Here, user reviews can be collected either through actual interviews or through user comments made to the app's releasing page on the app store.
Fortunately, such efforts can directly be leveraged to improve OpenHarmony apps as the identified requirements are often independent to mobile platforms.
Nevertheless, the proposed techniques could be also leveraged to mine user reviews that are specifically made to OpenHarmony apps.

$\longrightarrow$\textbf{Representative Works:}
Chen et al.~\cite{chen2022grounded} argue that it is possible to dig out user needs and preferences by analyzing user online comments, which can subsequently benefit app developers to make accurate market positioning and thereby increase the volume of app downloads. By using a set of NLP techniques such as semantic analysis, word frequency analysis, the authors demonstrate the possibility to obtain useful requirements.
Similarly, Palomba et al.~\cite{palomba2018crowdsourcing} propose to support the evolution of mobile apps via crowdsourcing user reviews. By surveying 73 developers, they have found that over 75\% of developers will take user reviews into consideration when updating their apps and such updates are often rewarded in terms of significant increases of user ratings.

\item \textbf{[Design] Sketch Recognizing}
App designers often use sketches to quickly draw the app's user interfaces so as to accelerate the iterative design process when designing apps.
Such sketches, however, cannot be directly used to build a prototype app that can be immediately tested to collect user feedback.
To bridge the gap, researchers have proposed techniques to automatically recognize sketches and subsequently transform them into UI components. In this way, app developers can focus on designing the user experience rather than building the prototypes with various tools.
Such approaches could be extremely beneficial to OpenHarmony developers when designing their apps.

$\longrightarrow$\textbf{Representative Works:}
Kim et al.~\cite{kim2018identifying} have presented to the community an approach to identify UI widgets of mobile apps directly from sketch images using geometric and text analysis features.
the extraction of graphic elements such as text or shapes from the input sketch image using the Optical Character Recognition (OCR) technique and edge detection.
Similarly, Li et al.~\cite{li2017xketch} have proposed to the community a sketch-based prototyping tool called Xketch for accelerating mobile app design processes.
They have demonstrated that Xketch is indeed useful and can benefit app developers in designing apps quickly on their tablets.


\item \textbf{[Design] Visual Search for Recommending Design Examples.}
Since it is non-trivial to design a beautiful user interface from scratch, developers often resort to relative UI design examples to gain inspiration and compare design alternatives.
However, finding such design examples is challenging as existing search systems only support text-based queries.
To mitigate this, our community has proposed to conduct a visual search, which takes as input a UI design image and outputs visually similar designs.
Since visual search is independent of mobile platforms, such efforts can directly be leveraged to benefit the OpenHarmony community as well.
Nevertheless, OpenHarmony apps may have specific preferences in their UI pages, there is also a need to invent dedicated visual search systems for supporting the design of OpenHarmony apps.

$\longrightarrow$\textbf{Representative Works:}
Bunian et al.~\cite{bunian2021vins} have proposed to the community a visual search system, which includes an object-detection based image retrieval framework that models the UI context and hierarchical structure.
Based on a large-scale UI dataset, the authors have shown that their visual search framework can achieve high performance in querying similar UI designs.

\item \textbf{[Development] Code Recommendation.} Mobile apps are developed based on an official SDK with thousands of APIs, and there are hundreds of thousands of APIs available in the wild through the so-called third-party libraries. there is hence a strong need to automatically recommend appropriate APIs (or libraries) for developers to choose when they implement their apps.
Furthermore, libraries have been demonstrated to be extremely useful for facilitating app development as they provide lots of existing function implementations that are reusable and are often high-quality (e.g., being already validated by their various usages).
It is not uncommon to encourage developers to leverage third-party libraries for implementing OpenHarmony apps. As the number of available libraries keeps growing, it is non-trivial for developers to search for the appropriate libraries.
Therefore, there is a strong need to automatically recommend the required libraries for OpenHarmony app developers.

$\longrightarrow$\textbf{Representative Work:} Zhao et al.~\cite{zhao2022apimatchmaker} have presented to the community a prototype tool called APIMatchmaker that automatically matches the correct APIs for supporting the development of Android apps. The recommended APIs are learned from other Android apps that are deemed similar to the one under development.
As another example, Yuan et al.~\cite{yuan2019api} propose to leverage API search engine to recommend function APIs and they go one step further to demonstrate the need to recommend event callbacks (that need to be overridden to contain function code) for developers.

\item \textbf{[Development] GUI Component Implementation.} Mobile apps involve lots of icons. To maintain the same look and feel, similar GUI components (icons or animations) across different mobile apps often reserve similar functionalities. Therefore, it is possible to learn the semantics behind popular GUI components and subsequently recommend code implementations to developers when relevant GUI components are used.

$\longrightarrow$\textbf{Representative Work:} Zhao et al.~\cite{zhao2021icon2code} have proposed an approach called Icon2Code that leverages an intelligent recommendation system for helping app developers efficiently and effectively implement the callback methods of Android icons. The recommendation system is built based on a large-scale dataset that contains mappings from icons to their code implementations.
Similarly, Wang et al.~\cite{wang2023animation2api} have proposed an approach to recommend APIs for implementing Android UI animations. This approach constructs a database containing mappings between UI animations in GIF/video format and their corresponding APIs and subsequently leverages it to achieve the recommendation.

\item \textbf{[Test] Random Testing (Test Case Generation).} 
Mobile apps, like any other software, need to be adequately tested before being released to app users. There is no difference when applied to OpenHarmony. 
There are at least two scenarios that require generating test cases to ensure that the apps behave as expected.
The first one is to generate test cases for unit tests, which only ensures the correctness of certain functions.
The second one is to generate inputs for apps that run on Mobile operating systems. The OpenHarmony community has similar requirements.
Random testing has been regarded as one of the most useful tools for generating test cases to explore mobile apps due to its easy-of-use and scalability. Our fellow researchers have shown that Monkey, a simple approach and tool for random testing of Android apps, is surprisingly effective, outperforming much more sophisticated tools by achieving higher code coverage.
In OpenHarmony, we believe this random testing approach is also required, which is also the base for supporting the implementation of many advanced app testing tools.

$\longrightarrow$\textbf{Representative Work:}
Amalfitano et al.~\cite{amalfitano2012using} have presented to the community a research prototype named AndroidRipper, which embeds an automated technique that tests Android apps via their Graphical User Interface (by automatically explores the app's GUI with the aim of exercising the application in a structured manner). Existing experimental results show that AndroidRipper outperforms the random testing approach, being capable of detecting severe and previously unknown faults in open-source Android apps.
Li et al.~\cite{li2017droidbot} have presented to the community another automated test input generator for Android apps. The tool, named DroidBot, is designed to be lightweight (no need to instrument the app under testing) and supports customization to fulfill dedicated testing GUI-guided strategies.
Similarly, in the field of iOS, Wu et al.~\cite{wu2023cydios} have presented to the community a model-based testing framework called CydiOS for randomly testing iOS apps.
The CydiOS tool has been made publicly available as an open-source project by its authors.

\item \textbf{[Test] Mock Testing.} Performing unit tests for mobile apps, including OpenHarmony apps, is non-trivial.
Indeed, certain functions under testing require the context that is a part of the app's lifecycle or the system.
This context information is only available when the app is running on the mobile system, which is contradictory to the fact that unit tests do not expect to have the apps actually run on mobile devices.

$\longrightarrow$\textbf{Representative Work:}
There are several well-known frameworks such as Mockito in MSE that are provided by practitioners to support mock unit testing.
Similar frameworks are highly demanded by the OpenHarmony community as well.
On the research side, Beresford et al.~\cite{beresford2011mockdroid} have proposed to the community a novel approach called MockDroid that allows a user to 'mock' an app's access to a resource.
The resource is subsequently reported as empty or unavailable whenever the app requests it. Their work is demonstrated to be useful for testing mobile apps w.r.t. their tolerance to resource failures.

\item \textbf{[Test] Targeted Fuzzing.} Nowadays, mobile apps are running on touch-sensitive displays involving many GUI pages, for which each of them involving various lifecycle methods and containing various widgets that are further associated with callback methods.
Such a complex setting makes it difficult to achieve highly efficient random testing.
To mitigate this, researchers propose to conduct targeted fuzzing, e.g.,  by generating test inputs that allow the app to reach a certain state.
OpenHarmony apps are GUI-intensive as well and the drawbacks of random testing would also apply to them. Hence, there is also a strong need to invent targeted fuzzing approaches to properly test OpenHarmony apps.

$\longrightarrow$\textbf{Representative Work:}
Rasthofer et al.~\cite{rasthofer2017making} present to the community a targeted fuzzing approach, namely FuzzDroid, for automatically generating an Android execution environment where an app exposes malicious behavior.
This objective is achieved by combining an extensible set of static and dynamic analyses through a search-based algorithm that steers the app toward a configurable target location.
As another example, Azim et al.~\cite{azim2013targeted} present another approach called $A^3E$ that uses static, taint-style, dataflow analysis to first conduct a high-level control flow graph that captures legal transitions among app screens.
$A^3E$ then performs targeted exploration to achieve fast, direct exploration of activities.

\item \textbf{[Test] Record-and-Replay.} Mobile apps, after being released to the public, need to be run on multiple devices that may run different framework versions.
Those devices may also have different screen sizes with customized framework versions.
To ensure the same app can be correctly run on all of those devices, researchers propose to conduct Record-and-Replay testing, which first records a testing scenario on a device and then applies it to other devices, with the expectation of achieving the same testing results.
As one of the most important features of OpenHarmony is to support the so-called ``1+8+N'' strategy (i.e., supporting one major device (i.e., smartphone), 8 important devices such as TV, Smartwatch, Pad, PC, etc., and many other user-customized devices, this record-and-replay technique is extremely important for OpenHarmony to ensure the ``1+8+N'' strategy to be successful.

$\longrightarrow$\textbf{Representative Work:}
Gomez et al.~\cite{gomez2013reran} present a prototype tool called RERAN that achieves timing- and touch-sensitive record-and-replay for Android. RERAN attempts to directly capture the low-level event stream on the phone and replay it later on with microsecond accuracy.
Since mobile apps may be run on different devices with diverse screen sizes, a record-and-replay tool may be applied to apps that could have different GUI layouts on different devices. To accommodate that, Guo et al.~\cite{guo2019sara} present another record-and-replay tool called Sara that achieves such a purpose through an adaptive replay mechanism with a dynamic instrumentation approach for taking into account the rich sources of inputs in current mobile apps.

\item \textbf{[Test] Crowdsourced Testing.} Automated app testing cannot achieve 100\% coverage and hence user commitments are always needed in order to ensure the quality of mobile apps.
However, manually exploring an app in a comprehensive way is difficult and time-consuming.
To alleviate that, researchers have proposed to leverage crowdsourcing efforts to achieve the aforementioned testing purpose.
Indeed, crowdsourced testing provides a promising way to conduct large-scale and user-oriented testing scenarios.
Such an approach could be also leveraged to comprehensively test OpenHarmony apps.

$\longrightarrow$\textbf{Representative Work:}
Ge et al.~\cite{ge2022leveraging} find that most crowdsourced app testing is of low quality as crowd workers are often unfamiliar with the app under test and do not know which part of the app should be tested.
To fill this gap, the authors propose to construct an Annotated Window Transition Graph (AWTG) model for the app under test by merging dynamic and static analysis results and subsequently leverage the AWTG model to implement a testing assistance pipeline that offers test task extraction, test task recommendation, and test task guidance for crowd workers.
Recently, Sun et al.~\cite{sun2023taming} present to the community a lightweight approach that aims to achieve fully automated crowdsourced app testing by only dispatching the app's partial code for crowdsourced execution.
The experimental results involving tests of API-related code only (of real-world apps) show that their approach is useful (as demonstrated by being able to find many API-induced compatibility issues) and welcome in practice.

\item \textbf{[Code Review] Static Analysis Framework.} Static analysis is a fundamental technical that has been frequently applied to resolve various Android app analysis problems. Such solutions are often implemented based on the so-called static analysis frameworks that offer implementations to core static analysis functions such as control-flow graph construction, call graph constructions, etc.
OpenHarmony takes a new program language called ArkTS to develop its apps. Therefore, an ArkTS-specific static analysis framework is required to support the implementation of other purpose-oriented static analysis approaches (e.g., vulnerability detection).

$\longrightarrow$\textbf{Representative Work:}
Soot~\cite{lam2011soot} is one of the most popular static analysis frameworks that are capable of analyzing Android apps.
Soot is initially designed for Java program analysis and is further extended for Android apps (which are written in Java) thanks to the Dexpler module contributed by Bartel et al.~\cite{bartel2012dexpler}.
Another popular static analysis framework should be the one named WALA~\cite{santos2022program}, which is developed and maintained by IBM.
In Android, both Soot and WALA have been recurrently adopted by our fellow researchers to support the implementation of static analysis approaches.

\item \textbf{[Code Review] GUI Modeling.}
Android apps are driven by graphical user interfaces (GUIs), which are known to be complex for static analysis approaches.
Indeed, a given GUI page may be composed of many UI widgets that are positioned with different layout strategies and each of them may accept various user events (such as clicks).
It is even more difficult for Android apps as the GUI pages of Android apps can be written both statically (via XML file) and dynamically (via Java code directly).
Hence, it is non-trivial to programmatically understand GUI pages. Therefore, researchers have to design dedicated
approaches to properly model the GUIs of apps for analyzing their behaviors.

$\longrightarrow$\textbf{Representative Work:}
ArchiDroid~\cite{liu2023ex} statically analyzes
the transition relationship among activities of apps and constructs the activity transition graph. It also models the
activity semantic and graph structure information via graph convolution network to automatically predict transitions between activities and augment the activity transition graph built by static analysis. Besides static analysis-based approaches, SceneDroid~\cite{zhang2023scene} explores activities and extracts the GUI scenes by a series of dynamic analysis techniques, and then presents the GUI scenes as a scene transition graph to model the GUI of apps.

\item \textbf{[Code Review] Static Taint Analysis (for Detecting Privacy Leaks).}
One of the most popular usages of static analysis is to perform static taint analysis for pinpointing sensitive data flows (also known as privacy leaks). Static taint analysis works by first coloring some variables that contain sensitive data such as the user's phone number and then tracking their flows in the code. A sensitive data flow is considered detected if such colored data eventually flows to sensitive operations (e.g., sending the colored data outside the device via SMS).
OpenHarmony apps will be run on mobile devices and hence will have similar requirements.
Therefore, it is also essential ed to invent static taint analysis approaches for examining OpenHarmony apps.

$\longrightarrow$\textbf{Representative Work:}
Arzt et al.~\cite{arzt2014flowdroid} have presented to the MSE community an open-source tool called FlowDroid, which performs context-, flow-, field-, and object-sensitive and lifecycle-aware taint analysis for Android apps.
The authors further provide on-demand algorithms for FlowDroid to achieve high efficiency and precision at the same time.
Based on FlowDroid, researchers further present to the community three extensions, namely IccTA~\cite{li2015iccta}, DroidRA~\cite{li2016droidra,li2016reflection,sun2020taming}, and SEEKER~\cite{sun2021characterizing}, that perform static taint analysis by additionally considering apps' inter-component communication, reflection, and sensor-related features, respectively.
In the field of iOS, Egele et al.~\cite{egele2011pios} present a tool called Pios, which leverages static taint analysis to detect privacy leaks in iOS apps.

\item \textbf{[Code Review] Code Similarity Analysis.}
Code similarity analysis is another common application of static analysis that has also been recurrently adopted by developers to achieve various functions, e.g., to detect code clones, the usage of third-party libraries, and repackaged (or piggybacked) apps.
Code similarity analysis is also essential to understand the difference between two code snippets, including the two timestamped versions of the same code snippet.
Such a difference can then be leveraged to support the implementation of various software engineering tasks such as automatically generating commit messages or inferring patches to given code defects, etc.

$\longrightarrow$\textbf{Representative Work:}
Russell et al.~\cite{crussell2013andarwin} have presented to the MSE community a prototype tool called AnDarwin for detecting semantically similar Android apps. 
AnDarwin leverages a clustering-based approach, for which it attempts to cluster similar apps into the same group based on semantic information extracted from the apps’ code.
More recently, Li et al.~\cite{li2017simidroid,li2019identifying} have developed another prototype tool called SimiDroid that aims at identifying similarities in Android apps through static code analysis.
The SimiDroid tool is designed to be a generic framework that can be easily extended to support multi-level comparisons of Android apps.
The authors have demonstrated the usefulness of SimiDroid by achieving efficient similarity analysis of Android apps in three scenarios (Resource, Component, and Method-level comparisons).

\item \textbf{[Code Review] App Lineage Analysis.}
Due to the fast evolution of the OS framework as well as the requirement to fix bugs or add new features, mobile apps are continuously updated by their developers (often over app stores).
Such updates will lead to a series of releases of the same app, which is referred to by the community as app lineages.
Because these app lineages have recorded all the app changes, our fellow researchers have proposed to mine them\footnote{Researchers have to focus on the app's released versions because it is often not possible to obtain its source code.} to learn why the mobile apps updated.
Similar approaches could also be applied to OpenHarmony, e.g., to mine knowledge for updating (or fixing) existing apps.

$\longrightarrow$\textbf{Representative Work:}
Gao et al.~\cite{gao2019understanding} present an experimental study about the evolution of Android app vulnerabilities. They first define the term ``app lineage'' (i.e., the series of a given app's historical versions).
Then, they collect a dataset of app lineages and subsequently leverage it to understand the vulnerability evolution by mining the updates between an app's two consecutive versions.
Their empirical study has revealed various interesting findings.
The authors further conduct another work to mine app lineages for understanding the evolution of Android app complexities~\cite{gao22019evolution}.
Their experimental results reveal a controversial finding where app developers might not really be aware of controlling the complexity of their apps.

\item \textbf{[Code Review] Automated Program Repair.}
Automated Program Repair (APR) has been a hot topic in the software engineering community for years. The idea of APR is for computers to automatically produce source code-level patches for bugs and vulnerabilities.
Our fellow researchers have also attempted to invent techniques to automatically repair mobile apps.
We argue that such techniques should also be explored to target OpenHarmony apps.

$\longrightarrow$\textbf{Representative Work:}
Marginean et al.~\cite{marginean2019sapfix} present an industry tool called SapFix that achieves end-to-end fault fixing, from test case design to deployed repairs in production code.
SapFix achieves its purpose by combining a number of different techniques, including mutation testing, search-based software testing, and fault localization.
Zhao et al.~\cite{zhao2022towards} have presented to the community another prototype tool called RepairDroid, which aims at automatically repairing compatibility issues directly in published Android apps (at the bytecode level).
To support flexible repair, the authors have introduced a generic app patch description language that allows users to create fix templates using IR code.

\item \textbf{[Code Review] Cross-language Static Analysis.}
Mobile apps are not always written in a single programming language.
Indeed, there are various apps that are implemented in multiple languages.
For example, the module requiring high performance in Android apps could be written in C or C++ while the main part is still written in Java, which is the default language to implement Android apps.
As another example, for such Android apps that leverage web-related components, certain functions could be written in Javascript, in order to supplement the main functions written in Java.
In order to properly analyze these apps involving multiple programming languages, we argue that there is a need to conduct cross-language static analysis, for which the data-flow analysis should propagate variables from one language to another.

$\longrightarrow$\textbf{Representative Work:}
Wei et al.~\cite{wei2018jn} and Zhou et al.~\cite{zhou2021finding} demonstrate that it is important to support inter-language static analysis in order to support security vetting of Android apps.
To do so, Samhi et al.~\cite{samhi2022jucify} present to the community a prototype tool called Jucify that aims to unify Android code (between Java and C/C++) to support static analysis. 
Their work is able to build a comprehensive call graph across all the methods written in the app, no matter they are written in Java or C/C++.
Xue et al.~\cite{xue2018ndroid} have also invented a prototype tool called NDroid for tracking information flows across multiple Android contexts, including the analysis of native code in Android apps~\cite{zhou2022ncscope}.

\item \textbf{[Code Review] Hybrid Analysis.}
As discussed previously, both testing (also known as dynamic analysis) and static analysis techniques are recurrently adopted by our fellow researchers to dissect mobile apps.
However, both of these two techniques are known to have drawbacks, e.g., testing approaches suffer from code coverage problems that eventually lead to false negative results, meanwhile, static analysis is known to likely yield false positive results.
To mitigate this, our fellow researchers have proposed to combine these two approaches to conduct the so-called hybrid analysis of mobile apps.
We believe there is also a need to invent hybrid approaches for analyzing OpenHarmony apps.

$\longrightarrow$\textbf{Representative Work:}
Wang et al.~\cite{wang2018automated} present an automated hybrid analysis of Android malware through augmenting fuzzing with forced execution. They propose an approach called DirectDroid, which aims to trigger hidden malicious behavior by bypassing some related checks when adopting fuzzing to feed the necessary program input.
Spreitzenbarth et al.~\cite{spreitzenbarth2013mobile} have developed another hybrid analysis approach called Mobile-Sandbox, for which static analysis is leveraged to reach higher code coverage during dynamic analysis (i.e., app testing).

\item \textbf{[Code Review] Machine/Deep Learning.}
Machine Learning has become one of the most popular techniques that are frequently adopted by our fellow researchers for reviewing apps’ logic code.
Indeed, a lot of research efforts are spent to find the best feature set that could closely represent the app's behavior.
Such a feature set is then leveraged to support two types of machine learning approaches: supervised learning and unsupervised learning.
Supervised learning requires knowing the labels of the training dataset, e.g., it is essential to collect a set of known malware in order to train a malware predictor.
On the contrary, unsupervised learning does not need to know the labels of the dataset. This type of approach is often used to cluster similar samples into the same group.
When deep learning is concerned, feature engineering is no longer needed.

$\longrightarrow$\textbf{Representative Work:}
Liu et al.~\cite{liu2022deep} have recently conducted a systematic literature review about deep learning approaches applied to defend Android malware.
The authors have surveyed papers published from 2014 to 2021 and have located 132 closely related papers.
The authors find that static analysis is the most used technique to obtain features from Android apps and there are 13 works that achieve malware classification by directly encoding the raw bytecode of Android apps into feature vectors.
Machine learning is not only applied to dissect malware but is also used for resolving other software engineering tasks.
For example, Rasthofer et al.~\cite{rasthofer2014machine} have presented to the community a machine learning-based approach for classifying and categorizing sources and sinks in Android, which can then be leveraged to support taint analysis of Android apps, so as to detect privacy leaks.

\item \textbf{[Deployment] Code Obfuscation.}
Because of the nature of mobile devices, mobile apps need to be downloaded to the devices before installation.
This, unfortunately, makes it possible for attackers to directly access the mobile apps.
Even worse, the attackers might be able to directly access the code implementations of the apps if reverse engineering techniques are applied.
To prevent attackers from easily understanding the code, the MSE community has adopted the practice to perform code obfuscation before assembling the app code to a release version.
Since OpenHarmony apps need to be installed on users' devices, it is also essential to invent code obfuscation techniques to prevent OpenHarmony apps from being exploited by attackers.

$\longrightarrow$\textbf{Representative Work:}
Aonzo et al.~\cite{aonzo2020obfuscapk} have developed an open-source black-box obfuscation tool for Android apps. The authors named their approach Obfuscapk and have designed a modular architecture for users to straightforwardly extend so as to support the implementation of new obfuscation strategies.
Dong et al.~\cite{dong2018understanding} conduct a large-scale empirical study of Android obfuscation techniques, with the hope of better understanding the usage of obfuscation. The authors have specifically looked into four popular obfuscation approaches: identifier renaming, string encryption, Java reflection, and packing, leading to various findings that could help developers select the most suitable obfuscation approach.

\item \textbf{[Deployment] App Hardening.}
Code obfuscation is a useful technique to prevent attackers from easily understanding the code.
It is nonetheless not possible to prevent attackers from obtaining the code.
With the help of deobfuscation approaches, attackers could still understand the implementation details.
To prevent that from happening, the MSE community further introduced to the community the so-called aWpp hardening technique, which aims to make it difficult to extract code implementation from the apps (e.g., will stop reverse engineering tools from disassembling released apps).

$\longrightarrow$\textbf{Representative Work:}
Russello et al.~\cite{russello2013firedroid} present to the MSE community a policy-based framework called FireDroid that enforces security policies without modifying Android OS or the actual applications.
FireDroid includes a novel mechanism to attach, monitor, and enforce policies for any process spawned by the Android’s mother process Zygote.
With that, FireDroid can be applied to block OS and app vulnerabilities, hardening security on Android phones.
Zhang et al.~\cite{zhang2015dexhunter} have conducted the first systematic investigation on Android packing services toward understanding the major techniques used by state-of-the-art packing services and their effects on apps.
They further find that the protection given by those packing services is not reliable, i.e., the Dex can be recovered.
To demonstrate that, the authors have designed and implemented a prototype tool called DexHunter for extracting Dex files from packed Android apps.
Following that, Xue et al.~\cite{xue2021happer,xue2021parema,xue2020packergrind} have gone steps further to achieve unpacking through various methods, e.g., Hardware-assisted approach, VM-based approach, etc. 

\end{enumerate}

\subsection{Gap in OS Framework-related Research}

As highlighted in Fig.~\ref{fig:openharmony_overview}, OS framework is the layer that connects the apps with the system capabilities. It provides all the necessary capabilities (including all the APIs offered by the SDK) to support apps running on mobile devices.
Since this part is closely related to apps, it has also been a frequent topic targeted by our SE researchers.
We now summarize the representative ones.

\begin{enumerate}

\setcounter{enumi}{19}

\item \textbf{[Static] Evolution Analysis.}
Like what has been done for mobile apps, our fellow researchers have also proposed approaches to study the evolution of OS frameworks.
They have shown that understanding the evolution of the framework could provide useful information for the mobile community.
However, unlike mobile apps, the studies related to the evolution of OS framework are mainly based on source code as the framework (mainly Android framework) is open-sourced.
Since the OpenHarmony's framework is also open-sourced, such techniques applied to study the evolution of the Android framework could be also applied to OpenHarmony.

$\longrightarrow$\textbf{Representative Work:}
Li et al.~\cite{li2020cda} have proposed to study the evolution of the Android framework to characterize deprecated APIs. Their empirical study has revealed various interesting findings including the inconsistency among the API's implementation, its comments, and annotations.
They have also found that the Android framework includes a lot of inaccessible APIs that are not designed to be invoked by client apps but have actually been accessed in practice~\cite{li2016accessing}.
As argued by Liu et al.~\cite{liu2021identifying}, by looking into the evolution of Android APIs, we could find the silently evolved APIs that could eventually lead to indiscoverable compatibility issues~\cite{sun2023taming} as the API's implementation is updated during the evolution while its comment remained the same.

\item \textbf{[Static] Permission Analysis.}
A lot of research efforts have been put in by our SE community to understand Android's permission system, which has been regarded as the primary mechanism to ensure the security of apps and the system.
Ideally, the apps should declare the permission they need to properly run on given mobile devices.
However, there is no clear mapping about permission to APIs provided to app developers.
As a result, app developers often declare more permissions than what the apps actually need, resulting in an enlarged attacking surface. 
Our fellow researchers have hence proposed to dissect the framework code to build such a mapping and subsequently to support more fine-grained permission analyses.
Since OpenHarmony also includes a permission system to ensure the security of apps, similar permission-related weaknesses that have been discovered in Android could also happen in OpenHarmony.
Hence, there is also a strong need to conduct similar research to ensure the proper use of permissions in OpenHarmony.

$\longrightarrow$\textbf{Representative Work:}
Au et al.~\cite{au2012pscout} present to the community a prototype tool called PScout that automatically extracts the permission specification from the Android OS source code (i.e., over a million lines of code) using static analysis.
Their approach has resolved several challenges including the one to take into account permission enforcement due to Android's use of IPC and Android's diverse permission-checking mechanisms.
Bartel et al.~\cite{bartel2014static} have conducted a similar study by leveraging static analysis for extracting permission checks from the Android framework.
Their approach is designed to be field-sensitive with an advanced class-hierarchy analysis strategy and uses novel domain-specific optimizations dedicated to Android.

\item \textbf{[Static] Access Control Enforcement.}
Security is not only the biggest problem in mobile apps, it is also one of the biggest problems in the OS framework side.
To ensure the security of the system, the OS framework often relies on access control mechanisms to achieve the purpose.
However, such access control mechanisms could be bypassed by malware so as to achieve unauthorized security-sensitive operations.
Therefore, there is a need to enforce the access control function being properly applied.

$\longrightarrow$\textbf{Representative Work:}
Zhou et al.~\cite{zhou2022uncovering} have presented to the community a prototype tool called IAceFinder that aims to extract and contrast the access control enforced in the Java and native contexts of Android and subsequently to discover cross-context inconsistencies, as a major means to stop access control functions from being bypassed.
The authors have applied their approach to analyzing 14 open-source Android OS frameworks (i.e., ROMs), from which they are able to disclose 23 inconsistencies that can be abused by attackers to compromise the device.

\item \textbf{[Static] Framework Customization.}
Due to the openness of Android and the requirement to provide vendor-specific user experience, the Android framework has been recurrently customized by smartphone vendors. For example, Xiaomi has done that and named the customized version MIUI. Similarly, Huawei has released EMUI to feature a more personalized user experience when using Huawei phones.
Unfortunately, such a wide range of customizations has introduced significant compatibility issues to the community, making it difficult for app developers to implement an app that is compatible with all the available mobile devices.
Our SE researchers have hence proposed approaches to mine the difference between the customized frameworks so as to mitigate the compatibility issues in the mobile community.
As an open-source system, OpenHarmony could face similar problems. Therefore, there is also a need to spend research efforts to control the customization and thereby keep such problems from happening in OpenHarmony.

$\longrightarrow$\textbf{Representative Work:}
Liu et al.~\cite{liu2022customized} have conducted an empirical study to understand whether customized Android frameworks keep pace with the official Android.
They have looked at the evolution of eight downstream frameworks (e.g., AOKP, AOSPA, LineageOS, SlimROMs, etc.) and discovered various interesting findings (e.g., Downstream projects perform merge operations only for a small portion of all the version releases in the upstream project and most of the downstream projects take more than 20 days to bring changes from their corresponding upstream projects).
The authors further look at the differences among the customized frameworks (including the ones modified by popular technical companies such as Xiaomi and Huawei) and find that this customization has led to serious compatibility issues (also known as the fragmentation problem) in the Android community~\cite{liu2023automatically}.
This result strongly suggests that more efforts are required to ensure framework customization is properly handled and managed.

\item \textbf{[Static/Dynamic] Vulnerability Detection.}
Due to the complexity and huge codebase of the Android system, vulnerable implementations commonly exist in
different aspects of the Android framework. 
There is hence a need to continuously scan for vulnerabilities so as to improve the system's security.
Our fellow researchers have hence proposed various approaches to achieve that, either statically or dynamically.
Note that mobile frameworks are often developed with multiple programming languages, vulnerability detection approaches are hence required to support cross-language analyses.

$\longrightarrow$\textbf{Representative Work:}
Luo et al.~\cite{luo2019tainting} have proposed a tool called CENTAUR that discovers the vulnerable interfaces of Android
system services that can be exploited by malicious apps to steal private data. In detail, CENTAUR leverages
symbolic execution and taint analysis to monitor the variables in the Android framework, which can be compromised
by malicious apps to steal private data. 
In dynamic analysis, Liu et al.~\cite{zhou2022uncovering} proposed an approach called FANS that employs fuzzing techniques to detect vulnerable system services.
It statically analyzes the data structure of each parameter of the interfaces of system services and then
randomly generates arguments to drive the execution of interfaces for triggering vulnerabilities in system services.

\item \textbf{[Dynamic] Runtime Instrumentation.}
Since it is not possible to resolve all the issues statically, researchers have also explored the possibility of dynamically analyzing the framework, e.g., to control the execution of the framework.
One of the representative works is to instrument the framework to add hook methods to interested functions. At runtime, such hook methods, when executed, will provide runtime information of the framework, which has been demonstrated to be useful for comprehending the behavior of the framework, so do the apps running on it.
Such a useful technique should be also provided to the OpenHarmony community so as to allow the implementation of many advanced framework/app analysis approaches.

$\longrightarrow$\textbf{Representative Work:}
One of the most famous runtime instrumentation approaches in Android is the Xposed framework, which allows developers to install little programs (called modules) to Android devices to customize their look and functionality.
On the research side, Costamagna et al.~\cite{costamagna2016artdroid} present a similar approach called ARTDroid that supports virtual method hooking on Android ART runtime.
As another example, the most representative work related to runtime instrumentation is the one proposed by Enck et al.~\cite{enck2014taintdroid}, who have presented to the MSE community one of the first approaches targeting runtime instrumentation in Android.
They have implemented an information-tracking system called TaintDroid, aiming to achieve real-time privacy monitoring on smartphones.
The runtime instrumentation of TaintDroid is enabled by leveraging Android's virtualized execution environment.

\end{enumerate}

\subsection{Gap in Ecosystem-related Research}
Except for the aforementioned research studies related to mobile apps and frameworks, there are also a significant number of studies focusing on the other aspects of MSE, for which we refer to in this work as ecosystem-related studies.
We now discuss some of the representative ones.

\begin{enumerate}
\setcounter{enumi}{23}

\item \textbf{[App Store] Consistency Check.}
App store has become a touchstone experience of modern living and has penetrated into many distinctive platforms.
The most famous app stores are the Google Play store and the Apple Store, which are set up to facilitate the discovery, purchasing, installation, and management of Android and iOS apps for Android phone and iPhone users, respectively.
These app stores essentially form a central repository that records a large list of available apps and their metadata, which is considered useful for helping users discover the app and subsequently decide whether to purchase it or not.
To keep the healthy of the ecosystem, app store maintainers will often set up a vetting system to filter low-quality apps (e.g., the app contains vulnerabilities or suffers from compatibility problems)  from entering the store.
Here, the metadata often contains two types of information: (1) the ones provided by the app authors such as the app's name and the app's description, and (2) the ones collected by the platform such as the app's user rating, etc.
In this section, since app analysis has been well discussed already, we will only focus on the metadata side and argue that the app and its metadata need to be kept consistent.
If not, the experience of using the app will be significantly impacted and that negative feeling could further be propagated to the experience of using the app store.

$\longrightarrow$\textbf{Representative Works:}
Gorla et al.~\cite{gorla2014checking} have proposed to check app behavior against app descriptions as they believe that there is no guarantee the code of the app does what it claims to do when uploaded to the app store. Their experimental results on a set of 22,500+ Android apps show that such inconsistency indeed exists in the community, confirming the hypothesis that the app store does not yet perform consistency checks at the time when apps are uploaded.
Another closely related example is the one proposed by Hu et al.~\ref{hu2020mobile} who introduce to the community a new type of attack called \emph{Mobile App Squatting}.
In ``App Squatting'', attackers release apps (on app stores) with identifiers (e.g., app name or package name) that are confusingly similar to those of popular apps or well-known Internet brands.
With such tricks, attackers hope to have their apps selected by app users who do not intend to use them.
All the aforementioned problems could be avoided if the app stores performed thorough consistency checks.

\item \textbf{[App Store] Compliance Check.}
Except for consistency checks, there is also a need to perform compliance checks before allowing mobile apps submitted to app stores.
There are various policies that mobile apps need to follow. 
Such policies include the ones made by the government (e.g., the General Data Protection Regulation (GDPR) by the European Union), by the app store itself (e.g., the Spam and Minimum Functionality policies by Google Play), as well as the ones made by certain libraries (the content policies and behavioral policies by AdMob \& AdSense.)
These compliance checks should be also conducted for vetting OpenHarmony apps and hence dedicated efforts are needed to implement such approaches.

$\longrightarrow$\textbf{Representative Works:}
Fan et al.~\cite{9251060} have conducted a study to explore the violations of GDPR compliance in Android eHealth apps.
Their experimental study shows that such violations (including the incompleteness of privacy policy, the inconsistency of data collection, and the insecurity of data transmission) are indeed widely presented in the Android community.
Zhao et al.~\cite{zhao2023mobile} have performed a study to check if mobile ads are in compliance with the app's age group.
Dong et al.~\cite{dong2018mobile} have conducted an exploratory study to understand how mobile apps violate the behavioral policies given by ad libraries.
All the aforementioned works have confirmed that there are lots of compliance violations in the current mobile community, which eventually lead to poor user experiences.
Therefore, to avoid such, we argue that the app store of OpenHarmony should support compliance checks so as to restrict the happening of such compliance violations.

\item \textbf{[App Review] Human Values.}
Mobile apps are essentially developed for users and it is necessary to consider the relationship between human values and the development and deployment of mobile apps.
Indeed, it has been demonstrated that violation of human values such as privacy, fairness, integrity, curiosity, honesty, or social justice by mobile apps (or technology in general) will lead to significant negative consequences.
If such violations could be identified earlier, developers can look to better address them and thereby mitigate them in the first place (e.g., before the apps are released to their users).
Since human values should also be the `first citizen' in OpenHarmony, such violation detection approaches should be also supported.

$\longrightarrow$\textbf{Representative Works:}
Obie et al.~\cite{obie2021first} have presented to the MSE community the first study about human values-violation in app reviews given by real-world app users.
Through 22,119 app reviews collected from the Google Play store, the authors find that 26.5\% of the reviews contained text indicating user-perceived violations of human values, with benevolence and self-direction as the most violated value categories.
Obie et al.~\ref{obie2022violation} further go one step deeper to look at the violation of honesty in mobile apps and subsequently propose approaches to automatically detect them.
Their study shows that honesty violation is quite common and top violation categories include unfair cancellation and refund policies, false advertisements, delusive subscriptions, cheating systems, etc.
These approaches highlight the need for proactive approaches taken by the community to better embed human values in OpenHarmony apps.

\item \textbf{[Other] Black Market Analysis.}
With the fast growth of the mobile ecosystem, attackers have attempted to explore various ways to gain illegal profits (e.g., through some hidden malicious behaviors).
For example, attackers have attempted to gain profits by injecting advertisements in benign apps or by sending SMS messages to premium-rate numbers.
Others have attempted to collect user private info (by leveraging single devices or by accumulating the info from multiple devices) and subsequently sell them to third parties for supporting other malicious activities.
The aforementioned types of activities are referred to as \emph{black market} by our fellow researchers who have spent lots of effort to understand and subsequently defend against them in the mobile ecosystem.
Unfortunately, as long as there are opportunities to gain profits, there will be malicious people to exploit it.
This applies to OpenHarmony as well.
Therefore, we argue that there is a strong need to mitigate the black market for OpenHarmony and we invite our fellow researchers to collaboratively explore this important research direction.

$\longrightarrow$\textbf{Representative Works:}
Gao et al.~\cite{gao2021demystifying} have conducted an exploratory study to demystify illegal mobile gambling apps, which have become one of the most popular and lucrative underground businesses.
Their study reveals that, in order to bypass the strict regulations from both government authorities and app markets, the devious app authors have developed a number of covert channels to distribute their apps and abused fourth-party payment services to gain profits.
As another example, Hu et al.~\cite{hu2019dating} have performed a thorough study to understand the ecosystem of fraudulent dating apps, which attempt to lure users into purchasing premium services to start conversations with other (likely fake female) accounts, i.e., chatbots.
All of such black market behaviors could happen to the OpenHarmony community as well so dedicated approaches are required to keep that from happening.

\end{enumerate}

\subsection{New Research Opportunities}

\begin{itemize}

\item \textbf{LLM-based SE Approaches for OpenHarmony.}
As summarized in Section~\ref{sec:mse}, the majority of Mobile Software Engineering research works focus on the analyzing phase. There are only a limited number of studies focusing on app development phases. 
This does make sense as Android app development has already been quite mature (with a lot of support from Google and the community) when our fellow researchers jumped into this field.
This is, however, not the case for OpenHarmony.
Indeed, OpenHarmony is still at a very early stage, with only a small number of apps developed and a limited number of third-party libraries made available to the community.
It will be extremely beneficial to the OpenHarmony community if there are more works proposed to facilitate the development of OpenHarmony apps.
Now, with the fast development of large language models (especially the development-focused ones such as Github's Copilot), we feel this is an even better opportunity to support that now.
LLMs could help developers quickly learn the basic knowledge of OpenHarmony, understand the usage of APIs, automatically generate  code (one line or multiple lines), generate unit test cases, recommend repair options, etc.

\item \textbf{Cross-platfrom Framework for Supporting OpenHarmony.}
To embrace the idea of developing once, running everywhere, the MSE community has invented the so-called cross-platform frameworks such as ReactNative and Flutter to support that.
These cross-platform frameworks by themselves have defined a way to develop the universal app. For example, with ReactNative, the codebase of the app is usually formed via Javascript. This codebase can then be compiled into both a native Android app and a native iOS app.
The best part of using cross-language platforms is that the app's maintenance is also unified. No matter it is to fix bugs or add new features, it only needs to be done once.
Considering this great benefit, we believe it will be extremely helpful to OpenHarmony's ecosystem if these cross-platform frameworks can support OpenHarmony.
In that case, all the existing apps that are developed via cross-platform frameworks can be directly running on OpenHarmony devices.
Therefore, we highly recommend our fellow researchers considering exploring this research direction.

\item \textbf{Learn from Android/iOS.}
In this work, we have summarized lots of Android/iOS-related approaches and believe it is necessary to learn from them by building dedicated approaches for OpenHarmony.
While that is certainly true, we also feel that there is a need to learn from the large number of artifacts accumulated in Android and iOS.
Indeed, the MSE community has gained a lot of artifacts, including millions of real-world apps, thousands of open-source apps, documentation, question-and-answer records, user reviews, etc.
Although harvested from different platforms, we argue that these artifacts could be still useful for supporting the implementation of OpenHarmony-related tasks.
For example, one possibility is to explore the direction of automatically transforming the Java-written Android apps (or Swift-written iOS apps) to ArkTS-written OpenHarmony apps.
In this work, we also invite our fellow researchers to explore this direction, flourishing the OpenHarmony ecosystem by standing on the shoulders of giants.

\end{itemize}

\section{Discussion}

OpenHarmony, as an emerging mobile platform, is still in its early stage, and so is OpenHarmony-focused software engineering research.
As summarized previously, although there are plenty of opportunities for our fellow researchers to explore in this field, there are still various challenges that need to be addressed.
In this section, we highlight some of the representative ones.

\subsection{Challenges in App/Library Development.}
In this work, we have highlighted the gaps that require to be filled in order to catch up with the popular mobile platforms (i.e., Android and iOS).
Towards filling the gaps we argue that there are still a number of challenges that need to be addressed.

\textbf{Lacking Data for (AI-based ) Learning.}
The rise of large language models has been demonstrated to be promising for automated code generation, automated test case generation, library API recommendation, etc.
However, it is not yet possible to directly achieve that for OpenHarmony as there is generally no data available for training (or fine-tuning).
Even with a set of OpenHarmony-related software data (e.g., ArkTS code and its comments), there is also a requirement to further distill high-quality ones in order to achieve a highly precise large language model, as the performance of large language models is known to be highly correlated with the quality of the training dataset.

\textbf{Lacking Third-party Libraries.}
At the moment, there are only a limited number of libraries (in ArkTS) available for supporting the implementation of OpenHarmony apps.
The lack of third-party libraries makes it difficult for developers to implement OpenHarmony apps as many of the functions need to be developed from scratch.
To fill this gap, the OpenHarmony community is currently encouraging practitioners and researchers to translate popular libraries in other languages to ArkTS.
However, this simple translation campaign will introduce another challenge, which is to keep updating the library following the updates of the original version.
To that end, we argue that dedicated efforts are required to ensure the maintainability of these libraries.

\subsection{Challenges in App/Library Analysis}
After app (or library) development, there is a strong need to ensure that the app/library satisfies the requirements and is of high quality.
The relevant challenges include the newly designed system architecture of OpenHarmony, the comprehensive GUI interactions, the newly introduced app programming language, etc.
We now summarize the representative ones.

\textbf{System-related Challenges.}
The Android system has introduced various challenges to the software engineering community in order to develop automated approaches to analyze Android apps.
First, Android takes components to construct apps, for which the components themselves are independently developed.
The components will not be directly connected at the code side and the actual invocation (via the so-called Inter-Component Communication (ICC) mechanism) will be done over the system.
This ICC mechanism could also be leveraged to implement inter-app communications, making it a challenge to perform inter-app analyses.
Second, the components in Android are designed to be run over a set of pre-defined methods (known as lifecycle methods) that will be triggered by the system following a certain order.
These lifecycle methods are not connected at the code site as well, making it also a challenge for static app analysis (from the analyzer's point of view, there is no relationship between two lifecycle methods, despite they may be continuously called by the system).
Third, similar to that of lifecycle methods, there are callback methods that are not directly connected to the app code as well.
These callback methods are directly invoked by the system when certain events (either system events such as receiving an SMS or UI events such as clicking a button) are triggered.
OpenHarmony generally shares the same challenges as that of Android.

\textbf{GUI-related Challenges.}
The GUI part has been known to be a challenge for precisely analyzing Android apps.
First of all, a given GUI page often contains a comprehensive view tree that includes various widgets with different types positioned via different layout strategies.
The widgets in the GUI page are further associated with interactive actions (e.g., a button is associated with a click event).
Furthermore, a given GUI page may contain different groups of widgets that will only be rendered if a certain condition is satisfied.
In OpenHarmony, the analysis of GUI pages is even more challenging as its design principle encourages to use of a single component (i.e., Ability) to implement multiple visual pages, which would be implemented via multiple components (i.e., Activities, one page per Activity) in Android.

\textbf{Language-induced Challenges.}
The language used to implement mobile apps per se may introduce challenges to the software engineering community.
For example, in the Android world, the reflection mechanism (inherited from Java) has been known to be a challenge for static analysis.
OpenHarmony takes a new language called ArkTS for developers to implement OpenHarmony apps and the ArkTS language per se may introduce various challenges to the software engineering community as well.
Indeed, ArkTS allows defining functions with optional parameters and default parameters, which may cause inconsistency between the function signature and its usage in practice.

\section{Related Work}

OpenHarmony software engineering is in its early stage and there are only limited works contributed to this field.
Indeed, as highlighted in Section~\ref{subsec:oh_research}, there are only 8 papers presented on this aspect.
In this section, we will not discuss these OpenHarmony-related works anymore.
Instead, we take this opportunity to highlight related works that provide a research roadmap or position statement for guiding a new research field, or a survey including literature reviews for summarizing a mature research direction.
We now highlight the representative ones.

\textbf{Research Roadmap.}
One of the most representative research roadmap reports is the one presented by Cheng et al.~\cite{chengsoftware} who have proposed to conduct software engineering research for self-adaptive systems.
After thorough discussions among the authors at a Dagstuhl seminar on Software Engineering for Self-Adaptive Systems, the authors have identified four views that are deemed essential to the software engineering of self-adaptive systems.
For each view, the authors then summarize the state-of-the-art and highlight the challenges that should be addressed in order to achieve the final goal, i.e., the software is able to automatically cope with the complexity of today’s software-intensive systems.
The authors have released another version (called the second research roadmap) five years later after the success of the first version.
The goal of this second roadmap paper~\cite{de2013software} remains the same, i.e., to summarize the state-of-the-art and to identify critical challenges for the systematic software engineering of self-adaptive systems.
Other representative research roadmap papers include the one proposed by France et al.~\cite{france2007model} who advocate model-driven development of complex software as well as the one proposed by Papazoglou et al.~\cite{papazoglou2008service} who advocate service-oriented computing as a new computing paradigm for supporting the development of rapid, low-cost and easy composition of distributed applications.
Both of these works have summarized the state-of-the-art and challenges faced by ongoing research activities.
More recently, McDermott et al.~\cite{mcdermott2020ai4se} present a research roadmap about Artificial Intelligence for Software Engineering (AI4SE) and Software Engineering for Artificial Intelligence (SE4AI), presenting key aspects aiming at enabling traditional systems engineering practice automation (AI4SE), and encourage new systems engineering practices supporting a new wave of automated, adaptive, and learning systems (SE4AI).

\textbf{Literature Review.}
A literature review involves surveying scholarly sources (mainly research publications) on a specific topic, aiming to provide an overview of the state-of-the-art that is further backed up with a critical evaluation of the material.
Except for providing a reflection on the past, it also gives a clear picture of the state of knowledge on the subject that is helpful for guiding future research directions.
Because of the aforementioned benefits, in this work, we have resorted to surveying the literature review papers (instead of the majority of primary publications) presented in the field of mobile software engineering.
Actually, conducting a survey of surveys is not new to the community.
Our fellow researchers have explored this type of study in various domains when the number of primary publications kept increasing until it became difficult to follow the growing body of literature papers in the field.
For example, AI-Zewairi et al.~\cite{al2017agile} have conducted a survey of surveys related to agile software development methodologies, which have gained rigorous attention in the software engineering community with an excessive number of research studies published.
As another example, McNabb et al.~\cite{mcnabb2017survey} have presented to the community a survey of surveys about information visualization, which has also become extremely popular and the number of publications has become increasingly difficult to follow.
Other representative works include the one proposed by Giraldo et al.~\cite{giraldo2017security} who have proposed a survey of surveys on the topic of security and privacy in Cyber-physical systems as well as the one proposed by Chatzimparmpas et al.~\cite{chatzimparmpas2020survey} who have conducted a survey of surveys on the use of visualization for interpreting machine learning models.

\section{Conclusion}
It has been evidenced that summarizing the research roadmap for a given topic is important as it highlights various research opportunities that communicate broad research goals to the community, connects researchers working on individual projects to larger impact opportunities, and helps professional societies and practitioners focus on more strategic goals.
Following this guidance, in this work, we propose to the community a research roadmap about software engineering for OpenHarmony, aiming at creating a synergy for the various stakeholders to work together to make OpenHarmony a successful mobile platform.
Specifically, we have summarized the status quo of OpenHarmony software engineering research, for which we show OpenHarmony research is still in its early stage.
We then highlight the research opportunities by summarizing the gap between OpenHarmony research and Mobile software engineering research, which is summarized through a survey of literature review papers.
After that, we briefly discuss the challenges in order to fill such a gap.

\bibliographystyle{ACM-Reference-Format}
\bibliography{main}


\begin{thebibliography}{129}


\ifx \showCODEN    \undefined \def \showCODEN     #1{\unskip}     \fi
\ifx \showDOI      \undefined \def \showDOI       #1{#1}\fi
\ifx \showISBNx    \undefined \def \showISBNx     #1{\unskip}     \fi
\ifx \showISBNxiii \undefined \def \showISBNxiii  #1{\unskip}     \fi
\ifx \showISSN     \undefined \def \showISSN      #1{\unskip}     \fi
\ifx \showLCCN     \undefined \def \showLCCN      #1{\unskip}     \fi
\ifx \shownote     \undefined \def \shownote      #1{#1}          \fi
\ifx \showarticletitle \undefined \def \showarticletitle #1{#1}   \fi
\ifx \showURL      \undefined \def \showURL       {\relax}        \fi
\providecommand\bibfield[2]{#2}
\providecommand\bibinfo[2]{#2}
\providecommand\natexlab[1]{#1}
\providecommand\showeprint[2][]{arXiv:#2}

\bibitem[Ahmad et~al\mbox{.}(2018)]%
        {ahmad2018perspectives}
\bibfield{author}{\bibinfo{person}{Naveed Ahmad}, \bibinfo{person}{Aimal
  Rextin}, {and} \bibinfo{person}{Um~E Kulsoom}.}
  \bibinfo{year}{2018}\natexlab{}.
\newblock \showarticletitle{Perspectives on usability guidelines for smartphone
  applications: An empirical investigation and systematic literature review}.
\newblock \bibinfo{journal}{\emph{Information and Software Technology}}
  \bibinfo{volume}{94} (\bibinfo{year}{2018}), \bibinfo{pages}{130--149}.
\newblock


\bibitem[Al-Subaihin et~al\mbox{.}(2021)]%
        {8606261}
\bibfield{author}{\bibinfo{person}{Afnan~A. Al-Subaihin},
  \bibinfo{person}{Federica Sarro}, \bibinfo{person}{Sue Black},
  \bibinfo{person}{Licia Capra}, {and} \bibinfo{person}{Mark Harman}.}
  \bibinfo{year}{2021}\natexlab{}.
\newblock \showarticletitle{App Store Effects on Software Engineering
  Practices}.
\newblock \bibinfo{journal}{\emph{IEEE Transactions on Software Engineering}}
  \bibinfo{volume}{47}, \bibinfo{number}{2} (\bibinfo{year}{2021}),
  \bibinfo{pages}{300--319}.
\newblock
\urldef\tempurl%
\url{https://doi.org/10.1109/TSE.2019.2891715}
\showDOI{\tempurl}


\bibitem[Al-Zewairi et~al\mbox{.}(2017)]%
        {al2017agile}
\bibfield{author}{\bibinfo{person}{Malek Al-Zewairi}, \bibinfo{person}{Mariam
  Biltawi}, \bibinfo{person}{Wael Etaiwi}, \bibinfo{person}{Adnan Shaout},
  {et~al\mbox{.}}} \bibinfo{year}{2017}\natexlab{}.
\newblock \showarticletitle{Agile software development methodologies: Survey of
  surveys}.
\newblock \bibinfo{journal}{\emph{Journal of Computer and Communications}}
  \bibinfo{volume}{5}, \bibinfo{number}{05} (\bibinfo{year}{2017}),
  \bibinfo{pages}{74}.
\newblock


\bibitem[Ali et~al\mbox{.}(2021)]%
        {ali2021self}
\bibfield{author}{\bibinfo{person}{Mughees Ali}, \bibinfo{person}{Saif
  Ur~Rehman Khan}, {and} \bibinfo{person}{Shahid Hussain}.}
  \bibinfo{year}{2021}\natexlab{}.
\newblock \showarticletitle{Self-adaptation in smartphone applications: Current
  state-of-the-art techniques, challenges, and future directions}.
\newblock \bibinfo{journal}{\emph{Data \& Knowledge Engineering}}
  \bibinfo{volume}{136} (\bibinfo{year}{2021}), \bibinfo{pages}{101929}.
\newblock


\bibitem[Amalfitano et~al\mbox{.}(2012)]%
        {amalfitano2012using}
\bibfield{author}{\bibinfo{person}{Domenico Amalfitano},
  \bibinfo{person}{Anna~Rita Fasolino}, \bibinfo{person}{Porfirio Tramontana},
  \bibinfo{person}{Salvatore De~Carmine}, {and} \bibinfo{person}{Atif~M
  Memon}.} \bibinfo{year}{2012}\natexlab{}.
\newblock \showarticletitle{Using GUI ripping for automated testing of Android
  applications}. In \bibinfo{booktitle}{\emph{Proceedings of the 27th IEEE/ACM
  International Conference on Automated Software Engineering}}.
  \bibinfo{pages}{258--261}.
\newblock


\bibitem[Aonzo et~al\mbox{.}(2020)]%
        {aonzo2020obfuscapk}
\bibfield{author}{\bibinfo{person}{Simone Aonzo},
  \bibinfo{person}{Gabriel~Claudiu Georgiu}, \bibinfo{person}{Luca Verderame},
  {and} \bibinfo{person}{Alessio Merlo}.} \bibinfo{year}{2020}\natexlab{}.
\newblock \showarticletitle{Obfuscapk: An open-source black-box obfuscation
  tool for Android apps}.
\newblock \bibinfo{journal}{\emph{SoftwareX}}  \bibinfo{volume}{11}
  (\bibinfo{year}{2020}), \bibinfo{pages}{100403}.
\newblock


\bibitem[Arzt et~al\mbox{.}(2014)]%
        {arzt2014flowdroid}
\bibfield{author}{\bibinfo{person}{Steven Arzt}, \bibinfo{person}{Siegfried
  Rasthofer}, \bibinfo{person}{Christian Fritz}, \bibinfo{person}{Eric Bodden},
  \bibinfo{person}{Alexandre Bartel}, \bibinfo{person}{Jacques Klein},
  \bibinfo{person}{Yves Le~Traon}, \bibinfo{person}{Damien Octeau}, {and}
  \bibinfo{person}{Patrick McDaniel}.} \bibinfo{year}{2014}\natexlab{}.
\newblock \showarticletitle{Flowdroid: Precise context, flow, field,
  object-sensitive and lifecycle-aware taint analysis for android apps}.
\newblock \bibinfo{journal}{\emph{Acm Sigplan Notices}} \bibinfo{volume}{49},
  \bibinfo{number}{6} (\bibinfo{year}{2014}), \bibinfo{pages}{259--269}.
\newblock


\bibitem[Au et~al\mbox{.}(2012)]%
        {au2012pscout}
\bibfield{author}{\bibinfo{person}{Kathy Wain~Yee Au}, \bibinfo{person}{Yi~Fan
  Zhou}, \bibinfo{person}{Zhen Huang}, {and} \bibinfo{person}{David Lie}.}
  \bibinfo{year}{2012}\natexlab{}.
\newblock \showarticletitle{Pscout: analyzing the android permission
  specification}. In \bibinfo{booktitle}{\emph{Proceedings of the 2012 ACM
  conference on Computer and communications security}}.
  \bibinfo{pages}{217--228}.
\newblock


\bibitem[Autili et~al\mbox{.}(2021)]%
        {autili2021software}
\bibfield{author}{\bibinfo{person}{Marco Autili}, \bibinfo{person}{Ivano
  Malavolta}, \bibinfo{person}{Alexander Perucci}, \bibinfo{person}{Gian~Luca
  Scoccia}, {and} \bibinfo{person}{Roberto Verdecchia}.}
  \bibinfo{year}{2021}\natexlab{}.
\newblock \showarticletitle{Software engineering techniques for statically
  analyzing mobile apps: research trends, characteristics, and potential for
  industrial adoption}.
\newblock \bibinfo{journal}{\emph{Journal of Internet Services and
  Applications}}  \bibinfo{volume}{12} (\bibinfo{year}{2021}),
  \bibinfo{pages}{1--60}.
\newblock


\bibitem[Azim and Neamtiu(2013)]%
        {azim2013targeted}
\bibfield{author}{\bibinfo{person}{Tanzirul Azim} {and} \bibinfo{person}{Iulian
  Neamtiu}.} \bibinfo{year}{2013}\natexlab{}.
\newblock \showarticletitle{Targeted and depth-first exploration for systematic
  testing of android apps}. In \bibinfo{booktitle}{\emph{Proceedings of the
  2013 ACM SIGPLAN international conference on Object oriented programming
  systems languages \& applications}}. \bibinfo{pages}{641--660}.
\newblock


\bibitem[Barmpatsalou et~al\mbox{.}(2018)]%
        {barmpatsalou2018current}
\bibfield{author}{\bibinfo{person}{Konstantia Barmpatsalou},
  \bibinfo{person}{Tiago Cruz}, \bibinfo{person}{Edmundo Monteiro}, {and}
  \bibinfo{person}{Paulo Simoes}.} \bibinfo{year}{2018}\natexlab{}.
\newblock \showarticletitle{Current and future trends in mobile device
  forensics: A survey}.
\newblock \bibinfo{journal}{\emph{ACM Computing Surveys (CSUR)}}
  \bibinfo{volume}{51}, \bibinfo{number}{3} (\bibinfo{year}{2018}),
  \bibinfo{pages}{1--31}.
\newblock


\bibitem[Bartel et~al\mbox{.}(2012)]%
        {bartel2012dexpler}
\bibfield{author}{\bibinfo{person}{Alexandre Bartel}, \bibinfo{person}{Jacques
  Klein}, \bibinfo{person}{Yves Le~Traon}, {and} \bibinfo{person}{Martin
  Monperrus}.} \bibinfo{year}{2012}\natexlab{}.
\newblock \showarticletitle{Dexpler: converting android dalvik bytecode to
  jimple for static analysis with soot}. In
  \bibinfo{booktitle}{\emph{Proceedings of the ACM SIGPLAN International
  Workshop on State of the Art in Java Program analysis}}.
  \bibinfo{pages}{27--38}.
\newblock


\bibitem[Bartel et~al\mbox{.}(2014)]%
        {bartel2014static}
\bibfield{author}{\bibinfo{person}{Alexandre Bartel}, \bibinfo{person}{Jacques
  Klein}, \bibinfo{person}{Martin Monperrus}, {and} \bibinfo{person}{Yves
  Le~Traon}.} \bibinfo{year}{2014}\natexlab{}.
\newblock \showarticletitle{Static Analysis for Extracting Permission Checks of
  a Large Scale Framework: The Challenges And Solutions for Analyzing Android}.
\newblock \bibinfo{journal}{\emph{IEEE Transactions on Software Engineering
  (TSE)}} (\bibinfo{year}{2014}).
\newblock


\bibitem[Beresford et~al\mbox{.}(2011)]%
        {beresford2011mockdroid}
\bibfield{author}{\bibinfo{person}{Alastair~R Beresford},
  \bibinfo{person}{Andrew Rice}, \bibinfo{person}{Nicholas Skehin}, {and}
  \bibinfo{person}{Ripduman Sohan}.} \bibinfo{year}{2011}\natexlab{}.
\newblock \showarticletitle{Mockdroid: trading privacy for application
  functionality on smartphones}. In \bibinfo{booktitle}{\emph{Proceedings of
  the 12th workshop on mobile computing systems and applications}}.
  \bibinfo{pages}{49--54}.
\newblock


\bibitem[Bi{\o}rn-Hansen et~al\mbox{.}(2018)]%
        {biorn2018survey}
\bibfield{author}{\bibinfo{person}{Andreas Bi{\o}rn-Hansen},
  \bibinfo{person}{Tor-Morten Gr{\o}nli}, {and} \bibinfo{person}{Gheorghita
  Ghinea}.} \bibinfo{year}{2018}\natexlab{}.
\newblock \showarticletitle{A survey and taxonomy of core concepts and research
  challenges in cross-platform mobile development}.
\newblock \bibinfo{journal}{\emph{ACM Computing Surveys (CSUR)}}
  \bibinfo{volume}{51}, \bibinfo{number}{5} (\bibinfo{year}{2018}),
  \bibinfo{pages}{1--34}.
\newblock


\bibitem[Brereton et~al\mbox{.}(2007)]%
        {brereton2007lessons}
\bibfield{author}{\bibinfo{person}{Pearl Brereton}, \bibinfo{person}{Barbara~A
  Kitchenham}, \bibinfo{person}{David Budgen}, \bibinfo{person}{Mark Turner},
  {and} \bibinfo{person}{Mohamed Khalil}.} \bibinfo{year}{2007}\natexlab{}.
\newblock \showarticletitle{Lessons from applying the systematic literature
  review process within the software engineering domain}.
\newblock \bibinfo{journal}{\emph{Journal of systems and software}}
  \bibinfo{volume}{80}, \bibinfo{number}{4} (\bibinfo{year}{2007}),
  \bibinfo{pages}{571--583}.
\newblock


\bibitem[Bunian et~al\mbox{.}(2021)]%
        {bunian2021vins}
\bibfield{author}{\bibinfo{person}{Sara Bunian}, \bibinfo{person}{Kai Li},
  \bibinfo{person}{Chaima Jemmali}, \bibinfo{person}{Casper Harteveld},
  \bibinfo{person}{Yun Fu}, {and} \bibinfo{person}{Magy~Seif Seif El-Nasr}.}
  \bibinfo{year}{2021}\natexlab{}.
\newblock \showarticletitle{Vins: Visual search for mobile user interface
  design}. In \bibinfo{booktitle}{\emph{Proceedings of the 2021 CHI Conference
  on Human Factors in Computing Systems}}. \bibinfo{pages}{1--14}.
\newblock


\bibitem[C. et~al\mbox{.}(2020)]%
        {10.1145/3417986}
\bibfield{author}{\bibinfo{person}{Marimuthu C.}, \bibinfo{person}{K.
  Chandrasekaran}, {and} \bibinfo{person}{Sridhar Chimalakonda}.}
  \bibinfo{year}{2020}\natexlab{}.
\newblock \showarticletitle{Energy Diagnosis of Android Applications: A
  Thematic Taxonomy and Survey}.
\newblock \bibinfo{journal}{\emph{ACM Comput. Surv.}} \bibinfo{volume}{53},
  \bibinfo{number}{6}, Article \bibinfo{articleno}{117} (\bibinfo{date}{dec}
  \bibinfo{year}{2020}), \bibinfo{numpages}{36}~pages.
\newblock
\showISSN{0360-0300}
\urldef\tempurl%
\url{https://doi.org/10.1145/3417986}
\showDOI{\tempurl}


\bibitem[Cao and Abdelzaher(2006)]%
        {cao2006liteos}
\bibfield{author}{\bibinfo{person}{Qing Cao} {and} \bibinfo{person}{Tarek
  Abdelzaher}.} \bibinfo{year}{2006}\natexlab{}.
\newblock \showarticletitle{LiteOS: a lightweight operating system for C++
  software development in sensor networks}. In
  \bibinfo{booktitle}{\emph{Proceedings of the 4th international conference on
  Embedded networked sensor systems}}. \bibinfo{pages}{361--362}.
\newblock


\bibitem[Chatzimparmpas et~al\mbox{.}(2020)]%
        {chatzimparmpas2020survey}
\bibfield{author}{\bibinfo{person}{Angelos Chatzimparmpas},
  \bibinfo{person}{Rafael~M Martins}, \bibinfo{person}{Ilir Jusufi}, {and}
  \bibinfo{person}{Andreas Kerren}.} \bibinfo{year}{2020}\natexlab{}.
\newblock \showarticletitle{A survey of surveys on the use of visualization for
  interpreting machine learning models}.
\newblock \bibinfo{journal}{\emph{Information Visualization}}
  \bibinfo{volume}{19}, \bibinfo{number}{3} (\bibinfo{year}{2020}),
  \bibinfo{pages}{207--233}.
\newblock


\bibitem[Chen et~al\mbox{.}(2022)]%
        {chen2022grounded}
\bibfield{author}{\bibinfo{person}{Tinggui Chen}, \bibinfo{person}{Chu Zhang},
  \bibinfo{person}{Jianjun Yang}, {and} \bibinfo{person}{Guodong Cong}.}
  \bibinfo{year}{2022}\natexlab{}.
\newblock \showarticletitle{Grounded Theory-Based User Needs Mining and Its
  Impact on APP Downloads: Exampled With WeChat APP}.
\newblock \bibinfo{journal}{\emph{Frontiers in Psychology}}
  \bibinfo{volume}{13} (\bibinfo{year}{2022}), \bibinfo{pages}{875310}.
\newblock


\bibitem[Cheng et~al\mbox{.}({[n.\,d.]})]%
        {chengsoftware}
\bibfield{author}{\bibinfo{person}{Betty~HC Cheng},
  \bibinfo{person}{Rog{\'e}rio de Lemos}, \bibinfo{person}{Holger Giese},
  \bibinfo{person}{Paola Inverardi}, {and} \bibinfo{person}{Jeff Magee}.}
  \bibinfo{year}{[n.\,d.]}\natexlab{}.
\newblock \showarticletitle{Software Engineering for Self-Adaptive Systems: A
  Research Roadmap}.
\newblock  (\bibinfo{year}{[n.\,d.]}).
\newblock


\bibitem[Costamagna and Zheng(2016)]%
        {costamagna2016artdroid}
\bibfield{author}{\bibinfo{person}{Valerio Costamagna} {and}
  \bibinfo{person}{Cong Zheng}.} \bibinfo{year}{2016}\natexlab{}.
\newblock \showarticletitle{Artdroid: A virtual-method hooking framework on
  android art runtime.}. In \bibinfo{booktitle}{\emph{IMPS@ ESSoS}}.
  \bibinfo{pages}{20--28}.
\newblock


\bibitem[Crussell et~al\mbox{.}(2013)]%
        {crussell2013andarwin}
\bibfield{author}{\bibinfo{person}{Jonathan Crussell}, \bibinfo{person}{Clint
  Gibler}, {and} \bibinfo{person}{Hao Chen}.} \bibinfo{year}{2013}\natexlab{}.
\newblock \showarticletitle{Andarwin: Scalable detection of semantically
  similar android applications}. In \bibinfo{booktitle}{\emph{Computer
  Security--ESORICS 2013: 18th European Symposium on Research in Computer
  Security, Egham, UK, September 9-13, 2013. Proceedings 18}}. Springer,
  \bibinfo{pages}{182--199}.
\newblock


\bibitem[De~Lemos et~al\mbox{.}(2013)]%
        {de2013software}
\bibfield{author}{\bibinfo{person}{Rog{\'e}rio De~Lemos},
  \bibinfo{person}{Holger Giese}, \bibinfo{person}{Hausi~A M{\"u}ller},
  \bibinfo{person}{Mary Shaw}, \bibinfo{person}{Jesper Andersson},
  \bibinfo{person}{Marin Litoiu}, \bibinfo{person}{Bradley Schmerl},
  \bibinfo{person}{Gabriel Tamura}, \bibinfo{person}{Norha~M Villegas},
  \bibinfo{person}{Thomas Vogel}, {et~al\mbox{.}}}
  \bibinfo{year}{2013}\natexlab{}.
\newblock \showarticletitle{Software engineering for self-adaptive systems: A
  second research roadmap}. In \bibinfo{booktitle}{\emph{Software Engineering
  for Self-Adaptive Systems II: International Seminar, Dagstuhl Castle,
  Germany, October 24-29, 2010 Revised Selected and Invited Papers}}. Springer,
  \bibinfo{pages}{1--32}.
\newblock


\bibitem[De~Munk and Malavolta(2021)]%
        {de2021measurement}
\bibfield{author}{\bibinfo{person}{Omar De~Munk} {and} \bibinfo{person}{Ivano
  Malavolta}.} \bibinfo{year}{2021}\natexlab{}.
\newblock \showarticletitle{Measurement-based experiments on the mobile web: A
  systematic mapping study}.
\newblock \bibinfo{journal}{\emph{Evaluation and Assessment in Software
  Engineering}} (\bibinfo{year}{2021}), \bibinfo{pages}{191--200}.
\newblock


\bibitem[Delgado-Santos et~al\mbox{.}(2022)]%
        {10.1145/3510579}
\bibfield{author}{\bibinfo{person}{Paula Delgado-Santos},
  \bibinfo{person}{Giuseppe Stragapede}, \bibinfo{person}{Ruben Tolosana},
  \bibinfo{person}{Richard Guest}, \bibinfo{person}{Farzin Deravi}, {and}
  \bibinfo{person}{Ruben Vera-Rodriguez}.} \bibinfo{year}{2022}\natexlab{}.
\newblock \showarticletitle{A Survey of Privacy Vulnerabilities of Mobile
  Device Sensors}.
\newblock \bibinfo{journal}{\emph{ACM Comput. Surv.}} \bibinfo{volume}{54},
  \bibinfo{number}{11s}, Article \bibinfo{articleno}{224} (\bibinfo{date}{sep}
  \bibinfo{year}{2022}), \bibinfo{numpages}{30}~pages.
\newblock
\showISSN{0360-0300}
\urldef\tempurl%
\url{https://doi.org/10.1145/3510579}
\showDOI{\tempurl}


\bibitem[Dong et~al\mbox{.}(2018b)]%
        {dong2018mobile}
\bibfield{author}{\bibinfo{person}{Feng Dong}, \bibinfo{person}{Haoyu Wang},
  \bibinfo{person}{Li Li}, \bibinfo{person}{Yao Guo}, \bibinfo{person}{Guoai
  Xu}, {and} \bibinfo{person}{Shaodong Zhang}.}
  \bibinfo{year}{2018}\natexlab{b}.
\newblock \showarticletitle{How Do Mobile Apps Violate the Behavioral Policy of
  Advertisement Libraries?}. In \bibinfo{booktitle}{\emph{The 19th Workshop on
  Mobile Computing Systems and Applications (HotMobile 2018)}}.
\newblock


\bibitem[Dong et~al\mbox{.}(2018a)]%
        {dong2018understanding}
\bibfield{author}{\bibinfo{person}{Shuaike Dong}, \bibinfo{person}{Menghao Li},
  \bibinfo{person}{Wenrui Diao}, \bibinfo{person}{Xiangyu Liu},
  \bibinfo{person}{Jian Liu}, \bibinfo{person}{Zhou Li},
  \bibinfo{person}{Fenghao Xu}, \bibinfo{person}{Kai Chen},
  \bibinfo{person}{Xiaofeng Wang}, {and} \bibinfo{person}{Kehuan Zhang}.}
  \bibinfo{year}{2018}\natexlab{a}.
\newblock \showarticletitle{Understanding android obfuscation techniques: A
  large-scale investigation in the wild}. In \bibinfo{booktitle}{\emph{Security
  and Privacy in Communication Networks: 14th International Conference,
  SecureComm 2018, Singapore, Singapore, August 8-10, 2018, Proceedings, Part
  I}}. Springer, \bibinfo{pages}{172--192}.
\newblock


\bibitem[Ebrahimi et~al\mbox{.}(2021)]%
        {EBRAHIMI2021106466}
\bibfield{author}{\bibinfo{person}{Fahimeh Ebrahimi}, \bibinfo{person}{Miroslav
  Tushev}, {and} \bibinfo{person}{Anas Mahmoud}.}
  \bibinfo{year}{2021}\natexlab{}.
\newblock \showarticletitle{Mobile app privacy in software engineering
  research: A systematic mapping study}.
\newblock \bibinfo{journal}{\emph{Information and Software Technology}}
  \bibinfo{volume}{133} (\bibinfo{year}{2021}), \bibinfo{pages}{106466}.
\newblock
\showISSN{0950-5849}
\urldef\tempurl%
\url{https://doi.org/10.1016/j.infsof.2020.106466}
\showDOI{\tempurl}


\bibitem[Egele et~al\mbox{.}(2011)]%
        {egele2011pios}
\bibfield{author}{\bibinfo{person}{Manuel Egele}, \bibinfo{person}{Christopher
  Kruegel}, \bibinfo{person}{Engin Kirda}, {and} \bibinfo{person}{Giovanni
  Vigna}.} \bibinfo{year}{2011}\natexlab{}.
\newblock \showarticletitle{Pios: Detecting privacy leaks in ios
  applications.}. In \bibinfo{booktitle}{\emph{NDSS}}.
  \bibinfo{pages}{177--183}.
\newblock


\bibitem[Enck et~al\mbox{.}(2014)]%
        {enck2014taintdroid}
\bibfield{author}{\bibinfo{person}{William Enck}, \bibinfo{person}{Peter
  Gilbert}, \bibinfo{person}{Seungyeop Han}, \bibinfo{person}{Vasant
  Tendulkar}, \bibinfo{person}{Byung-Gon Chun}, \bibinfo{person}{Landon~P Cox},
  \bibinfo{person}{Jaeyeon Jung}, \bibinfo{person}{Patrick McDaniel}, {and}
  \bibinfo{person}{Anmol~N Sheth}.} \bibinfo{year}{2014}\natexlab{}.
\newblock \showarticletitle{Taintdroid: an information-flow tracking system for
  realtime privacy monitoring on smartphones}.
\newblock \bibinfo{journal}{\emph{ACM Transactions on Computer Systems (TOCS)}}
  \bibinfo{volume}{32}, \bibinfo{number}{2} (\bibinfo{year}{2014}),
  \bibinfo{pages}{1--29}.
\newblock


\bibitem[Fan et~al\mbox{.}(2020)]%
        {9251060}
\bibfield{author}{\bibinfo{person}{Ming Fan}, \bibinfo{person}{Le Yu},
  \bibinfo{person}{Sen Chen}, \bibinfo{person}{Hao Zhou},
  \bibinfo{person}{Xiapu Luo}, \bibinfo{person}{Shuyue Li},
  \bibinfo{person}{Yang Liu}, \bibinfo{person}{Jun Liu}, {and}
  \bibinfo{person}{Ting Liu}.} \bibinfo{year}{2020}\natexlab{}.
\newblock \showarticletitle{An Empirical Evaluation of GDPR Compliance
  Violations in Android mHealth Apps}. In \bibinfo{booktitle}{\emph{2020 IEEE
  31st International Symposium on Software Reliability Engineering (ISSRE)}}.
  \bibinfo{pages}{253--264}.
\newblock
\urldef\tempurl%
\url{https://doi.org/10.1109/ISSRE5003.2020.00032}
\showDOI{\tempurl}


\bibitem[France and Rumpe(2007)]%
        {france2007model}
\bibfield{author}{\bibinfo{person}{Robert France} {and}
  \bibinfo{person}{Bernhard Rumpe}.} \bibinfo{year}{2007}\natexlab{}.
\newblock \showarticletitle{Model-driven development of complex software: A
  research roadmap}. In \bibinfo{booktitle}{\emph{Future of Software
  Engineering (FOSE'07)}}. IEEE, \bibinfo{pages}{37--54}.
\newblock


\bibitem[Francese et~al\mbox{.}(2017)]%
        {francese2017mobile}
\bibfield{author}{\bibinfo{person}{Rita Francese}, \bibinfo{person}{Carmine
  Gravino}, \bibinfo{person}{Michele Risi}, \bibinfo{person}{Giuseppe
  Scanniello}, {and} \bibinfo{person}{Genoveffa Tortora}.}
  \bibinfo{year}{2017}\natexlab{}.
\newblock \showarticletitle{Mobile app development and management: results from
  a qualitative investigation}. In \bibinfo{booktitle}{\emph{2017 IEEE/ACM 4th
  International Conference on Mobile Software Engineering and Systems
  (MOBILESoft)}}. IEEE, \bibinfo{pages}{133--143}.
\newblock


\bibitem[Gao et~al\mbox{.}(2019a)]%
        {gao22019evolution}
\bibfield{author}{\bibinfo{person}{Jun Gao}, \bibinfo{person}{Li Li},
  \bibinfo{person}{Tegawend{\'e}~F Bissyand{\'e}}, {and}
  \bibinfo{person}{Jacques Klein}.} \bibinfo{year}{2019}\natexlab{a}.
\newblock \showarticletitle{On the Evolution of Mobile App Complexity}. In
  \bibinfo{booktitle}{\emph{The 24th International Conference on Engineering of
  Complex Computer Systems (ICECCS 2019)}}.
\newblock


\bibitem[Gao et~al\mbox{.}(2019b)]%
        {gao2019understanding}
\bibfield{author}{\bibinfo{person}{Jun Gao}, \bibinfo{person}{Li Li},
  \bibinfo{person}{Pingfan Kong}, \bibinfo{person}{Tegawend{\'e}~F
  Bissyand{\'e}}, {and} \bibinfo{person}{Jacques Klein}.}
  \bibinfo{year}{2019}\natexlab{b}.
\newblock \showarticletitle{Understanding the Evolution of Android App
  Vulnerabilities}.
\newblock \bibinfo{journal}{\emph{IEEE Transactions on Reliability (TRel)}}
  (\bibinfo{year}{2019}).
\newblock


\bibitem[Gao et~al\mbox{.}(2021)]%
        {gao2021demystifying}
\bibfield{author}{\bibinfo{person}{Yuhao Gao}, \bibinfo{person}{Haoyu Wang},
  \bibinfo{person}{Li Li}, \bibinfo{person}{Xiapu Luo},
  \bibinfo{person}{Xuanzhe Liu}, {and} \bibinfo{person}{Guoai Xu}.}
  \bibinfo{year}{2021}\natexlab{}.
\newblock \showarticletitle{Demystifying Illegal Mobile Gambling Apps}. In
  \bibinfo{booktitle}{\emph{The Web Conference 2021 (WWW 2021)}}.
\newblock


\bibitem[Ge et~al\mbox{.}(2022)]%
        {ge2022leveraging}
\bibfield{author}{\bibinfo{person}{Xiuting Ge}, \bibinfo{person}{Shengcheng
  Yu}, \bibinfo{person}{Chunrong Fang}, \bibinfo{person}{Qi Zhu}, {and}
  \bibinfo{person}{Zhihong Zhao}.} \bibinfo{year}{2022}\natexlab{}.
\newblock \showarticletitle{Leveraging android automated testing to assist
  crowdsourced testing}.
\newblock \bibinfo{journal}{\emph{IEEE Transactions on Software Engineering}}
  \bibinfo{volume}{49}, \bibinfo{number}{4} (\bibinfo{year}{2022}),
  \bibinfo{pages}{2318--2336}.
\newblock


\bibitem[Genc-Nayebi and Abran(2017)]%
        {genc2017systematic}
\bibfield{author}{\bibinfo{person}{Necmiye Genc-Nayebi} {and}
  \bibinfo{person}{Alain Abran}.} \bibinfo{year}{2017}\natexlab{}.
\newblock \showarticletitle{A systematic literature review: Opinion mining
  studies from mobile app store user reviews}.
\newblock \bibinfo{journal}{\emph{Journal of Systems and Software}}
  \bibinfo{volume}{125} (\bibinfo{year}{2017}), \bibinfo{pages}{207--219}.
\newblock


\bibitem[Giraldo et~al\mbox{.}(2017)]%
        {giraldo2017security}
\bibfield{author}{\bibinfo{person}{Jairo Giraldo}, \bibinfo{person}{Esha
  Sarkar}, \bibinfo{person}{Alvaro~A Cardenas}, \bibinfo{person}{Michail
  Maniatakos}, {and} \bibinfo{person}{Murat Kantarcioglu}.}
  \bibinfo{year}{2017}\natexlab{}.
\newblock \showarticletitle{Security and privacy in cyber-physical systems: A
  survey of surveys}.
\newblock \bibinfo{journal}{\emph{IEEE Design \& Test}} \bibinfo{volume}{34},
  \bibinfo{number}{4} (\bibinfo{year}{2017}), \bibinfo{pages}{7--17}.
\newblock


\bibitem[Gomez et~al\mbox{.}(2013)]%
        {gomez2013reran}
\bibfield{author}{\bibinfo{person}{Lorenzo Gomez}, \bibinfo{person}{Iulian
  Neamtiu}, \bibinfo{person}{Tanzirul Azim}, {and} \bibinfo{person}{Todd
  Millstein}.} \bibinfo{year}{2013}\natexlab{}.
\newblock \showarticletitle{Reran: Timing-and touch-sensitive record and replay
  for android}. In \bibinfo{booktitle}{\emph{2013 35th International Conference
  on Software Engineering (ICSE)}}. IEEE, \bibinfo{pages}{72--81}.
\newblock


\bibitem[Gorla et~al\mbox{.}(2014)]%
        {gorla2014checking}
\bibfield{author}{\bibinfo{person}{Alessandra Gorla}, \bibinfo{person}{Ilaria
  Tavecchia}, \bibinfo{person}{Florian Gross}, {and} \bibinfo{person}{Andreas
  Zeller}.} \bibinfo{year}{2014}\natexlab{}.
\newblock \showarticletitle{Checking App Behavior against App Descriptions}. In
  \bibinfo{booktitle}{\emph{Proceedings of the 36th International Conference on
  Software Engineering}} (Hyderabad, India) \emph{(\bibinfo{series}{ICSE
  2014})}. \bibinfo{publisher}{Association for Computing Machinery},
  \bibinfo{address}{New York, NY, USA}, \bibinfo{pages}{1025–1035}.
\newblock
\showISBNx{9781450327565}
\urldef\tempurl%
\url{https://doi.org/10.1145/2568225.2568276}
\showDOI{\tempurl}


\bibitem[Guo et~al\mbox{.}(2019)]%
        {guo2019sara}
\bibfield{author}{\bibinfo{person}{Jiaqi Guo}, \bibinfo{person}{Shuyue Li},
  \bibinfo{person}{Jian-Guang Lou}, \bibinfo{person}{Zijiang Yang}, {and}
  \bibinfo{person}{Ting Liu}.} \bibinfo{year}{2019}\natexlab{}.
\newblock \showarticletitle{Sara: self-replay augmented record and replay for
  android in industrial cases}. In \bibinfo{booktitle}{\emph{Proceedings of the
  28th acm sigsoft international symposium on software testing and analysis}}.
  \bibinfo{pages}{90--100}.
\newblock


\bibitem[Hort et~al\mbox{.}(2021)]%
        {hort2021survey}
\bibfield{author}{\bibinfo{person}{Max Hort}, \bibinfo{person}{Maria Kechagia},
  \bibinfo{person}{Federica Sarro}, {and} \bibinfo{person}{Mark Harman}.}
  \bibinfo{year}{2021}\natexlab{}.
\newblock \showarticletitle{A survey of performance optimization for mobile
  applications}.
\newblock \bibinfo{journal}{\emph{IEEE Transactions on Software Engineering}}
  \bibinfo{volume}{48}, \bibinfo{number}{8} (\bibinfo{year}{2021}),
  \bibinfo{pages}{2879--2904}.
\newblock


\bibitem[Hoseini-Tabatabaei et~al\mbox{.}(2013)]%
        {hoseini2013survey}
\bibfield{author}{\bibinfo{person}{Seyed~Amir Hoseini-Tabatabaei},
  \bibinfo{person}{Alexander Gluhak}, {and} \bibinfo{person}{Rahim Tafazolli}.}
  \bibinfo{year}{2013}\natexlab{}.
\newblock \showarticletitle{A survey on smartphone-based systems for
  opportunistic user context recognition}.
\newblock \bibinfo{journal}{\emph{ACM Computing Surveys (CSUR)}}
  \bibinfo{volume}{45}, \bibinfo{number}{3} (\bibinfo{year}{2013}),
  \bibinfo{pages}{1--51}.
\newblock


\bibitem[Hu et~al\mbox{.}(2019)]%
        {hu2019dating}
\bibfield{author}{\bibinfo{person}{Yangyu Hu}, \bibinfo{person}{Haoyu Wang},
  \bibinfo{person}{Yajin Zhou}, \bibinfo{person}{Yao Guo}, \bibinfo{person}{Li
  Li}, \bibinfo{person}{Bingxuan Luo}, {and} \bibinfo{person}{Fangren Xu}.}
  \bibinfo{year}{2019}\natexlab{}.
\newblock \showarticletitle{Dating with Scambots: Understanding the Ecosystem
  of Fraudulent Dating Applications}.
\newblock \bibinfo{journal}{\emph{IEEE Transactions on Dependable and Secure
  Computing (TDSC)}} (\bibinfo{year}{2019}).
\newblock


\bibitem[Jabangwe et~al\mbox{.}(2018)]%
        {jabangwe2018software}
\bibfield{author}{\bibinfo{person}{Ronald Jabangwe}, \bibinfo{person}{Henry
  Edison}, {and} \bibinfo{person}{Anh~Nguyen Duc}.}
  \bibinfo{year}{2018}\natexlab{}.
\newblock \showarticletitle{Software engineering process models for mobile app
  development: A systematic literature review}.
\newblock \bibinfo{journal}{\emph{Journal of Systems and Software}}
  \bibinfo{volume}{145} (\bibinfo{year}{2018}), \bibinfo{pages}{98--111}.
\newblock


\bibitem[Joorabchi et~al\mbox{.}(2013)]%
        {joorabchi2013real}
\bibfield{author}{\bibinfo{person}{Mona~Erfani Joorabchi}, \bibinfo{person}{Ali
  Mesbah}, {and} \bibinfo{person}{Philippe Kruchten}.}
  \bibinfo{year}{2013}\natexlab{}.
\newblock \showarticletitle{Real challenges in mobile app development}. In
  \bibinfo{booktitle}{\emph{2013 ACM/IEEE International Symposium on Empirical
  Software Engineering and Measurement}}. IEEE, \bibinfo{pages}{15--24}.
\newblock


\bibitem[J\'{u}nior et~al\mbox{.}(2022)]%
        {10.1145/3507903}
\bibfield{author}{\bibinfo{person}{Misael~C. J\'{u}nior},
  \bibinfo{person}{Domenico Amalfitano}, \bibinfo{person}{Lina Garc\'{e}s},
  \bibinfo{person}{Anna~Rita Fasolino}, \bibinfo{person}{Stev\~{a}o~A.
  Andrade}, {and} \bibinfo{person}{M\'{a}rcio Delamaro}.}
  \bibinfo{year}{2022}\natexlab{}.
\newblock \showarticletitle{Dynamic Testing Techniques of Non-Functional
  Requirements in Mobile Apps: A Systematic Mapping Study}.
\newblock \bibinfo{journal}{\emph{ACM Comput. Surv.}} \bibinfo{volume}{54},
  \bibinfo{number}{10s}, Article \bibinfo{articleno}{214} (\bibinfo{date}{sep}
  \bibinfo{year}{2022}), \bibinfo{numpages}{38}~pages.
\newblock
\showISSN{0360-0300}
\urldef\tempurl%
\url{https://doi.org/10.1145/3507903}
\showDOI{\tempurl}


\bibitem[Kaur and Kaur(2019)]%
        {kaur2019investigation}
\bibfield{author}{\bibinfo{person}{Anureet Kaur} {and} \bibinfo{person}{Kulwant
  Kaur}.} \bibinfo{year}{2019}\natexlab{}.
\newblock \showarticletitle{Investigation on test effort estimation of mobile
  applications: Systematic literature review and survey}.
\newblock \bibinfo{journal}{\emph{Information and Software technology}}
  \bibinfo{volume}{110} (\bibinfo{year}{2019}), \bibinfo{pages}{56--77}.
\newblock


\bibitem[Kim et~al\mbox{.}(2018b)]%
        {kim2018identifying}
\bibfield{author}{\bibinfo{person}{Seoyeon Kim}, \bibinfo{person}{Jisu Park},
  \bibinfo{person}{Jinman Jung}, \bibinfo{person}{Seongbae Eun},
  \bibinfo{person}{Y-S Yun}, \bibinfo{person}{S So}, \bibinfo{person}{B Kim},
  \bibinfo{person}{H Min}, {and} \bibinfo{person}{J Heo}.}
  \bibinfo{year}{2018}\natexlab{b}.
\newblock \showarticletitle{Identifying UI widgets of mobile applications from
  sketch images}.
\newblock  (\bibinfo{year}{2018}).
\newblock


\bibitem[Kim et~al\mbox{.}(2018a)]%
        {8330039}
\bibfield{author}{\bibinfo{person}{Young~Geun Kim}, \bibinfo{person}{Joonho
  Kong}, {and} \bibinfo{person}{Sung~Woo Chung}.}
  \bibinfo{year}{2018}\natexlab{a}.
\newblock \showarticletitle{A Survey on Recent OS-Level Energy Management
  Techniques for Mobile Processing Units}.
\newblock \bibinfo{journal}{\emph{IEEE Transactions on Parallel and Distributed
  Systems}} \bibinfo{volume}{29}, \bibinfo{number}{10} (\bibinfo{year}{2018}),
  \bibinfo{pages}{2388--2401}.
\newblock
\urldef\tempurl%
\url{https://doi.org/10.1109/TPDS.2018.2822683}
\showDOI{\tempurl}


\bibitem[Kong et~al\mbox{.}(2018)]%
        {kong2018automated}
\bibfield{author}{\bibinfo{person}{Pingfan Kong}, \bibinfo{person}{Li Li},
  \bibinfo{person}{Jun Gao}, \bibinfo{person}{Kui Liu},
  \bibinfo{person}{Tegawend{\'e}~F Bissyand{\'e}}, {and}
  \bibinfo{person}{Jacques Klein}.} \bibinfo{year}{2018}\natexlab{}.
\newblock \showarticletitle{Automated Testing of Android Apps: A Systematic
  Literature Review}.
\newblock \bibinfo{journal}{\emph{IEEE Transactions on Reliability}}
  (\bibinfo{year}{2018}).
\newblock


\bibitem[Kong et~al\mbox{.}(2019)]%
        {8453877}
\bibfield{author}{\bibinfo{person}{Pingfan Kong}, \bibinfo{person}{Li Li},
  \bibinfo{person}{Jun Gao}, \bibinfo{person}{Kui Liu},
  \bibinfo{person}{Tegawendé~F. Bissyandé}, {and} \bibinfo{person}{Jacques
  Klein}.} \bibinfo{year}{2019}\natexlab{}.
\newblock \showarticletitle{Automated Testing of Android Apps: A Systematic
  Literature Review}.
\newblock \bibinfo{journal}{\emph{IEEE Transactions on Reliability}}
  \bibinfo{volume}{68}, \bibinfo{number}{1} (\bibinfo{year}{2019}),
  \bibinfo{pages}{45--66}.
\newblock
\urldef\tempurl%
\url{https://doi.org/10.1109/TR.2018.2865733}
\showDOI{\tempurl}


\bibitem[Lam et~al\mbox{.}(2011)]%
        {lam2011soot}
\bibfield{author}{\bibinfo{person}{Patrick Lam}, \bibinfo{person}{Eric Bodden},
  \bibinfo{person}{Ondrej Lhot{\'a}k}, {and} \bibinfo{person}{Laurie Hendren}.}
  \bibinfo{year}{2011}\natexlab{}.
\newblock \showarticletitle{The Soot framework for Java program analysis: a
  retrospective}. In \bibinfo{booktitle}{\emph{Cetus Users and Compiler
  Infastructure Workshop (CETUS 2011)}}, Vol.~\bibinfo{volume}{15}.
\newblock


\bibitem[Lee et~al\mbox{.}(2022)]%
        {10.1145/3530814}
\bibfield{author}{\bibinfo{person}{Hansoo Lee}, \bibinfo{person}{Joonyoung
  Park}, {and} \bibinfo{person}{Uichin Lee}.} \bibinfo{year}{2022}\natexlab{}.
\newblock \showarticletitle{A Systematic Survey on Android API Usage for
  Data-Driven Analytics with Smartphones}.
\newblock \bibinfo{journal}{\emph{ACM Comput. Surv.}} \bibinfo{volume}{55},
  \bibinfo{number}{5}, Article \bibinfo{articleno}{104} (\bibinfo{date}{dec}
  \bibinfo{year}{2022}), \bibinfo{numpages}{38}~pages.
\newblock
\showISSN{0360-0300}
\urldef\tempurl%
\url{https://doi.org/10.1145/3530814}
\showDOI{\tempurl}


\bibitem[Li et~al\mbox{.}(2015)]%
        {li2015iccta}
\bibfield{author}{\bibinfo{person}{Li Li}, \bibinfo{person}{Alexandre Bartel},
  \bibinfo{person}{Tegawend{\'e}~F Bissyand{\'e}}, \bibinfo{person}{Jacques
  Klein}, \bibinfo{person}{Yves Le~Traon}, \bibinfo{person}{Steven Arzt},
  \bibinfo{person}{Siegfried Rasthofer}, \bibinfo{person}{Eric Bodden},
  \bibinfo{person}{Damien Octeau}, {and} \bibinfo{person}{Patrick Mcdaniel}.}
  \bibinfo{year}{2015}\natexlab{}.
\newblock \showarticletitle{{IccTA: Detecting Inter-Component Privacy Leaks in
  Android Apps}}. In \bibinfo{booktitle}{\emph{Proceedings of the 37th
  International Conference on Software Engineering (ICSE 2015)}}.
\newblock


\bibitem[Li et~al\mbox{.}(2017a)]%
        {li2017simidroid}
\bibfield{author}{\bibinfo{person}{Li Li}, \bibinfo{person}{Tegawend{\'e}~F
  Bissyand{\'e}}, {and} \bibinfo{person}{Jacques Klein}.}
  \bibinfo{year}{2017}\natexlab{a}.
\newblock \showarticletitle{SimiDroid: Identifying and Explaining Similarities
  in Android Apps}. In \bibinfo{booktitle}{\emph{The 16th IEEE International
  Conference On Trust, Security And Privacy In Computing And Communications
  (TrustCom 2017)}}.
\newblock


\bibitem[Li et~al\mbox{.}(2019a)]%
        {li2019rebooting}
\bibfield{author}{\bibinfo{person}{Li Li}, \bibinfo{person}{Tegawend{\'e}~F
  Bissyand{\'e}}, {and} \bibinfo{person}{Jacques Klein}.}
  \bibinfo{year}{2019}\natexlab{a}.
\newblock \showarticletitle{Rebooting research on detecting repackaged android
  apps: Literature review and benchmark}.
\newblock \bibinfo{journal}{\emph{IEEE Transactions on Software Engineering}}
  \bibinfo{volume}{47}, \bibinfo{number}{4} (\bibinfo{year}{2019}),
  \bibinfo{pages}{676--693}.
\newblock


\bibitem[Li et~al\mbox{.}(2016a)]%
        {li2016accessing}
\bibfield{author}{\bibinfo{person}{Li Li}, \bibinfo{person}{Tegawend{\'e}~F
  Bissyand{\'e}}, \bibinfo{person}{Yves Le~Traon}, {and}
  \bibinfo{person}{Jacques Klein}.} \bibinfo{year}{2016}\natexlab{a}.
\newblock \showarticletitle{Accessing Inaccessible Android APIs: An Empirical
  Study}. In \bibinfo{booktitle}{\emph{The 32nd International Conference on
  Software Maintenance and Evolution (ICSME 2016)}}.
\newblock


\bibitem[Li et~al\mbox{.}(2016b)]%
        {li2016droidra}
\bibfield{author}{\bibinfo{person}{Li Li}, \bibinfo{person}{Tegawend{\'e}~F
  Bissyand{\'e}}, \bibinfo{person}{Damien Octeau}, {and}
  \bibinfo{person}{Jacques Klein}.} \bibinfo{year}{2016}\natexlab{b}.
\newblock \showarticletitle{DroidRA: Taming Reflection to Support Whole-Program
  Analysis of Android Apps}. In \bibinfo{booktitle}{\emph{The 2016
  International Symposium on Software Testing and Analysis (ISSTA 2016)}}.
\newblock


\bibitem[Li et~al\mbox{.}(2016c)]%
        {li2016reflection}
\bibfield{author}{\bibinfo{person}{Li Li}, \bibinfo{person}{Tegawend{\'e}~F
  Bissyand{\'e}}, \bibinfo{person}{Damien Octeau}, {and}
  \bibinfo{person}{Jacques Klein}.} \bibinfo{year}{2016}\natexlab{c}.
\newblock \showarticletitle{Reflection-Aware Static Analysis of Android Apps}.
  In \bibinfo{booktitle}{\emph{The 31st IEEE/ACM International Conference on
  Automated Software Engineering, Demo Track (ASE 2016)}}.
\newblock


\bibitem[Li et~al\mbox{.}(2017b)]%
        {li2017static}
\bibfield{author}{\bibinfo{person}{Li Li}, \bibinfo{person}{Tegawend{\'e}~F
  Bissyand{\'e}}, \bibinfo{person}{Mike Papadakis}, \bibinfo{person}{Siegfried
  Rasthofer}, \bibinfo{person}{Alexandre Bartel}, \bibinfo{person}{Damien
  Octeau}, \bibinfo{person}{Jacques Klein}, {and} \bibinfo{person}{Yves
  Le~Traon}.} \bibinfo{year}{2017}\natexlab{b}.
\newblock \showarticletitle{Static Analysis of Android Apps: A Systematic
  Literature Review}.
\newblock \bibinfo{journal}{\emph{Information and Software Technology}}
  (\bibinfo{year}{2017}).
\newblock


\bibitem[Li et~al\mbox{.}(2019b)]%
        {li2019identifying}
\bibfield{author}{\bibinfo{person}{Li Li}, \bibinfo{person}{Tegawend{\'e}~F
  Bissyand{\'e}}, \bibinfo{person}{Haoyu Wang}, {and} \bibinfo{person}{Jacques
  Klein}.} \bibinfo{year}{2019}\natexlab{b}.
\newblock \showarticletitle{On Identifying and Explaining Similarities in
  Android Apps}.
\newblock \bibinfo{journal}{\emph{Journal of Computer Science and Technology
  (JCST)}} (\bibinfo{year}{2019}).
\newblock


\bibitem[Li et~al\mbox{.}(2020)]%
        {li2020cda}
\bibfield{author}{\bibinfo{person}{Li Li}, \bibinfo{person}{Jun Gao},
  \bibinfo{person}{Tegawend{\'e}~F Bissyand{\'e}}, \bibinfo{person}{Lei Ma},
  \bibinfo{person}{Xin Xia}, {and} \bibinfo{person}{Jacques Klein}.}
  \bibinfo{year}{2020}\natexlab{}.
\newblock \showarticletitle{CDA: Characterising Deprecated Android APIs}.
\newblock \bibinfo{journal}{\emph{Empirical Software Engineering (EMSE)}}
  (\bibinfo{year}{2020}).
\newblock


\bibitem[Li et~al\mbox{.}(2017c)]%
        {li2017androzoo++}
\bibfield{author}{\bibinfo{person}{Li Li}, \bibinfo{person}{Jun Gao},
  \bibinfo{person}{M{\'e}d{\'e}ric Hurier}, \bibinfo{person}{Pingfan Kong},
  \bibinfo{person}{Tegawend{\'e}~F Bissyand{\'e}}, \bibinfo{person}{Alexandre
  Bartel}, \bibinfo{person}{Jacques Klein}, {and} \bibinfo{person}{Yves
  Le~Traon}.} \bibinfo{year}{2017}\natexlab{c}.
\newblock \showarticletitle{AndroZoo++: Collecting Millions of Android Apps and
  Their Metadata for the Research Community}.
\newblock \bibinfo{journal}{\emph{arXiv preprint arXiv:1709.05281}}
  (\bibinfo{year}{2017}).
\newblock


\bibitem[Li et~al\mbox{.}(2017d)]%
        {li2017xketch}
\bibfield{author}{\bibinfo{person}{Shu-Hui Li}, \bibinfo{person}{Jia-Jyun Hsu},
  \bibinfo{person}{Chih-Ya Chang}, \bibinfo{person}{Pin-Hsuan Chen}, {and}
  \bibinfo{person}{Neng-Hao Yu}.} \bibinfo{year}{2017}\natexlab{d}.
\newblock \showarticletitle{Xketch: A sketch-based prototyping tool to
  accelerate mobile app design process}. In
  \bibinfo{booktitle}{\emph{Proceedings of the 2017 ACM Conference Companion
  Publication on Designing Interactive Systems}}. \bibinfo{pages}{301--304}.
\newblock


\bibitem[Li et~al\mbox{.}(2017e)]%
        {li2017droidbot}
\bibfield{author}{\bibinfo{person}{Yuanchun Li}, \bibinfo{person}{Ziyue Yang},
  \bibinfo{person}{Yao Guo}, {and} \bibinfo{person}{Xiangqun Chen}.}
  \bibinfo{year}{2017}\natexlab{e}.
\newblock \showarticletitle{Droidbot: a lightweight ui-guided test input
  generator for android}. In \bibinfo{booktitle}{\emph{2017 IEEE/ACM 39th
  International Conference on Software Engineering Companion (ICSE-C)}}. IEEE,
  \bibinfo{pages}{23--26}.
\newblock


\bibitem[Liu et~al\mbox{.}(2022a)]%
        {liu2022customized}
\bibfield{author}{\bibinfo{person}{Pei Liu}, \bibinfo{person}{Mattia Fazzini},
  \bibinfo{person}{John Grundy}, {and} \bibinfo{person}{Li Li}.}
  \bibinfo{year}{2022}\natexlab{a}.
\newblock \showarticletitle{Do Customized Android Frameworks Keep Pace with
  Android?}. In \bibinfo{booktitle}{\emph{The 19th International Conference on
  Mining Software Repositories (MSR 2022)}}.
\newblock


\bibitem[Liu et~al\mbox{.}(2021)]%
        {liu2021identifying}
\bibfield{author}{\bibinfo{person}{Pei Liu}, \bibinfo{person}{Li Li},
  \bibinfo{person}{Yichun Yan}, \bibinfo{person}{Mattia Fazzini}, {and}
  \bibinfo{person}{John Grundy}.} \bibinfo{year}{2021}\natexlab{}.
\newblock \showarticletitle{Identifying and Characterizing Silently-Evolved
  Methods in the Android API}. In \bibinfo{booktitle}{\emph{The 43rd ACM/IEEE
  International Conference on Software Engineering, SEIP Track (ICSE-SEIP
  2021)}}.
\newblock


\bibitem[Liu et~al\mbox{.}(2020)]%
        {liu2020androzooopen}
\bibfield{author}{\bibinfo{person}{Pei Liu}, \bibinfo{person}{Li Li},
  \bibinfo{person}{Yanjie Zhao}, \bibinfo{person}{Xiaoyu Sun}, {and}
  \bibinfo{person}{John Grundy}.} \bibinfo{year}{2020}\natexlab{}.
\newblock \showarticletitle{AndroZooOpen: Collecting Large-scale Open Source
  Android Apps for the Research Community}. In \bibinfo{booktitle}{\emph{The
  2020 International Conference on Mining Software Repositories, Data Track
  (MSR 2020)}}.
\newblock


\bibitem[LIU et~al\mbox{.}(2023)]%
        {liu2023automatically}
\bibfield{author}{\bibinfo{person}{PEI LIU}, \bibinfo{person}{YANJIE ZHAO},
  \bibinfo{person}{MATTIA FAZZINI}, \bibinfo{person}{HAIPENG CAI},
  \bibinfo{person}{JOHN GRUNDY}, {and} \bibinfo{person}{LI LI}.}
  \bibinfo{year}{2023}\natexlab{}.
\newblock \showarticletitle{Automatically Detecting Incompatible Android APIs}.
\newblock \bibinfo{journal}{\emph{IEEE Transactions on Software Engineering}}
  (\bibinfo{year}{2023}).
\newblock


\bibitem[Liu et~al\mbox{.}(2022b)]%
        {10.1145/3544968}
\bibfield{author}{\bibinfo{person}{Yue Liu}, \bibinfo{person}{Chakkrit
  Tantithamthavorn}, \bibinfo{person}{Li Li}, {and} \bibinfo{person}{Yepang
  Liu}.} \bibinfo{year}{2022}\natexlab{b}.
\newblock \showarticletitle{Deep Learning for Android Malware Defenses: A
  Systematic Literature Review}.
\newblock \bibinfo{journal}{\emph{ACM Comput. Surv.}} \bibinfo{volume}{55},
  \bibinfo{number}{8}, Article \bibinfo{articleno}{153} (\bibinfo{date}{dec}
  \bibinfo{year}{2022}), \bibinfo{numpages}{36}~pages.
\newblock
\showISSN{0360-0300}
\urldef\tempurl%
\url{https://doi.org/10.1145/3544968}
\showDOI{\tempurl}


\bibitem[Liu et~al\mbox{.}(2022c)]%
        {liu2022deep}
\bibfield{author}{\bibinfo{person}{Yue Liu}, \bibinfo{person}{Chakkrit
  Tantithamthavorn}, \bibinfo{person}{Li Li}, {and} \bibinfo{person}{Yepang
  Liu}.} \bibinfo{year}{2022}\natexlab{c}.
\newblock \showarticletitle{Deep Learning for Android Malware Defenses: a
  Systematic Literature Review}.
\newblock \bibinfo{journal}{\emph{ACM Computing Surveys (CSUR)}}
  (\bibinfo{year}{2022}).
\newblock


\bibitem[Liu et~al\mbox{.}(2023)]%
        {liu2023ex}
\bibfield{author}{\bibinfo{person}{Zhe Liu}, \bibinfo{person}{Chunyang Chen},
  \bibinfo{person}{Junjie Wang}, \bibinfo{person}{Yuhui Su},
  \bibinfo{person}{Yuekai Huang}, \bibinfo{person}{Jun Hu}, {and}
  \bibinfo{person}{Qing Wang}.} \bibinfo{year}{2023}\natexlab{}.
\newblock \showarticletitle{Ex pede Herculem: Augmenting Activity Transition
  Graph for Apps via Graph Convolution Network}. In
  \bibinfo{booktitle}{\emph{2023 IEEE/ACM 45th International Conference on
  Software Engineering (ICSE)}}. IEEE, \bibinfo{pages}{1983--1995}.
\newblock


\bibitem[Luo et~al\mbox{.}(2020)]%
        {10.1145/3372788}
\bibfield{author}{\bibinfo{person}{Chu Luo}, \bibinfo{person}{Jorge Goncalves},
  \bibinfo{person}{Eduardo Velloso}, {and} \bibinfo{person}{Vassilis
  Kostakos}.} \bibinfo{year}{2020}\natexlab{}.
\newblock \showarticletitle{A Survey of Context Simulation for Testing Mobile
  Context-Aware Applications}.
\newblock \bibinfo{journal}{\emph{ACM Comput. Surv.}} \bibinfo{volume}{53},
  \bibinfo{number}{1}, Article \bibinfo{articleno}{21} (\bibinfo{date}{feb}
  \bibinfo{year}{2020}), \bibinfo{numpages}{39}~pages.
\newblock
\showISSN{0360-0300}
\urldef\tempurl%
\url{https://doi.org/10.1145/3372788}
\showDOI{\tempurl}


\bibitem[Luo et~al\mbox{.}(2019)]%
        {luo2019tainting}
\bibfield{author}{\bibinfo{person}{Lannan Luo}, \bibinfo{person}{Qiang Zeng},
  \bibinfo{person}{Chen Cao}, \bibinfo{person}{Kai Chen}, \bibinfo{person}{Jian
  Liu}, \bibinfo{person}{Limin Liu}, \bibinfo{person}{Neng Gao},
  \bibinfo{person}{Min Yang}, \bibinfo{person}{Xinyu Xing}, {and}
  \bibinfo{person}{Peng Liu}.} \bibinfo{year}{2019}\natexlab{}.
\newblock \showarticletitle{Tainting-assisted and context-migrated symbolic
  execution of Android framework for vulnerability discovery and exploit
  generation}.
\newblock \bibinfo{journal}{\emph{IEEE Transactions on Mobile Computing}}
  \bibinfo{volume}{19}, \bibinfo{number}{12} (\bibinfo{year}{2019}),
  \bibinfo{pages}{2946--2964}.
\newblock


\bibitem[Marginean et~al\mbox{.}(2019)]%
        {marginean2019sapfix}
\bibfield{author}{\bibinfo{person}{Alexandru Marginean},
  \bibinfo{person}{Johannes Bader}, \bibinfo{person}{Satish Chandra},
  \bibinfo{person}{Mark Harman}, \bibinfo{person}{Yue Jia}, \bibinfo{person}{Ke
  Mao}, \bibinfo{person}{Alexander Mols}, {and} \bibinfo{person}{Andrew
  Scott}.} \bibinfo{year}{2019}\natexlab{}.
\newblock \showarticletitle{Sapfix: Automated end-to-end repair at scale}. In
  \bibinfo{booktitle}{\emph{2019 IEEE/ACM 41st International Conference on
  Software Engineering: Software Engineering in Practice (ICSE-SEIP)}}. IEEE,
  \bibinfo{pages}{269--278}.
\newblock


\bibitem[Martin et~al\mbox{.}(2016)]%
        {martin2016survey}
\bibfield{author}{\bibinfo{person}{William Martin}, \bibinfo{person}{Federica
  Sarro}, \bibinfo{person}{Yue Jia}, \bibinfo{person}{Yuanyuan Zhang}, {and}
  \bibinfo{person}{Mark Harman}.} \bibinfo{year}{2016}\natexlab{}.
\newblock \showarticletitle{A survey of app store analysis for software
  engineering}.
\newblock \bibinfo{journal}{\emph{IEEE transactions on software engineering}}
  \bibinfo{volume}{43}, \bibinfo{number}{9} (\bibinfo{year}{2016}),
  \bibinfo{pages}{817--847}.
\newblock


\bibitem[McDermott et~al\mbox{.}(2020)]%
        {mcdermott2020ai4se}
\bibfield{author}{\bibinfo{person}{Tom McDermott}, \bibinfo{person}{Dan
  DeLaurentis}, \bibinfo{person}{Peter Beling}, \bibinfo{person}{Mark
  Blackburn}, {and} \bibinfo{person}{Mary Bone}.}
  \bibinfo{year}{2020}\natexlab{}.
\newblock \showarticletitle{AI4SE and SE4AI: A research roadmap}.
\newblock \bibinfo{journal}{\emph{Insight}} \bibinfo{volume}{23},
  \bibinfo{number}{1} (\bibinfo{year}{2020}), \bibinfo{pages}{8--14}.
\newblock


\bibitem[McNabb and Laramee(2017)]%
        {mcnabb2017survey}
\bibfield{author}{\bibinfo{person}{Liam McNabb} {and} \bibinfo{person}{Robert~S
  Laramee}.} \bibinfo{year}{2017}\natexlab{}.
\newblock \showarticletitle{Survey of Surveys (SoS)-mapping the landscape of
  survey papers in information visualization}. In
  \bibinfo{booktitle}{\emph{computer graphics forum}},
  Vol.~\bibinfo{volume}{36}. Wiley Online Library, \bibinfo{pages}{589--617}.
\newblock


\bibitem[Nakamura et~al\mbox{.}(2022)]%
        {nakamura2022factors}
\bibfield{author}{\bibinfo{person}{Walter~T Nakamura},
  \bibinfo{person}{Edson~Cesar de Oliveira}, \bibinfo{person}{Elaine~HT de
  Oliveira}, \bibinfo{person}{David Redmiles}, {and} \bibinfo{person}{Tayana
  Conte}.} \bibinfo{year}{2022}\natexlab{}.
\newblock \showarticletitle{What factors affect the UX in mobile apps? A
  systematic mapping study on the analysis of app store reviews}.
\newblock \bibinfo{journal}{\emph{Journal of Systems and Software}}
  \bibinfo{volume}{193} (\bibinfo{year}{2022}), \bibinfo{pages}{111462}.
\newblock


\bibitem[Obie et~al\mbox{.}(2021)]%
        {obie2021first}
\bibfield{author}{\bibinfo{person}{Humphrey Obie}, \bibinfo{person}{Waqar
  Hussain}, \bibinfo{person}{Xin Xia}, \bibinfo{person}{John Grundy},
  \bibinfo{person}{Li Li}, \bibinfo{person}{Burak Turhan}, \bibinfo{person}{Jon
  Whittle}, {and} \bibinfo{person}{Mojtaba Shahin}.}
  \bibinfo{year}{2021}\natexlab{}.
\newblock \showarticletitle{A First Look at Human Values-Violation in App
  Reviews}. In \bibinfo{booktitle}{\emph{The 43rd ACM/IEEE International
  Conference on Software Engineering, SEIS Track (ICSE-SEIS 2021)}}.
\newblock


\bibitem[Obie et~al\mbox{.}(2022)]%
        {obie2022violation}
\bibfield{author}{\bibinfo{person}{Humphrey Obie}, \bibinfo{person}{Idowu
  Ilekura}, \bibinfo{person}{Hung Du}, \bibinfo{person}{Mojtaba Shahin},
  \bibinfo{person}{John Grundy}, \bibinfo{person}{Li Li}, \bibinfo{person}{Jon
  Whittle}, {and} \bibinfo{person}{Burak Turhan}.}
  \bibinfo{year}{2022}\natexlab{}.
\newblock \showarticletitle{On the Violation of Honesty in Mobile Apps:
  Automated Detection and Categories}. In \bibinfo{booktitle}{\emph{The 19th
  International Conference on Mining Software Repositories (MSR 2022)}}.
\newblock


\bibitem[Palomba et~al\mbox{.}(2018)]%
        {palomba2018crowdsourcing}
\bibfield{author}{\bibinfo{person}{Fabio Palomba}, \bibinfo{person}{Mario
  Linares-V{\'a}squez}, \bibinfo{person}{Gabriele Bavota},
  \bibinfo{person}{Rocco Oliveto}, \bibinfo{person}{Massimiliano Di~Penta},
  \bibinfo{person}{Denys Poshyvanyk}, {and} \bibinfo{person}{Andrea De~Lucia}.}
  \bibinfo{year}{2018}\natexlab{}.
\newblock \showarticletitle{Crowdsourcing user reviews to support the evolution
  of mobile apps}.
\newblock \bibinfo{journal}{\emph{Journal of Systems and Software}}
  \bibinfo{volume}{137} (\bibinfo{year}{2018}), \bibinfo{pages}{143--162}.
\newblock


\bibitem[Papazoglou et~al\mbox{.}(2008)]%
        {papazoglou2008service}
\bibfield{author}{\bibinfo{person}{Michael~P Papazoglou},
  \bibinfo{person}{Paolo Traverso}, \bibinfo{person}{Schahram Dustdar}, {and}
  \bibinfo{person}{Frank Leymann}.} \bibinfo{year}{2008}\natexlab{}.
\newblock \showarticletitle{Service-oriented computing: a research roadmap}.
\newblock \bibinfo{journal}{\emph{International Journal of Cooperative
  Information Systems}} \bibinfo{volume}{17}, \bibinfo{number}{02}
  (\bibinfo{year}{2008}), \bibinfo{pages}{223--255}.
\newblock


\bibitem[Pereira and Rodrigues(2013)]%
        {pereira2013survey}
\bibfield{author}{\bibinfo{person}{Orlando~RE Pereira} {and}
  \bibinfo{person}{Joel~JPC Rodrigues}.} \bibinfo{year}{2013}\natexlab{}.
\newblock \showarticletitle{Survey and analysis of current mobile learning
  applications and technologies}.
\newblock \bibinfo{journal}{\emph{ACM Computing Surveys (CSUR)}}
  \bibinfo{volume}{46}, \bibinfo{number}{2} (\bibinfo{year}{2013}),
  \bibinfo{pages}{1--35}.
\newblock


\bibitem[Qiu et~al\mbox{.}(2022)]%
        {qiu2022re}
\bibfield{author}{\bibinfo{person}{Fangze Qiu}, \bibinfo{person}{Huaxiao
  Huang}, {and} \bibinfo{person}{Yuji Dong}.} \bibinfo{year}{2022}\natexlab{}.
\newblock \showarticletitle{A Re-configurable Interaction Model in Distributed
  IoT Environment}. In \bibinfo{booktitle}{\emph{2022 International Conference
  on Cyber-Enabled Distributed Computing and Knowledge Discovery (CyberC)}}.
  IEEE, \bibinfo{pages}{80--86}.
\newblock


\bibitem[Qiu et~al\mbox{.}(2020)]%
        {10.1145/3417978}
\bibfield{author}{\bibinfo{person}{Junyang Qiu}, \bibinfo{person}{Jun Zhang},
  \bibinfo{person}{Wei Luo}, \bibinfo{person}{Lei Pan}, \bibinfo{person}{Surya
  Nepal}, {and} \bibinfo{person}{Yang Xiang}.} \bibinfo{year}{2020}\natexlab{}.
\newblock \showarticletitle{A Survey of Android Malware Detection with Deep
  Neural Models}.
\newblock \bibinfo{journal}{\emph{ACM Comput. Surv.}} \bibinfo{volume}{53},
  \bibinfo{number}{6}, Article \bibinfo{articleno}{126} (\bibinfo{date}{dec}
  \bibinfo{year}{2020}), \bibinfo{numpages}{36}~pages.
\newblock
\showISSN{0360-0300}
\urldef\tempurl%
\url{https://doi.org/10.1145/3417978}
\showDOI{\tempurl}


\bibitem[Rasthofer et~al\mbox{.}(2014)]%
        {rasthofer2014machine}
\bibfield{author}{\bibinfo{person}{Siegfried Rasthofer},
  \bibinfo{person}{Steven Arzt}, {and} \bibinfo{person}{Eric Bodden}.}
  \bibinfo{year}{2014}\natexlab{}.
\newblock \showarticletitle{A Machine-learning Approach for Classifying and
  Categorizing Android Sources and Sinks}.
\newblock \bibinfo{journal}{\emph{NDSS}} (\bibinfo{year}{2014}).
\newblock


\bibitem[Rasthofer et~al\mbox{.}(2017)]%
        {rasthofer2017making}
\bibfield{author}{\bibinfo{person}{Siegfried Rasthofer},
  \bibinfo{person}{Steven Arzt}, \bibinfo{person}{Stefan Triller}, {and}
  \bibinfo{person}{Michael Pradel}.} \bibinfo{year}{2017}\natexlab{}.
\newblock \showarticletitle{Making malory behave maliciously: Targeted fuzzing
  of android execution environments}. In \bibinfo{booktitle}{\emph{2017
  IEEE/ACM 39th International Conference on Software Engineering (ICSE)}}.
  IEEE, \bibinfo{pages}{300--311}.
\newblock


\bibitem[Russello et~al\mbox{.}(2013)]%
        {russello2013firedroid}
\bibfield{author}{\bibinfo{person}{Giovanni Russello},
  \bibinfo{person}{Arturo~Blas Jimenez}, \bibinfo{person}{Habib Naderi}, {and}
  \bibinfo{person}{Wannes van~der Mark}.} \bibinfo{year}{2013}\natexlab{}.
\newblock \showarticletitle{Firedroid: Hardening security in almost-stock
  android}. In \bibinfo{booktitle}{\emph{Proceedings of the 29th Annual
  Computer Security Applications Conference}}. \bibinfo{pages}{319--328}.
\newblock


\bibitem[Samhi et~al\mbox{.}(2022)]%
        {samhi2022jucify}
\bibfield{author}{\bibinfo{person}{Jordan Samhi}, \bibinfo{person}{Jun Gao},
  \bibinfo{person}{Nadia Daoudi}, \bibinfo{person}{Pierre Graux},
  \bibinfo{person}{Henri Hoyez}, \bibinfo{person}{Xiaoyu Sun},
  \bibinfo{person}{Kevin Allix}, \bibinfo{person}{Tegawend{\'e}~F
  Bissyand{\'e}}, {and} \bibinfo{person}{Jacques Klein}.}
  \bibinfo{year}{2022}\natexlab{}.
\newblock \showarticletitle{Jucify: A step towards android code unification for
  enhanced static analysis}. In \bibinfo{booktitle}{\emph{Proceedings of the
  44th International Conference on Software Engineering}}.
  \bibinfo{pages}{1232--1244}.
\newblock


\bibitem[Santos and Dolby(2022)]%
        {santos2022program}
\bibfield{author}{\bibinfo{person}{Joanna Cecilia Da~Silva Santos} {and}
  \bibinfo{person}{Julian Dolby}.} \bibinfo{year}{2022}\natexlab{}.
\newblock \showarticletitle{Program Analysis using WALA}. In
  \bibinfo{booktitle}{\emph{ACM Joint Meeting European Software Engineering
  Conference and Symposium on the Foundations of Software Engineering}}.
\newblock


\bibitem[Senanayake et~al\mbox{.}(2023)]%
        {10.1145/3556974}
\bibfield{author}{\bibinfo{person}{Janaka Senanayake}, \bibinfo{person}{Harsha
  Kalutarage}, \bibinfo{person}{Mhd~Omar Al-Kadri}, \bibinfo{person}{Andrei
  Petrovski}, {and} \bibinfo{person}{Luca Piras}.}
  \bibinfo{year}{2023}\natexlab{}.
\newblock \showarticletitle{Android Source Code Vulnerability Detection: A
  Systematic Literature Review}.
\newblock \bibinfo{journal}{\emph{ACM Comput. Surv.}} \bibinfo{volume}{55},
  \bibinfo{number}{9}, Article \bibinfo{articleno}{187} (\bibinfo{date}{jan}
  \bibinfo{year}{2023}), \bibinfo{numpages}{37}~pages.
\newblock
\showISSN{0360-0300}
\urldef\tempurl%
\url{https://doi.org/10.1145/3556974}
\showDOI{\tempurl}


\bibitem[Shahzad et~al\mbox{.}(2017)]%
        {shahzad2017socio}
\bibfield{author}{\bibinfo{person}{Basit Shahzad},
  \bibinfo{person}{Abdullatif~M Abdullatif}, \bibinfo{person}{Kashif Saleem},
  {and} \bibinfo{person}{Wasif Jameel}.} \bibinfo{year}{2017}\natexlab{}.
\newblock \showarticletitle{Socio-technical challenges and mitigation
  guidelines in developing mobile healthcare applications}.
\newblock \bibinfo{journal}{\emph{Journal of Medical Imaging and Health
  Informatics}} \bibinfo{volume}{7}, \bibinfo{number}{3}
  (\bibinfo{year}{2017}), \bibinfo{pages}{704--712}.
\newblock


\bibitem[Shamsujjoha et~al\mbox{.}(2021)]%
        {shamsujjoha2021developing}
\bibfield{author}{\bibinfo{person}{Md Shamsujjoha}, \bibinfo{person}{John
  Grundy}, \bibinfo{person}{Li Li}, \bibinfo{person}{Hourieh Khalajzadeh},
  {and} \bibinfo{person}{Qinghua Lu}.} \bibinfo{year}{2021}\natexlab{}.
\newblock \showarticletitle{Developing mobile applications via model driven
  development: a systematic literature review}.
\newblock \bibinfo{journal}{\emph{Information and Software Technology}}
  \bibinfo{volume}{140} (\bibinfo{year}{2021}), \bibinfo{pages}{106693}.
\newblock


\bibitem[Silva et~al\mbox{.}(2022)]%
        {silva2022mapping}
\bibfield{author}{\bibinfo{person}{Henrique~Neves Silva},
  \bibinfo{person}{Jackson Prado~Lima}, \bibinfo{person}{Silvia~Regina
  Vergilio}, {and} \bibinfo{person}{Andre~Takeshi Endo}.}
  \bibinfo{year}{2022}\natexlab{}.
\newblock \showarticletitle{A mapping study on mutation testing for mobile
  applications}.
\newblock \bibinfo{journal}{\emph{Software Testing, Verification and
  Reliability}} \bibinfo{volume}{32}, \bibinfo{number}{8}
  (\bibinfo{year}{2022}), \bibinfo{pages}{e1801}.
\newblock


\bibitem[Su et~al\mbox{.}(2017)]%
        {su2017guided}
\bibfield{author}{\bibinfo{person}{Ting Su}, \bibinfo{person}{Guozhu Meng},
  \bibinfo{person}{Yuting Chen}, \bibinfo{person}{Ke Wu},
  \bibinfo{person}{Weiming Yang}, \bibinfo{person}{Yao Yao},
  \bibinfo{person}{Geguang Pu}, \bibinfo{person}{Yang Liu}, {and}
  \bibinfo{person}{Zhendong Su}.} \bibinfo{year}{2017}\natexlab{}.
\newblock \showarticletitle{Guided, stochastic model-based GUI testing of
  Android apps}. In \bibinfo{booktitle}{\emph{Proceedings of the 2017 11th
  Joint Meeting on Foundations of Software Engineering}}.
  \bibinfo{pages}{245--256}.
\newblock


\bibitem[Sufatrio et~al\mbox{.}(2015)]%
        {sufatrio2015securing}
\bibfield{author}{\bibinfo{person}{Sufatrio}, \bibinfo{person}{Darell~JJ Tan},
  \bibinfo{person}{Tong-Wei Chua}, {and} \bibinfo{person}{Vrizlynn~LL Thing}.}
  \bibinfo{year}{2015}\natexlab{}.
\newblock \showarticletitle{Securing android: a survey, taxonomy, and
  challenges}.
\newblock \bibinfo{journal}{\emph{ACM Computing Surveys (CSUR)}}
  \bibinfo{volume}{47}, \bibinfo{number}{4} (\bibinfo{year}{2015}),
  \bibinfo{pages}{1--45}.
\newblock


\bibitem[Sun et~al\mbox{.}(2021)]%
        {sun2021characterizing}
\bibfield{author}{\bibinfo{person}{Xiaoyu Sun}, \bibinfo{person}{Xiao Chen},
  \bibinfo{person}{Kui Liu}, \bibinfo{person}{Sheng Wen}, \bibinfo{person}{Li
  Li}, {and} \bibinfo{person}{John Grundy}.} \bibinfo{year}{2021}\natexlab{}.
\newblock \showarticletitle{Characterizing Sensor Leaks in Android Apps}. In
  \bibinfo{booktitle}{\emph{The 32nd International Symposium on Software
  Reliability Engineering (ISSRE 2021)}}.
\newblock


\bibitem[Sun et~al\mbox{.}({[n.\,d.]})]%
        {sun2023taming}
\bibfield{author}{\bibinfo{person}{Xiaoyu Sun}, \bibinfo{person}{Xiao Chen},
  \bibinfo{person}{Yonghui Liu}, \bibinfo{person}{John Grundy}, {and}
  \bibinfo{person}{year={2023}~publisher={IEEE} Li, journal={IEEE Transactions
  on Software Engineering}}.} \bibinfo{year}{[n.\,d.]}\natexlab{}.
\newblock \showarticletitle{Taming Android Fragmentation through Lightweight
  Crowdsourced Testing}.
\newblock  (\bibinfo{year}{[n.\,d.]}).
\newblock


\bibitem[Sun et~al\mbox{.}(2020)]%
        {sun2020taming}
\bibfield{author}{\bibinfo{person}{Xiaoyu Sun}, \bibinfo{person}{Li Li},
  \bibinfo{person}{Tegawend{\'e}~F Bissyand{\'e}}, \bibinfo{person}{Jacques
  Klein}, \bibinfo{person}{Damien Octeau}, {and} \bibinfo{person}{John
  Grundy}.} \bibinfo{year}{2020}\natexlab{}.
\newblock \showarticletitle{Taming Reflection: An Essential Step Towards
  Whole-Program Analysis of Android Apps}.
\newblock \bibinfo{journal}{\emph{ACM Transactions on Software Engineering and
  Methodology (TOSEM)}} (\bibinfo{year}{2020}).
\newblock


\bibitem[Wang et~al\mbox{.}(2018)]%
        {wang2018automated}
\bibfield{author}{\bibinfo{person}{Xiaolei Wang}, \bibinfo{person}{Yuexiang
  Yang}, {and} \bibinfo{person}{Sencun Zhu}.} \bibinfo{year}{2018}\natexlab{}.
\newblock \showarticletitle{Automated hybrid analysis of android malware
  through augmenting fuzzing with forced execution}.
\newblock \bibinfo{journal}{\emph{IEEE Transactions on Mobile Computing}}
  \bibinfo{volume}{18}, \bibinfo{number}{12} (\bibinfo{year}{2018}),
  \bibinfo{pages}{2768--2782}.
\newblock


\bibitem[Wang et~al\mbox{.}(2023)]%
        {wang2023animation2api}
\bibfield{author}{\bibinfo{person}{Yihui Wang}, \bibinfo{person}{Huaxiao Liu},
  \bibinfo{person}{Shanquan Gao}, {and} \bibinfo{person}{Xiao Tang}.}
  \bibinfo{year}{2023}\natexlab{}.
\newblock \showarticletitle{Animation2API: API Recommendation for the
  Implementation of Android UI Animations}.
\newblock \bibinfo{journal}{\emph{IEEE Transactions on Software Engineering}}
  (\bibinfo{year}{2023}).
\newblock


\bibitem[Wang et~al\mbox{.}(2022)]%
        {wang2022runtime}
\bibfield{author}{\bibinfo{person}{Ying Wang}, \bibinfo{person}{Yibo Wang},
  \bibinfo{person}{Sinan Wang}, \bibinfo{person}{Yepang Liu},
  \bibinfo{person}{Chang Xu}, \bibinfo{person}{Shing-Chi Cheung},
  \bibinfo{person}{Hai Yu}, {and} \bibinfo{person}{Zhiliang Zhu}.}
  \bibinfo{year}{2022}\natexlab{}.
\newblock \showarticletitle{Runtime permission issues in android apps:
  Taxonomy, practices, and ways forward}.
\newblock \bibinfo{journal}{\emph{IEEE Transactions on Software Engineering}}
  \bibinfo{volume}{49}, \bibinfo{number}{1} (\bibinfo{year}{2022}),
  \bibinfo{pages}{185--210}.
\newblock


\bibitem[Wei et~al\mbox{.}(2018)]%
        {wei2018jn}
\bibfield{author}{\bibinfo{person}{Fengguo Wei}, \bibinfo{person}{Xingwei Lin},
  \bibinfo{person}{Xinming Ou}, \bibinfo{person}{Ting Chen}, {and}
  \bibinfo{person}{Xiaosong Zhang}.} \bibinfo{year}{2018}\natexlab{}.
\newblock \showarticletitle{Jn-saf: Precise and efficient ndk/jni-aware
  inter-language static analysis framework for security vetting of android
  applications with native code}. In \bibinfo{booktitle}{\emph{Proceedings of
  the 2018 ACM SIGSAC Conference on Computer and Communications Security}}.
  \bibinfo{pages}{1137--1150}.
\newblock


\bibitem[Wimalasooriya et~al\mbox{.}(2022)]%
        {wimalasooriya2022systematic}
\bibfield{author}{\bibinfo{person}{Chathrie Wimalasooriya},
  \bibinfo{person}{Sherlock~A Licorish}, \bibinfo{person}{Daniel~Alencar da
  Costa}, {and} \bibinfo{person}{Stephen~G MacDonell}.}
  \bibinfo{year}{2022}\natexlab{}.
\newblock \showarticletitle{A systematic mapping study addressing the
  reliability of mobile applications: The need to move beyond testing
  reliability}.
\newblock \bibinfo{journal}{\emph{Journal of Systems and Software}}
  \bibinfo{volume}{186} (\bibinfo{year}{2022}), \bibinfo{pages}{111166}.
\newblock


\bibitem[Wu et~al\mbox{.}(2023b)]%
        {wu2023cydios}
\bibfield{author}{\bibinfo{person}{Shuohan Wu}, \bibinfo{person}{Jianfeng Li},
  \bibinfo{person}{Hao Zhou}, \bibinfo{person}{Yongsheng Fang},
  \bibinfo{person}{Kaifa Zhao}, \bibinfo{person}{Haoyu Wang},
  \bibinfo{person}{Chenxiong Qian}, {and} \bibinfo{person}{Xiapu Luo}.}
  \bibinfo{year}{2023}\natexlab{b}.
\newblock \showarticletitle{CydiOS: A Model-Based Testing Framework for iOS
  Apps}. In \bibinfo{booktitle}{\emph{Proceedings of the 32nd ACM SIGSOFT
  International Symposium on Software Testing and Analysis}}.
  \bibinfo{pages}{1--13}.
\newblock


\bibitem[Wu et~al\mbox{.}(2023a)]%
        {wu2023systematic}
\bibfield{author}{\bibinfo{person}{Zhiqiang Wu}, \bibinfo{person}{Xin Chen},
  {and} \bibinfo{person}{Scott Uk-Jin Lee}.} \bibinfo{year}{2023}\natexlab{a}.
\newblock \showarticletitle{A systematic literature review on Android-specific
  smells}.
\newblock \bibinfo{journal}{\emph{Journal of Systems and Software}}
  \bibinfo{volume}{201} (\bibinfo{year}{2023}), \bibinfo{pages}{111677}.
\newblock


\bibitem[Xu et~al\mbox{.}(2016)]%
        {xu2016toward}
\bibfield{author}{\bibinfo{person}{Meng Xu}, \bibinfo{person}{Chengyu Song},
  \bibinfo{person}{Yang Ji}, \bibinfo{person}{Ming-Wei Shih},
  \bibinfo{person}{Kangjie Lu}, \bibinfo{person}{Cong Zheng},
  \bibinfo{person}{Ruian Duan}, \bibinfo{person}{Yeongjin Jang},
  \bibinfo{person}{Byoungyoung Lee}, \bibinfo{person}{Chenxiong Qian},
  {et~al\mbox{.}}} \bibinfo{year}{2016}\natexlab{}.
\newblock \showarticletitle{Toward engineering a secure android ecosystem: A
  survey of existing techniques}.
\newblock \bibinfo{journal}{\emph{ACM Computing Surveys (CSUR)}}
  \bibinfo{volume}{49}, \bibinfo{number}{2} (\bibinfo{year}{2016}),
  \bibinfo{pages}{1--47}.
\newblock


\bibitem[Xue et~al\mbox{.}(2018)]%
        {xue2018ndroid}
\bibfield{author}{\bibinfo{person}{Lei Xue}, \bibinfo{person}{Chenxiong Qian},
  \bibinfo{person}{Hao Zhou}, \bibinfo{person}{Xiapu Luo},
  \bibinfo{person}{Yajin Zhou}, \bibinfo{person}{Yuru Shao}, {and}
  \bibinfo{person}{Alvin~TS Chan}.} \bibinfo{year}{2018}\natexlab{}.
\newblock \showarticletitle{NDroid: Toward tracking information flows across
  multiple Android contexts}.
\newblock \bibinfo{journal}{\emph{IEEE Transactions on Information Forensics
  and Security}} \bibinfo{volume}{14}, \bibinfo{number}{3}
  (\bibinfo{year}{2018}), \bibinfo{pages}{814--828}.
\newblock


\bibitem[Xue et~al\mbox{.}(2021a)]%
        {xue2021parema}
\bibfield{author}{\bibinfo{person}{Lei Xue}, \bibinfo{person}{Yuxiao Yan},
  \bibinfo{person}{Luyi Yan}, \bibinfo{person}{Muhui Jiang},
  \bibinfo{person}{Xiapu Luo}, \bibinfo{person}{Dinghao Wu}, {and}
  \bibinfo{person}{Yajin Zhou}.} \bibinfo{year}{2021}\natexlab{a}.
\newblock \showarticletitle{Parema: an unpacking framework for demystifying
  VM-based Android packers}. In \bibinfo{booktitle}{\emph{Proceedings of the
  30th ACM SIGSOFT International Symposium on Software Testing and Analysis}}.
  \bibinfo{pages}{152--164}.
\newblock


\bibitem[Xue et~al\mbox{.}(2020)]%
        {xue2020packergrind}
\bibfield{author}{\bibinfo{person}{Lei Xue}, \bibinfo{person}{Hao Zhou},
  \bibinfo{person}{Xiapu Luo}, \bibinfo{person}{Le Yu},
  \bibinfo{person}{Dinghao Wu}, \bibinfo{person}{Yajin Zhou}, {and}
  \bibinfo{person}{Xiaobo Ma}.} \bibinfo{year}{2020}\natexlab{}.
\newblock \showarticletitle{Packergrind: An adaptive unpacking system for
  android apps}.
\newblock \bibinfo{journal}{\emph{IEEE Transactions on Software Engineering}}
  \bibinfo{volume}{48}, \bibinfo{number}{2} (\bibinfo{year}{2020}),
  \bibinfo{pages}{551--570}.
\newblock


\bibitem[Xue et~al\mbox{.}(2021b)]%
        {xue2021happer}
\bibfield{author}{\bibinfo{person}{Lei Xue}, \bibinfo{person}{Hao Zhou},
  \bibinfo{person}{Xiapu Luo}, \bibinfo{person}{Yajin Zhou},
  \bibinfo{person}{Yang Shi}, \bibinfo{person}{Guofei Gu},
  \bibinfo{person}{Fengwei Zhang}, {and} \bibinfo{person}{Man~Ho Au}.}
  \bibinfo{year}{2021}\natexlab{b}.
\newblock \showarticletitle{Happer: Unpacking android apps via a
  hardware-assisted approach}. In \bibinfo{booktitle}{\emph{2021 IEEE Symposium
  on Security and Privacy (SP)}}. IEEE, \bibinfo{pages}{1641--1658}.
\newblock


\bibitem[Yasuda et~al\mbox{.}(2020)]%
        {10.1145/3368961}
\bibfield{author}{\bibinfo{person}{Yuri D.~V. Yasuda}, \bibinfo{person}{Luiz
  Eduardo~G. Martins}, {and} \bibinfo{person}{Fabio A.~M. Cappabianco}.}
  \bibinfo{year}{2020}\natexlab{}.
\newblock \showarticletitle{Autonomous Visual Navigation for Mobile Robots: A
  Systematic Literature Review}.
\newblock \bibinfo{journal}{\emph{ACM Comput. Surv.}} \bibinfo{volume}{53},
  \bibinfo{number}{1}, Article \bibinfo{articleno}{13} (\bibinfo{date}{feb}
  \bibinfo{year}{2020}), \bibinfo{numpages}{34}~pages.
\newblock
\showISSN{0360-0300}
\urldef\tempurl%
\url{https://doi.org/10.1145/3368961}
\showDOI{\tempurl}


\bibitem[Yuan et~al\mbox{.}(2019)]%
        {yuan2019api}
\bibfield{author}{\bibinfo{person}{Weizhao Yuan}, \bibinfo{person}{Hoang~H
  Nguyen}, \bibinfo{person}{Lingxiao Jiang}, \bibinfo{person}{Yuting Chen},
  \bibinfo{person}{Jianjun Zhao}, {and} \bibinfo{person}{Haibo Yu}.}
  \bibinfo{year}{2019}\natexlab{}.
\newblock \showarticletitle{API recommendation for event-driven Android
  application development}.
\newblock \bibinfo{journal}{\emph{Information and Software Technology}}
  \bibinfo{volume}{107} (\bibinfo{year}{2019}), \bibinfo{pages}{30--47}.
\newblock


\bibitem[Zein et~al\mbox{.}(2016)]%
        {DBLP:journals/jss/ZeinSG16}
\bibfield{author}{\bibinfo{person}{Samer Zein}, \bibinfo{person}{Norsaremah
  Salleh}, {and} \bibinfo{person}{John Grundy}.}
  \bibinfo{year}{2016}\natexlab{}.
\newblock \showarticletitle{A systematic mapping study of mobile application
  testing techniques}.
\newblock \bibinfo{journal}{\emph{J. Syst. Softw.}}  \bibinfo{volume}{117}
  (\bibinfo{year}{2016}), \bibinfo{pages}{334--356}.
\newblock
\urldef\tempurl%
\url{https://doi.org/10.1016/j.jss.2016.03.065}
\showDOI{\tempurl}


\bibitem[Zhan et~al\mbox{.}(2021)]%
        {zhan2021research}
\bibfield{author}{\bibinfo{person}{Xian Zhan}, \bibinfo{person}{Tianming Liu},
  \bibinfo{person}{Lingling Fan}, \bibinfo{person}{Li Li}, \bibinfo{person}{Sen
  Chen}, \bibinfo{person}{Xiapu Luo}, {and} \bibinfo{person}{Yang Liu}.}
  \bibinfo{year}{2021}\natexlab{}.
\newblock \showarticletitle{Research on third-party libraries in android apps:
  A taxonomy and systematic literature review}.
\newblock \bibinfo{journal}{\emph{IEEE Transactions on Software Engineering}}
  (\bibinfo{year}{2021}).
\newblock


\bibitem[Zhang et~al\mbox{.}(2023)]%
        {zhang2023scene}
\bibfield{author}{\bibinfo{person}{Xiangyu Zhang}, \bibinfo{person}{Lingling
  Fan}, \bibinfo{person}{Sen Chen}, \bibinfo{person}{Yucheng Su}, {and}
  \bibinfo{person}{Boyuan Li}.} \bibinfo{year}{2023}\natexlab{}.
\newblock \showarticletitle{Scene-Driven Exploration and GUI Modeling for
  Android Apps}.
\newblock \bibinfo{journal}{\emph{arXiv preprint arXiv:2308.10228}}
  (\bibinfo{year}{2023}).
\newblock


\bibitem[Zhang et~al\mbox{.}(2015)]%
        {zhang2015dexhunter}
\bibfield{author}{\bibinfo{person}{Yueqian Zhang}, \bibinfo{person}{Xiapu Luo},
  {and} \bibinfo{person}{Haoyang Yin}.} \bibinfo{year}{2015}\natexlab{}.
\newblock \showarticletitle{Dexhunter: toward extracting hidden code from
  packed android applications}. In \bibinfo{booktitle}{\emph{Computer
  Security--ESORICS 2015: 20th European Symposium on Research in Computer
  Security, Vienna, Austria, September 21-25, 2015, Proceedings, Part II 20}}.
  Springer, \bibinfo{pages}{293--311}.
\newblock


\bibitem[Zhao et~al\mbox{.}(2022a)]%
        {zhao2022towards}
\bibfield{author}{\bibinfo{person}{Yanjie Zhao}, \bibinfo{person}{Li Li},
  \bibinfo{person}{Kui Liu}, {and} \bibinfo{person}{John Grundy}.}
  \bibinfo{year}{2022}\natexlab{a}.
\newblock \showarticletitle{Towards Automatically Repairing Compatibility
  Issues in Published Android Apps}. In \bibinfo{booktitle}{\emph{The 44th
  International Conference on Software Engineering (ICSE 2022)}}.
\newblock


\bibitem[Zhao et~al\mbox{.}(2021)]%
        {zhao2021icon2code}
\bibfield{author}{\bibinfo{person}{Yanjie Zhao}, \bibinfo{person}{Li Li},
  \bibinfo{person}{Xiaoyu Sun}, \bibinfo{person}{Pei Liu}, {and}
  \bibinfo{person}{John Grundy}.} \bibinfo{year}{2021}\natexlab{}.
\newblock \showarticletitle{Icon2Code: Recommending code implementations for
  Android GUI components}.
\newblock \bibinfo{journal}{\emph{Information and Software Technology}}
  \bibinfo{volume}{138} (\bibinfo{year}{2021}), \bibinfo{pages}{106619}.
\newblock


\bibitem[Zhao et~al\mbox{.}(2022b)]%
        {zhao2022apimatchmaker}
\bibfield{author}{\bibinfo{person}{Yanjie Zhao}, \bibinfo{person}{Li Li},
  \bibinfo{person}{Haoyu Wang}, \bibinfo{person}{Qiang He}, {and}
  \bibinfo{person}{John Grundy}.} \bibinfo{year}{2022}\natexlab{b}.
\newblock \showarticletitle{APIMatchmaker: Matching the Right APIs for
  Supporting the Development of Android Apps}.
\newblock \bibinfo{journal}{\emph{IEEE Transactions on Software Engineering
  (TSE)}} (\bibinfo{year}{2022}).
\newblock


\bibitem[Zhao et~al\mbox{.}(2023)]%
        {zhao2023mobile}
\bibfield{author}{\bibinfo{person}{Yanjie Zhao}, \bibinfo{person}{Tianming
  Liu}, \bibinfo{person}{Haoyu Wang}, \bibinfo{person}{Yepang Liu},
  \bibinfo{person}{John Grundy}, {and} \bibinfo{person}{Li Li}.}
  \bibinfo{year}{2023}\natexlab{}.
\newblock \showarticletitle{Are Mobile Advertisements in Compliance with
  App’s Age Group?}. In \bibinfo{booktitle}{\emph{Proceedings of the ACM Web
  Conference 2023}}. \bibinfo{pages}{3132--3141}.
\newblock


\bibitem[Zhou et~al\mbox{.}(2022a)]%
        {zhou2022uncovering}
\bibfield{author}{\bibinfo{person}{Hao Zhou}, \bibinfo{person}{Xiapu Luo},
  \bibinfo{person}{Haoyu Wang}, {and} \bibinfo{person}{Haipeng Cai}.}
  \bibinfo{year}{2022}\natexlab{a}.
\newblock \showarticletitle{Uncovering Intent based Leak of Sensitive Data in
  Android Framework}. In \bibinfo{booktitle}{\emph{Proceedings of the 2022 ACM
  SIGSAC Conference on Computer and Communications Security}}.
  \bibinfo{pages}{3239--3252}.
\newblock


\bibitem[Zhou et~al\mbox{.}(2021)]%
        {zhou2021finding}
\bibfield{author}{\bibinfo{person}{Hao Zhou}, \bibinfo{person}{Haoyu Wang},
  \bibinfo{person}{Shuohan Wu}, \bibinfo{person}{Xiapu Luo},
  \bibinfo{person}{Yajin Zhou}, \bibinfo{person}{Ting Chen}, {and}
  \bibinfo{person}{Ting Wang}.} \bibinfo{year}{2021}\natexlab{}.
\newblock \showarticletitle{Finding the missing piece: permission specification
  analysis for Android NDK}. In \bibinfo{booktitle}{\emph{2021 36th IEEE/ACM
  International Conference on Automated Software Engineering (ASE)}}. IEEE,
  \bibinfo{pages}{505--516}.
\newblock


\bibitem[Zhou et~al\mbox{.}(2022b)]%
        {zhou2022ncscope}
\bibfield{author}{\bibinfo{person}{Hao Zhou}, \bibinfo{person}{Shuohan Wu},
  \bibinfo{person}{Xiapu Luo}, \bibinfo{person}{Ting Wang},
  \bibinfo{person}{Yajin Zhou}, \bibinfo{person}{Chao Zhang}, {and}
  \bibinfo{person}{Haipeng Cai}.} \bibinfo{year}{2022}\natexlab{b}.
\newblock \showarticletitle{NCScope: hardware-assisted analyzer for native code
  in Android apps}. In \bibinfo{booktitle}{\emph{Proceedings of the 31st ACM
  SIGSOFT International Symposium on Software Testing and Analysis}}.
  \bibinfo{pages}{629--641}.
\newblock


\end{thebibliography}

\end{document}